\documentclass[a4paper,fleqn,usenatbib]{mnras}

\usepackage{newtxtext,newtxmath}
\usepackage[utf8]{inputenc} 

\usepackage[T1]{fontenc}
\usepackage{ae,aecompl}
\usepackage{subfigure}

\usepackage{graphicx}	
\usepackage{amsmath}	
\usepackage{amssymb}	
\usepackage{longtable}
\usepackage{lscape}

\newcommand{ \Msun } {M$_{\odot}$}
\newcommand{ \Herschel } {\textit{Herschel}}

\newcommand{ \Ntot } {192}

\newcommand{\stud}{Student's \textit{t}-distribution}
\newcommand{\stan}{{\tt Stan}}

\title[JINGLE V: Dust properties]{JINGLE V: Dust properties of nearby galaxies derived from hierarchical Bayesian SED fitting}

\author[I. Lamperti et al.]{\Large \parbox{\textwidth}{
Isabella Lamperti$^{1}$\thanks{E-mail: i.lamperti.16@ucl.ac.uk},
Am{\'e}lie Saintonge$^{1}$,
Ilse De Looze$^{1,2}$,
Gioacchino Accurso$^{1}$,
Christopher J. R. Clark$^{3}$,
Matthew W. L. Smith$^{4}$,
Christine D. Wilson$^{5}$,  
Elias Brinks$^{6}$,
Toby Brown$^{5}$,
Martin Bureau$^{7}$,
David L. Clements$^{8}$,
Stephen Eales$^{4}$,
David H. W. Glass$^{9}$, 
Ho Seong Hwang$^{10}$, 
Jong Chul Lee$^{10}$,
Lihwai Lin$^{11}$,
Michał J. Michałowski$^{12}$,
Mark Sargent$^{13}$,
Thomas G. Williams$^{4}$,
Ting Xiao$^{14}$, and
Chentao Yang$^{15,16,17}$
}
\\
\\
$^{1}$Department of Physics and Astronomy, University College London, Gower Street, London WC1E 6BT, UK\\
$^{2}$Sterrenkundig Observatorium, Universiteit Gent, Krijgslaan 281 S9, B-9000 Gent, Belgium\\
$^{3}$Space Telescope Science Institute, 3700 San Martin Drive, Baltimore, Maryland 21218, USA\\
$^{4}$School of Physics and Astronomy, Cardiff University, Queens Buildings, The Parade, Cardiff, CF24 3AA, UK \\
$^{5}$Department of Physics and Astronomy, McMaster University, Hamilton, ON L8S 4M1, Canada\\
$^{6}$Centre for Astrophysics Research, University of Hertfordshire, College Lane, AL10 9AB, UK \\
$^{7}$Sub-department of Astrophysics, University of Oxford, Denys Wilkinson Building, Keble Road, Oxford, OX1 3RH, UK\\
$^{8}$Blackett Laboratory, Physics Department, Imperial College, London, SW7 2AZ, UK \\
$^{9}$Jeremiah Horrocks Institute, School of Physical Sciences and Computing, University of Central Lancashire, Preston, Lancashire, PR1 2HE, UK\\
$^{10}$Korea Astronomy and Space Science Institute, 776 Daedeokdae-ro, Yuseong-gu, Daejeon 34055, Republic of Korea\\
$^{11}$Institute of Astronomy and Astrophysics, Academia Sinica, P.O. Box 23-141, Taipei 10617, Taiwan \\
$^{12}$Astronomical Observatory Institute, Faculty of Physics, Adam Mickiewicz University, ul. Sloneczna 36, 60-286 Pozna{\'n}, Poland\\
$^{13}$Astronomy Centre, Department of Physics and Astronomy, University of Sussex, Brighton BN1 9QH, England \\
$^{14}$Shanghai Astronomical Observatory, 80 Nandan Road, Xuhui District, Shanghai, China 200030\\
$^{15}$European Southern Observatory, Alonso de C{\'o}rdova 3107, Casilla 19001, Vitacura, Santiago, Chile \\
$^{16}$Purple Mountain Observatory, Chinese Academy of Sciences, Nanjing 210008, China \\
$^{17}$Key Laboratory of Radio Astronomy, Chinese Academy of Sciences, Nanjing 210008, China
}

\date{Accepted 2019 August 15. Received 2019 August 1; in original form 2019 May 24}

\pubyear{2019}

\begin{document}
\label{firstpage}
\pagerange{\pageref{firstpage}--\pageref{lastpage}}
\maketitle

\begin{abstract}
We study the dust properties of \Ntot\ nearby galaxies from the JINGLE  survey using photometric data in the 22-850\micron\ range. We derive the total dust mass, temperature $T$  and emissivity index $\beta$ of the galaxies through the fitting of their spectral energy distribution (SED) using a single modified black-body model (SMBB).
We apply a hierarchical Bayesian approach that reduces the known degeneracy between $T$ and $\beta$. 
Applying the hierarchical approach, the strength of the $T$-$\beta$ anti-correlation is reduced from a Pearson correlation coefficient $R=-0.79$ to $R=-0.52$.
For the JINGLE galaxies we measure dust temperatures in the range $17-30$~K and dust emissivity indices $\beta$ in the range $0.6-2.2$.
  We compare the  SMBB model with the broken emissivity modified black-body (BMBB) and the two modified black-bodies (TMBB) models.
The results derived with the SMBB and TMBB are in good agreement, thus applying the SMBB, which comes with fewer free parameters, does not penalize the measurement of the cold dust properties in the JINGLE sample.
 We investigate the relation between $T$ and $\beta$ and other global galaxy properties in the JINGLE and \Herschel\  Reference Survey (HRS) sample.
  We find that $\beta$ correlates with the stellar mass surface density ($R=0.62$) and anti-correlates with the HI mass fraction ($M_{HI}/M_*$, $R=-0.65$), whereas the dust temperature correlates strongly with the SFR normalized by the dust mass ($R=0.73$). These relations can be used to  estimate $T$ and $\beta$ in galaxies with insufficient photometric data available to measure them directly through SED fitting. 
\end{abstract}

\begin{keywords}
galaxies: evolution -- galaxies: ISM -- ISM: dust, extinction -- submillimetre: ISM 
\end{keywords}



\section{Introduction}
Interstellar dust plays an important role in galaxies: it helps to balance gas heating and cooling and the surface of dust grains provides a favourable place for chemical reactions to occur.
Dust contributes only a small fraction of the mass of the interstellar medium (ISM), but in normal star-forming galaxies it can re-radiate  up to $\sim30\%$ of the stellar light in the infrared \citep[e.g.][]{Clements1996}.

The two main places where dust is formed is in the ejecta of core-collapse supernovae and in the envelopes of asymptotic giant branch (AGB) stars \citep{Galliano2018b}. These two production mechanisms alone however can not account for the amount of dust observed in high redshift galaxies \citep{Bertoldi2003, Priddey2003, Rowlands2014, Watson2015, Michalowski2015}. Grain growth is another mechanism that can  increase the dust content of a galaxy, but it is not well understood how much this process can contribute to the total dust production \citep{Barlow1978c, Ferrara2016, Ceccarelli2018}. In order to resolve this tension, we need first to improve our understanding of all the mechanisms of dust production and growth. Second, it is the necessary to have tools to accurately measure the dust content of distant galaxies and have a good understanding of the uncertainties on these measurements; this is the question this paper tackles.
 
Dust masses are measured by fitting the spectral energy distribution (SED) of galaxies in the far-infrared/sub-millimeter spectral range.
 The standard model used is a modified black-body function (MBB), which depends on the dust mass, temperature ($T$) and emissivity index $\beta$.
An anti-correlation between temperature and $\beta$ has been observed in galactic sources and luminous infrared galaxies \citep{Dupac2003, Desert2008, Yang2007}. However, it has been shown  that noise in the data can introduce an artificial anti-correlation between $T$ and $\beta$ \citep[e.g.][]{Shetty2009a, Shetty2009b}. An incorrect estimate of $T$ and $\beta$ would consequently bias the measurement of the dust mass.
A way to overcome this problem and break the $T-\beta$ degeneracy is to use a hierarchical Bayesian approach
\citep{Kelly2012, Juvela2013, Veneziani2013, Galliano2018}. 
The hierarchical approach uses  the information from the parameter distribution of the entire sample of galaxies to better constrain temperature and $\beta$ for each single galaxy.
 The hierarchical method has the advantage that it does not require knowing the prior distribution of the parameters before the fitting, but can infer the parameters describing the prior directly during the fitting procedure, after assuming the shape of the distribution. The limitation of this is that the prior is only valid for the sample of galaxies under consideration, i.e. the prior depends on the population that one is considering.

The \Herschel\ Space Observatory\footnote{\textit{Herschel} is an ESA space observatory with science instruments provided by European-led Principal Investigator consortia and with important participation from NASA.} \citep{Pilbratt2010} has been key for the study of dust in nearby galaxies, providing photometric observations in the wavelength range 100-500\micron, that allowed to characterize the shape of their far-infrared SED. 
 The \Herschel\ Reference Survey \citep[HRS, ][]{Boselli2010b} is a  guaranteed time program  that  measured the far-infrared SED of $\sim$ 300 nearby galaxies. 
Using HRS galaxies, \cite{Cortese2014} show that their far-infrared and submm colors are inconsistent with a single modified black-body model with the same emissivity index $\beta$ for all galaxies.

Dust continuum observations can also be used to infer the molecular gas mass of a galaxy. 
It has been shown that the dust continuum luminosity of galaxies correlates with the CO luminosity \citep{Hildebrand1983, Magdis2012, Eales2012, Scoville2014, Groves2015} and this relation can be used to infer the molecular gas mass of a galaxy by applying a molecular gas-to-dust ratio. 
 This method can be extremely useful for faint or high-redshift galaxies, since the dust emission is brighter and therefore easier to observe than the CO line emission.
 This method can therefore be beneficial for measuring the molecular gas content of large samples of galaxies.

The JINGLE (JCMT dust and gas In Nearby Galaxies Legacy Exploration),  survey is a large program on the James Clerk Maxwell Telescope (JCMT) which aims to characterize the dust and molecular gas in nearby galaxies and study the relation between the two \citep{Saintonge2018}. 
JINGLE combines dust observations from the SCUBA-2 camera on the JCMT (and from \Herschel), with the cold gas measurements obtained with the JCMT RxA instrument. 
 With both measurements of the dust and cold gas properties for a statistical sample of nearby galaxies, we can study the variations in the dust-to-gas mass ratio as a function of galaxy and dust properties.

One of the objectives of the survey is to benchmark dust scaling relations with other galaxy properties such as stellar mass, metallicity, and star-formation rate.
 These relations can be used to estimate the dust temperature and dust emissivity index in galaxies for which there are not enough photometric data available to measure them directly through SED fitting. This can be useful especially for high redshift galaxies.

An excess of emission at wavelengths $\geq500$\micron\ with respect to the modified black-body model has been observed in numerous dwarf galaxies \citep[e.g.][]{Galametz2011, Remy-Ruyer2013, Remy-Ruyer2015},  in late-type galaxies \citep{Dumke2004, Bendo2006, Galametz2009}, in the Magellanic Clouds \citep{Israel2010, Bot2010b},  and in M33 \citep{Hermelo2016, Relano2018}. The origin of this `submm' excess is still an open question.
The SCUBA-2 observations at 850\micron\ can help to place better constraints on the submm slope and investigate the presence of this excess in the JINGLE sample.

In this paper we take advantage of the large and homogeneous JINGLE sample  and apply a hierarchical Bayesian approach to reduce the $T-\beta$ degeneracy and obtain more accurate measurements of the dust parameters using MBB models.
 The hierarchical approach is crucial to disentangle dust temperature $T$ and emissivity index $\beta$ and allows us for the first time to study the independent relations of these two dust quantities with other galaxy global properties. 

This paper is organised as follows. In Section \ref{sec:sample} we present the sample and  the data used in this paper. Then we describe the classical and hierarchical Bayesian SED fitting methods and compare the two methods using simulated SEDs (Section \ref{sec:method}). Section \ref{sec:results} illustrates the results of the SED fitting of the JINGLE sample, the $T$-$\beta$ relation, and  comparison of different modified black-body models.
In Section \ref{sec:dust_scal_relation} we derive scaling relations between dust quantities and other global galaxy properties. Finally in Section \ref{sec:conclusions} we summarize the main results and our conclusions. Readers who are less interested in the statistical methods and tests of the fitting methods may wish to skip ahead to Section \ref{sec:results}.

\section{Sample and data}
\label{sec:sample}

\subsection{JINGLE sample}
The \Ntot\ galaxies in the JINGLE sample have stellar masses in the range $\log M_*/M_\odot=9-11.3$ and are in the redshift range $0.01 < z < 0.05$. 
The targets were selected from the \textit{H}-ATLAS survey \citep{Eales2010, Maddox2018} with the requirement to have a detection $\geq 3\sigma$ in the 250\micron\ and 350\micron\ SPIRE bands.
Additionally, they have been selected to have a flat logarithmic stellar mass distribution.
Due to these requirements, they are mainly main-sequence star-forming galaxies with $-1.5 <\ \log \text{SFR}/[$\Msun\ yr$^{-1}] <\ 1.5$ (see Figure~\ref{fig:SFR_Mstar_JINGLE_HRS}).  A detailed description of the selection criteria is provided in \cite{Saintonge2018}. 
Most of the JINGLE objects are late-type galaxies, with only seven classified as early-type galaxies \citep{Saintonge2018}.

Properties of the JINGLE galaxies used in this work (such as SFR, metallicity, distances,...) are taken from the JINGLE catalog \citep{Saintonge2018}. In particular, we use the star-formation rates and stellar masses measured with {\tt MAGPHYS}   \citep{daCunha2008}. In this paper we refer to JINGLE galaxies using their corresponding JINGLE ID, as described in the JINGLE catalog \citep{Saintonge2018}.

\subsection{HRS sample}

 To extend our analysis to a larger range in galaxy properties, we include in our analysis also galaxies from the \Herschel\ Reference Survey \citep[HRS, ][]{Boselli2010b}.
 The HRS is a volume-limited sample (15 Mpc $\leq D \leq$ 25 Mpc) of 323 galaxies, with flux limits in the $K$-band to minimize selection effects due to dust and young high-mass stars. 
 A large fraction of HRS galaxies lie in clusters, with 47\% of the HRS galaxies listed in the Virgo Cluster Catalogue alone. They have stellar masses in the range $\log M_*/M_\odot= 8.4-11.3$.
Galaxies from the HRS have been observed in the five \Herschel\ bands (at 100\micron, 160\micron, 250\micron, 350\micron, and 500\micron), but do not have observations at 850\micron. In our analysis we use the SFR and stellar masses measured with {{\tt MAGPHYS}} by \cite{DeVis2017}, to be consistent with the JINGLE measurements.

\begin{figure}
\centering
\subfigure{\includegraphics[width=0.45\textwidth]
{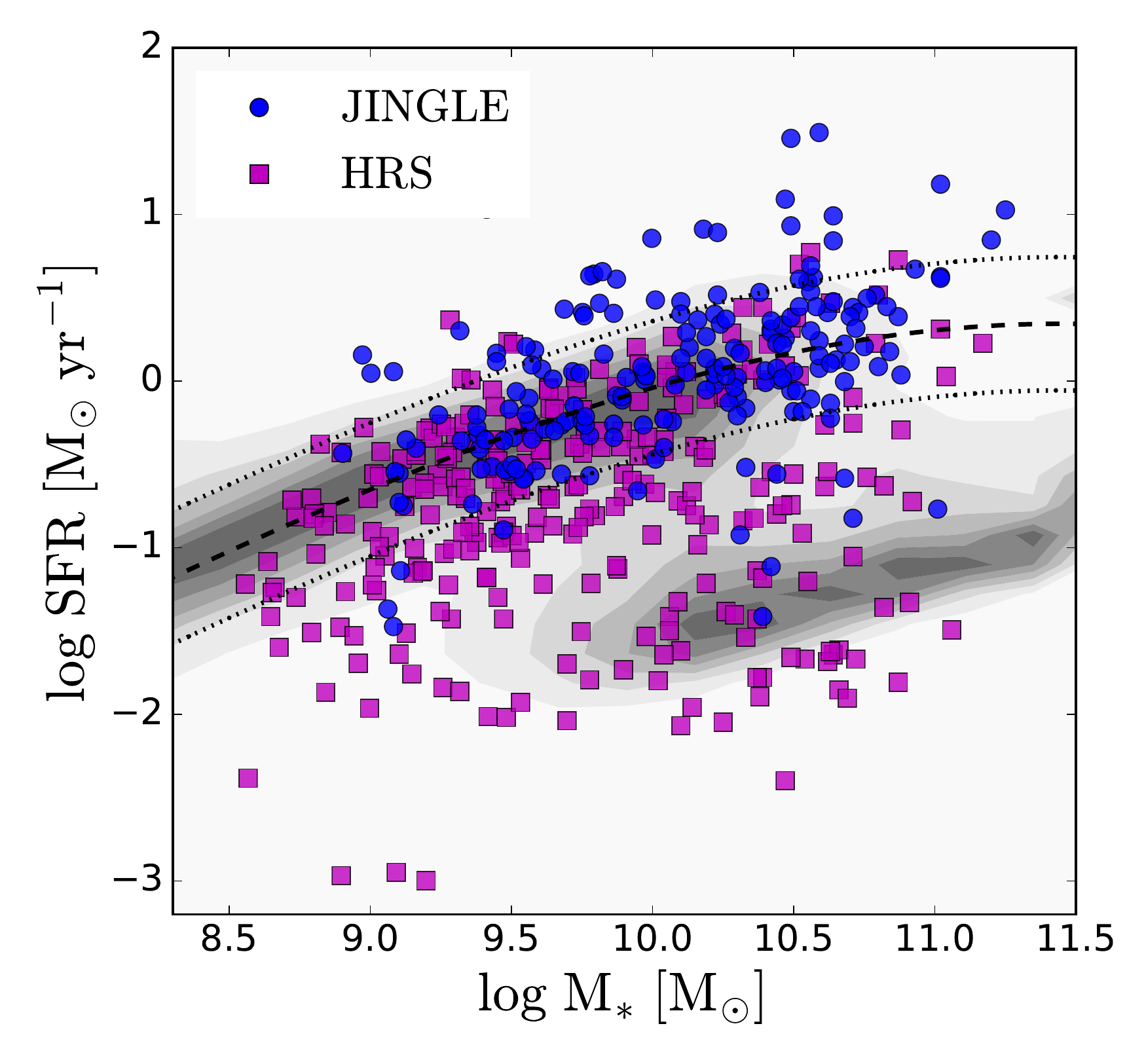}}
\caption{Distribution of the JINGLE and HRS sample in the SFR-$M_*$ plane. The position of the star formation main sequence \citep{Saintonge2016} is shown as a dashed line, the 0.4 dex dispersion is shown by dotted lines. The grey contours show the distribution of SDSS galaxies at redshift $z < 0.05$. }
\label{fig:SFR_Mstar_JINGLE_HRS} 
\end{figure}

Figure \ref{fig:SFR_Mstar_JINGLE_HRS} shows the JINGLE and HRS galaxies on the SFR-$M_*$ plane.
With respect to the JINGLE galaxies, the HRS sample includes galaxies which are less massive ($\log M_* < 9$) and with lower SFR ($-2 < \log$ (SFR/[\Msun\ yr$^{-1}]) < 0.6$, mean $\log$ (SFR/[\Msun\ yr$^{-1}] )= -0.71$) compared to JINGLE, which has a mean $\log (\text{SFR}$/[\Msun\ yr$^{-1}]) = 0.04$. HRS galaxies are also less dusty than JINGLE targets (De Looze et al., in prep.), since  contrary to JINGLE they have not been selected based on detection in the infrared bands. 
 The HRS sample includes also a large number of early-type galaxies  \citep[62/323, ][]{Smith2012}, which are not well represented in the JINGLE sample (7/\Ntot).
Therefore by including this sample in our analysis, we can test whether the dust scaling relations that we find with the JINGLE sample hold also for other types of galaxies.
 Additionally, increasing the  dynamical range of galaxy properties will help to  constrain better the dust scaling relations.

\subsection{Data}

\subsubsection{JINGLE}
Our data set consists of photometric points at 22\micron\ (WISE), 60\micron\ (IRAS), 100\micron, 160\micron\  (\Herschel /PACS), 250\micron, 350\micron, 500\micron\ (\Herschel /SPIRE), and 850\micron\ (SCUBA-2).
A detailed description of the JINGLE photometric data set is given in \cite{Smith2019} and De Looze et al. (in prep.). Here we summarize the most important points.
The fluxes of the WISE, \Herschel, and SCUBA-2 bands have been extracted from matched apertures based on the SPIRE 250\micron\ band. The flux extraction is described in detail by \cite{Smith2019}. One galaxy (JINGLE 62) has been removed from the sample since it is not detected in the 250\micron\ band and therefore it is not listed in the release version of the \textit{H}-ATLAS DR2 catalogue \citep{Maddox2018}. Thus the sample analysed in this work consists of \Ntot\ galaxies.

We consider upper limits for fluxes with peak signal-to-noise ratio (S/N) < 3. 
Since the CO(3-2) 345.79 GHz line emits in the 850\micron\ band, we corrected the SCUBA-2 flux by subtracting the estimated contribution of the CO(3-2) line  \citep[for details see][]{Smith2019}.
After subtracting the CO(3-2) emission, some of the fluxes become negative, due to the uncertainties in the 850\micron\ fluxes and in the CO(3-2) predictions. These fluxes are consistent with zero within the uncertainties and are considered as upper limits. 
In our sample, there are 66 galaxies with peak S/N<3 and additionally 4 galaxies have negative 850\micron\ flux, even though their peak S/N> 3 before subtraction of the  CO(3-2) contribution. 
 For all these cases, we use conservative upper limits equal to five times the flux uncertainty in that band. 

 The IRAS 60\micron\ fluxes are derived using the  Scan Processing and Integration Tool (SCANPI\footnote{http://irsa.ipac.caltech.edu/applications/Scanpi/}), following the strategy of \cite{Sanders2003}.
In our sample, 69/\Ntot\ galaxies have 5$\sigma$ upper limits for the 60\micron\ flux and 22/\Ntot\ do not have IRAS 60\micron\ observations.

\subsubsection{HRS}
For the HRS sample, we have flux measurements  in the \Herschel /PACS \citep{Cortese2014} and \Herschel /SPIRE bands \citep{Ciesla2012}, from 100\micron\ to 500\micron. We note that, contrary to JINGLE, this sample does not have observations at 850\micron, therefore the long-wavelength slope of the SED can be constrained only by the 500\micron\ point. 
In the case of non-detections, we consider upper limits equal to five times the flux uncertainties as we do for the JINGLE sample.

We exclude from the sample 39 galaxies which are not detected in all of the \Herschel\ bands, and therefore do not have constraints on their dust properties.  We also exclude four galaxies which do not have SFR and stellar mass measurements from \cite{DeVis2017}. They were excluded from the sample  because their SEDs show signs of contamination from dust heated by an active galactic nucleus or a hot X-ray halo or from synchrotron radiation emission  \citep{Eales2017}.
 The final sample consists of 41 early-type  and 239 late-type galaxies, for a total of 280 galaxies.

\section{Method}
\label{sec:method}
	
\subsection{Models}
\label{sec:models}
To describe the far-infrared and sub-millimeter spectral energy distribution (SED) we adopt the three models employed by \cite{Gordon2014} for the SED fit of the Magellanic Clouds: single modified black-body (SMBB),  broken emissivity law modified black-body (BMBB), and two modified black-bodies (TMBB). We describe below the analytic functions and the parameters used for the three models:

\begin{itemize}
\item \textbf{SMBB:} The single modified black-body model describes the dust emission $F_{\lambda}$ (in units of W m$^{-2}$ Hz$^{-1}$ sr$^{-1}$) at each wavelength $\lambda$  in the following way \citep{Hildebrand1983}:
	\begin{equation}
	F_{\lambda} = \frac{M_{\text{dust}}}{D^{2}} 		\kappa_{\lambda} B_{\lambda}(T)
	\end{equation}
where $M_{\text{dust}}$ is the dust mass in the galaxy and $D$ is the distance of the galaxy. $B_{\lambda}$($T$) is the Planck function for the emission of a black-body with a dust temperature $T$ given by:
\begin{equation}
B_{\lambda}(T) = \frac{2hc^2}{\lambda^{5}} \frac{1}{\exp\left( \frac{hc}{k_{B}T\lambda} \right)-1} .
\end{equation}
The dust mass absorption coefficient $\kappa$ describes which dust mass gives rise to an observed luminosity. The value of $\kappa$ depends on the physical properties of the dust, such as the mass density of the constituent materials, the efficiency with which they emit, the grain surface-to-volume ratio, and the grain size distribution \citep{Koehler2015, Ysard2018}.
 The SMBB applies a dust emissivity power law to characterise the behaviour of $\kappa$ as a function of wavelength:
\begin{equation}
 \kappa_{\lambda}= \kappa_{0}\left( \frac{\lambda_{0}}{\lambda}\right)^{\beta}
\end{equation} 
where $\kappa_{0}$ is the reference dust mass absorption coefficient. Laboratory studies found that the absorption coefficient depends also on the dust temperature and dust  emissivity index $\beta$, with higher $\kappa$ values observed for higher temperatures and lower $\beta$ values \citep{Coupeaud2011}. For simplicity, here we assume a constant  value $\kappa_0$= $\kappa(500\mu\text{m} ) = 0.051\ \text{m}^2 \ \text{kg}^{-1}$ from \cite{Clark2016}. 

This model has three free parameters ($M_{dust}$, $T$, and $\beta$), and assumes that the dust emission can be described by a dust component with a single temperature. At  wavelengths shorter than 100 $\mu$m, a second warmer dust component can contribute to the FIR emission \citep[e.g.][]{Relano2018}. Therefore for this model, we  use only the flux bands with wavelengths $\geq 100 \mu$m. 
 Additionally, we use the 60$\mu$m point as an upper limit, in order to better constrain the dust temperature.\\

\item \textbf{BMBB:} When fitting the FIR SED with a SMBB model, some galaxies show an excess in the flux at wavelengths $\geq $ 500\micron, called `sub-millimeter' excess 
\citep{Lisenfeld2002, Galliano2003, Dumke2004, Bendo2006, Galametz2009,Israel2010, Bot2010b, Hermelo2016}.
 The broken emissivity law modified black-body (BMBB) model assumes that the submm excess is due to variations in the wavelength dependence of the dust emissivity law.
 These variations are parametrized by a broken power law:
\begin{equation}
\kappa_{\lambda}=
	\Bigg \{\begin{array}{ll}
		\kappa_0\left(\frac{\lambda_0}{\lambda}\right)^{\beta_1} & \text{if } \lambda < \lambda_b \\
		\kappa_0\left(\frac{\lambda_0}{\lambda_b} \right)^{\beta_1}  \left(\frac{\lambda_b}{\lambda}\right)^{\beta_2}& \text{if } \lambda > \lambda_b \\
	\end{array}
\end{equation} 
where $\lambda_b$ is the wavelength of the break.
This model has five free parameters: $M_{\text{dust}}$, $T$,  $\beta_1$, $\beta_2$, and  $\lambda_b$.
Also for this model, we use only the flux bands with wavelengths $\geq 100$\micron. In order to have good constraints on the fitting parameters, it is crucial to have a detection of the 850\micron\ flux. If the SCUBA-2 point is not detected, an upper limit is not enough to constrain the parameters of this model. Without the 850\micron\ flux point, the 500\micron\ flux point is the only one that can be used to determine $\beta_2$ and $\lambda_b$, leading to large uncertainties on their values.\\

\item \textbf{TMBB:} The two modified black-body model assumes that the FIR SED is emitted by two dust populations with different temperatures. The dust emission is parametrized by two modified black-bodies: one for the cold dust (indicatively $T < 40$ K) and one for the warm dust (indicatively $T > 40$ K):
\begin{equation}
F_{\lambda} = F_{\lambda}^{\text{SMBB}_{cold}}+F_{\lambda}^{\text{SMBB}_{warm}}
\end{equation}
where the two SMBB components are defined as above.
 In order to reduce the number of free parameters, we fix the $\beta$ value of the warm component to 1.5 \citep{Coupeaud2011, Boselli2012}, while we leave the $\beta$ value of the cold component as a free parameter.
 So in this model we have five free parameters: $M_{\text{cold}}$, $T_{\text{cold}}$, $\beta_{\text{cold}}$, $M_{\text{warm}}$, and $T_{\text{warm}}$. For the fitting, we use the fluxes in all available bands from 22 to 850\micron. 
\end{itemize}

All these models assume that dust grains are optically thin. According to dust models, this assumption holds for wavelengths $\geq 100$\micron, while at shorter wavelengths it is possible that dust is optically thick \citep{Draine2007}. \cite{Casey2012} modelled the SED of 65 luminous infrared galaxies from the GOALS survey \citep{Armus2009} and found that  even if the dust is optically thick, the difference in the SED shape at 22\micron\ would be small. \cite{Utomo2019} studied the dust emission at resolved scales in four nearby galaxies (Small and Large Magellanic Clouds, M31, and M33) and found that most of the dust emitting at wavelengths longer than 100\micron\ is optically thin.  They observe that at wavelengths $\sim$20\micron\ some regions of the galaxies become optically thick, but on global galaxy scales we do not expect these regions to dominate the emission.

\begin{figure*}
\centering
\subfigure{\includegraphics[width=0.32\textwidth]
{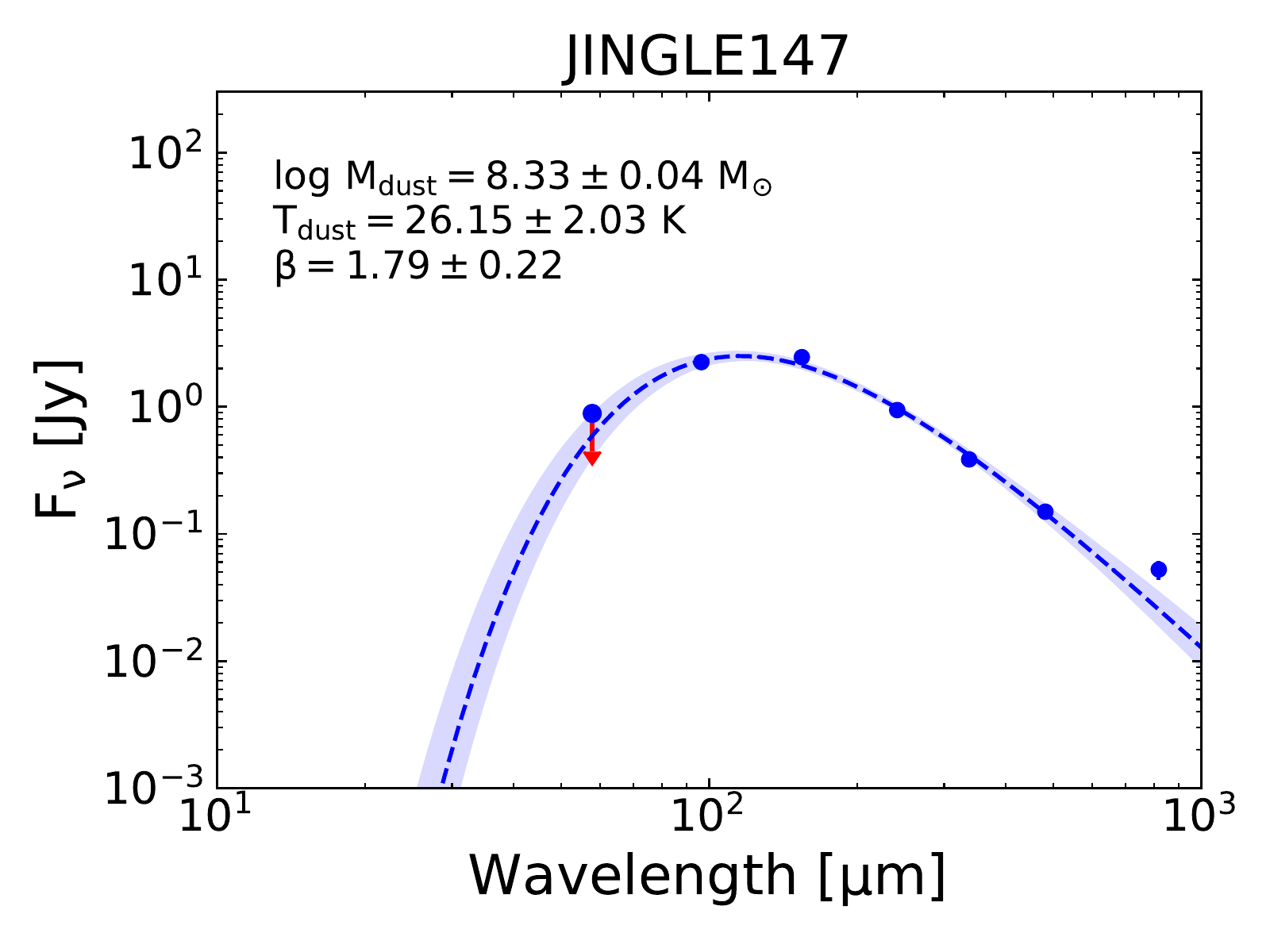}}
\subfigure{\includegraphics[width=0.32\textwidth]
{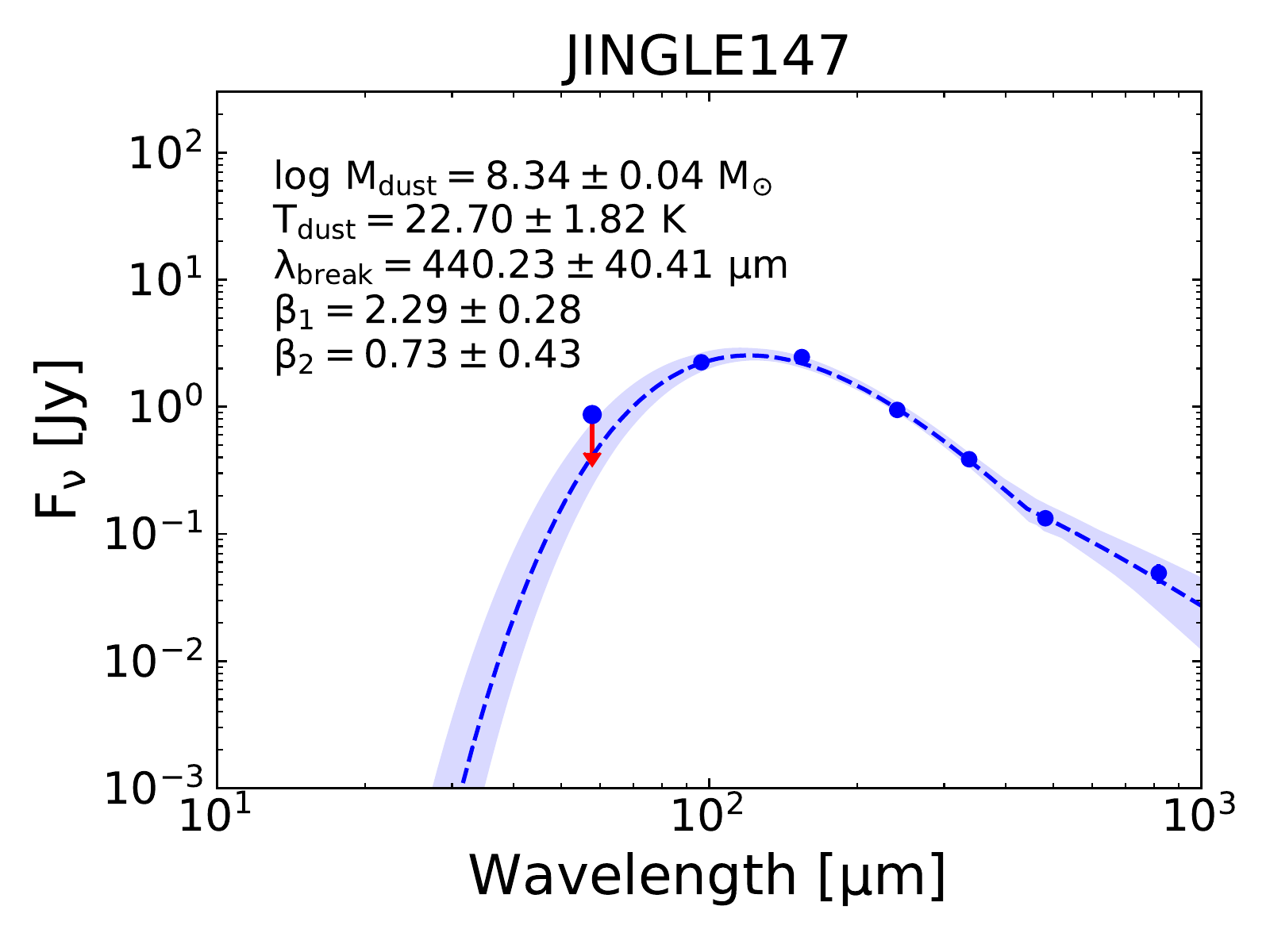}}
\subfigure{\includegraphics[width=0.32\textwidth]
{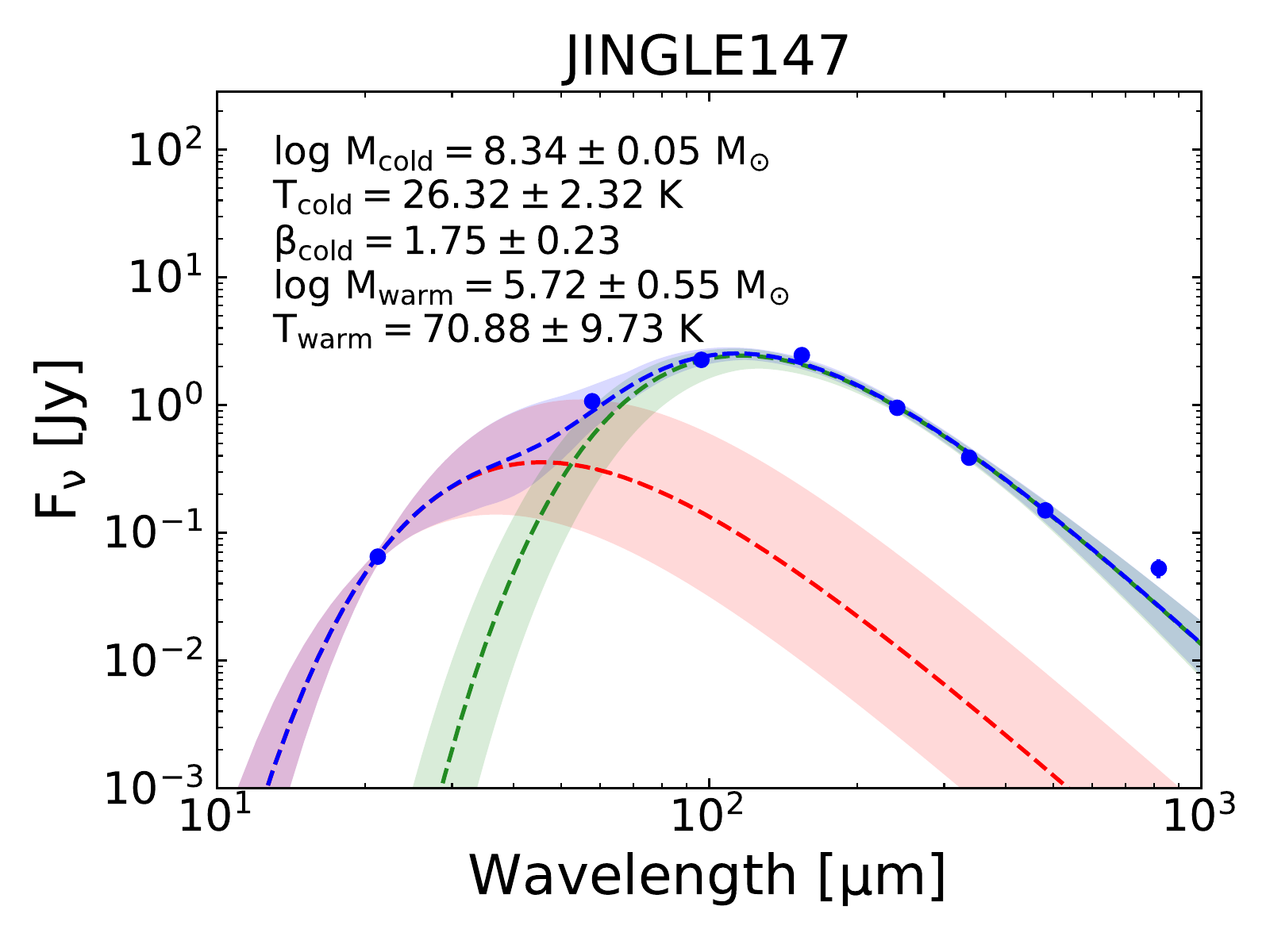}}
\caption{Example of FIR SED of one galaxy from the JINGLE sample, fitted with the non-hierarchical approach using the three models: single modified black-body (SMBB, left panel), broken emissivity law modified black-body (BMBB, middle panel) and two modified black-bodies (TMBB, right panel). The shaded regions show the lower and upper 1-$\sigma$ uncertainties on the SED models, defined by taking the maximum and minimum flux values of the models with likelihood values in the highest 68th percentile.}
\label{fig:example_SED_models} 
\end{figure*}

We apply the SMBB model to both the JINGLE and HRS sample, while we apply the BMBB and TMBB models only to the JINGLE sample. We make this decision because for the HRS sample we do not have the 850\micron\ flux point, and therefore we do not have enough flux points for models with a large number of free parameters.
Additionally, for the BMBB model it is very important to have the 850\micron\ point to constrain the emissivity index $\beta_2$ after the break.
Fig. \ref{fig:example_SED_models} shows an example of the SED fitting of one galaxy from the JINGLE sample using the three models.

\subsection{Introduction to the Bayesian SED fitting method}
\label{sec:Bayesian_method} 
In this section we briefly describe the Bayesian approach
 used for the SED fitting  \citep[we follow the same notation as in][]{Galliano2018}.
 Readers who are less interested in the statistical methods may wish to go directly to the results presented in Section \ref{sec:results}.
The observed SED of a galaxy ($F^{obs}$) can be described in the following way:
\begin{equation}
F^{obs}(\lambda_j) = F^{mod}(\lambda_j, \vec{\theta})+\epsilon(\lambda_j)\cdot F^{err}(\lambda_j)
\end{equation}
where $F^{obs}(\lambda_j)$ is the flux observed at the wavelength $\lambda_j$ and $F^{mod}(\lambda_j,\vec{\theta})$ is the flux described by our model with parameters $\vec{\theta}$. The last term describes the deviation of the observed flux from  the model due to random noise: $F^{err}(\lambda_j)$ is the amplitude of the noise and $\epsilon(\lambda_j)$ is a random variable with mean $< \epsilon> = 0$ and standard deviation $\sigma(\epsilon) =1$. 
We can reverse the previous formula to express $\epsilon(\lambda_j)$ as a function of the other quantities:
\begin{equation}
\epsilon(\lambda_j) = \frac{F^{obs}(\lambda_j) - F^{mod}(\lambda_j, \vec{\theta})}{F^{err}(\lambda_j)}.
\label{eq:eps}
\end{equation}

The goal is to find the best parameters to fit the data by minimising the offset between the model and the data. From a Bayesian point of view, this is equivalent to maximising the likelihood of the model, given the data.
The probability of the data given the model parameters $\vec{\theta}$ can be expressed as:
\begin{equation}
p(\vec{F}^{obs}|\vec{\theta}) = \prod_{j=1}^m p(\epsilon(\lambda_j, \vec{\theta}))
\end{equation}
where $\vec{F}^{obs} = \left(F^{obs}(\lambda_1), . . . , F^{obs}(\lambda_m)\right)$ is the vector containing the flux emission at each waveband $j=1,...,m$.
We are interested in the probability of the model parameters, knowing the observations. Thus we can use the Bayes' theorem to write the expression:
\begin{equation}
p(\vec{\theta}|\vec{F}^{obs}) = \frac{p(\vec{F}^{obs}|\vec{\theta}) \cdot p(\vec{\theta})}{p(\vec{F}^{obs})} \propto  p(\vec{F}^{obs}|\vec{\theta}) \cdot p(\vec{\theta}),
\label{eq:ptheta_given_data}
\end{equation}
where $p(\vec{\theta})$ is the `prior' distribution, and $p(\vec{\theta}|\vec{F}_{obs})$ is the `posterior' distribution. The denominator $p(\vec{F}^{obs})$ can be neglected since it is constant for a given set of observed fluxes. By sampling the posterior distribution in the parameter space  we can construct the posterior probability density function (PDF).
 Examples of posterior probability density functions (PDF) are shown in the appendix (Fig. \ref{fig:post_triangle_plot}). The figure shows the PDFs obtained from the SED fit of one galaxy using the SMBB, BMBB, and TMBB models.

\subsection{Hierarchical Bayesian method}
\label{sec:Hier_method}
The difference between the classical and hierarchical Bayesian method is that in the former the prior distribution is an assumption and in the latter it is defined by the data sample \citep[e.g.][]{Gelman2004, Galliano2018}. Hierarchical methods require therefore a population of objects, which are used to define the prior distributions. In the case of SED fitting, the sample can be formed by multiple spatially resolved regions of the same galaxy or by a sample of galaxies with similar properties.
 The entire sample is then fitted simultaneously, in order to extract both the information about the prior distribution of the sample and the posterior distribution of the single elements of the sample.
 
 \cite{Kelly2012} showed that the hierarchical method can be used to reduce the degeneracy between $T$ and $\beta$. This approach has subsequently been used in other studies to reduce the $T$-$\beta$ degeneracy  \citep{Juvela2013, Veneziani2013,Galliano2018}. The key assumption behind the hierarchical approach is that the dust parameters of our sample of galaxies follow a common distribution. In our case we assume that they follow a \stud. Thanks to this assumption, we are able to better constrain model parameters, especially for galaxies with low S/N, where a large range of combinations of $T$ and $\beta$ provide reasonably good fits to the data. In those cases, the prior helps to constrain the range of possible $T$ and $\beta$. The key point of the hierarchical approach is that we do not need to specify the mean and standard deviation of the prior distribution before doing the fit, but they can be inferred by the data.

The new parameters describing the prior distribution of the parameters $\vec{\theta}$ are called \textit{hyper-parameters}. The commonly used hyper-parameters are: 
\begin{itemize}
\item $\vec{\mu}$: the average of the parameter vector $\vec{\theta}$;
\item $\Sigma$: the covariance matrix describing the standard deviation and correlation of $\vec{\theta}$.
\end{itemize}
Using this formalism, the posterior distribution of the parameters given the data $p(\vec{\theta}|\vec{F}_{obs})$ for the $i$-th galaxy in the sample becomes:
\begin{equation}
p(\vec{\theta_i}|\vec{F_i}^{obs}, \vec{\mu}, \Sigma) \propto  p(\vec{F_i}^{obs}|\vec{\theta_i}) \cdot p(\vec{\theta_i}|\vec{\mu}, \Sigma).
\label{eq:ptheta_given_data_hier_one_galaxy}
\end{equation}
This is the hierarchical equivalent of eq. (\ref{eq:ptheta_given_data}).
The posterior distribution of the parameters and hyper-parameters for the entire sample of $n$ galaxies is:
\begin{multline}
p(\vec{\theta_1}, ..., \vec{\theta_n}, \vec{\mu}, \Sigma| \vec{F_1}^{obs},...,\vec{F_n}^{obs})
\propto   \prod_{i=1}^n p(\vec{\theta_i}|\vec{F_i}^{obs}, \vec{\mu}, \Sigma) \cdot p(\vec{\mu}) \cdot p(\Sigma) \\
\propto \prod_{i=1}^n p(\vec{F_i}^{obs}|\vec{\theta_i}) \cdot p(\vec{\theta_i}|\vec{\mu}, \Sigma)
\cdot p(\vec{\mu}) \cdot p(\Sigma),
\end{multline}
\label{eq:ptheta_given_data_hier}
where $p(\vec{\mu})$ and $p(\Sigma)$ are the prior distributions of the hyper-parameters. When compared to the classical Bayesian method, the hierarchical method is able to recover the distribution of parameters with better precision, especially if the noise in the data is high \citep{Kelly2012, Galliano2018}. 
In that case, the hierarchical approach uses the information about the parameter distribution obtained from the rest of the sample to better constrain the parameters for the particular objects where the quality of the data is low.
 The hierarchical method will not necessarily perform better in measuring the parameter of a single object, but it will be less biased when measuring the distribution of parameters of the entire population.

\subsection{Noise distribution}
\label{sec:noise_distr}
In this section we describe the functions used to model the noise distribution for both the non-hierarchical and hierarchical approaches. 
The noise is usually modelled with a normal distribution or a \stud . The \stud\ has a higher probability in the tails with respect to the normal distribution, allowing for more outliers.  Its shape is described by  the number of degrees of freedom $f$:  as $f$ decreases, more probability will be in the tails of the distribution.
 The normal distribution is a special case of the \textit{t}-distribution with the number of the degrees of freedom that goes to infinity, $f \rightarrow \infty$.

The probability density of a normal distribution is defined as:
\begin{equation}
\text{Normal}(y|\mu, \sigma) = \frac{1}{\sqrt{2 \pi} \sigma}\exp \left(-\frac{1}{2}\left(\frac{y-\mu}{\sigma} \right)^2 \right),
\end{equation} 
where $\mu$ is the mean and $\sigma$ is the standard deviation. 
The multivariate normal distribution is the generalization of the one-dimensional normal distribution to a higher dimension $m$:
\begin{multline}
\text{MultiNormal}(\vec{y}|\vec{\mu}, \Sigma) = \\
\frac{1}{(2 \pi)^{m/2}} \frac{1}{\sqrt{|\Sigma|}}\exp \left(-\frac{1}{2} (\vec{y}-\vec{\mu})^T \Sigma^{-1} (\vec{y}-\vec{\mu}) \right),
\end{multline}
where $m$ is the dimension of the vector $\vec{y}$, $\Sigma$ is the $m\times m$ covariance matrix, and $(\vec{y}-\vec{\mu})^T$ indicates the transpose of the vector $(\vec{y}-\vec{\mu})$.

The \stud\ is defined as:
\begin{equation}
\text{Student}(y|\mu, \sigma, f) =  \frac{\Gamma((f+1)/2)}{\Gamma (f/2)}
\frac{1}{\sqrt{f \pi} \sigma} \left( 1+\frac{1}{f}\left( \frac{y-\mu}{\sigma} \right)^2 \right) ^{-\frac{f+1}{2}},
\end{equation} 
where $f$ is the number of degrees of freedom. The multivariate \stud\ is the generalization of the one-dimensional distribution to a higher dimension $m$:
\begin{multline}
\text{MultiStudent}(\vec{y}|\vec{\mu}, \Sigma, f) = \\
 \frac{\Gamma((f+m)/2)}{\Gamma (f/2)}
\frac{1}{(f \pi)^{m/2}} \frac{1}{\sqrt{|\Sigma|}}
\left( 1+\frac{1}{f} (\vec{y}-\vec{\mu})^T \Sigma^{-1} (\vec{y}-\vec{\mu}) \right)^{-\frac{f+m}{2}},
\end{multline}
where $m$ is the dimension of the vector $\vec{y}$.

We expect to observe a flux excess at 850\micron\ for some galaxies, given the fact that the submm excess has been reported in numerous studies \citep[e.g.][]{Galametz2011, Remy-Ruyer2013, Remy-Ruyer2015, Hermelo2016}. Since the 850\micron\ fluxes have usually larger uncertainties than the other points, if we use a \stud, the SMBB model will assume that every change in slope at 850\micron\ is due to the error being underestimated, rather than to a physical effect. The model will then `ignore' the 850\micron\ point, and produce a fit considering only the \Herschel\ points. Since we believe that there is information in the longer wavelength points, we therefore decide to use a normal distribution for the error.
 In Section \ref{sec:normal_vs_student} of the appendix we compare the results obtained using the Student and normal distribution.
 
In both the non-hierarchical and hierarchical case, we model the noise as:
\begin{equation}
p(\vec{F}^{obs}|\vec{F}^{mod}(\vec{\theta}), C) = \text{MultiNormal}(\vec{F}^{obs}|\vec{F}^{mod}(\vec{\theta}), C),
\end{equation}
where $C$ is the covariance matrix, which describes the uncertainties associated with the flux densities in the different wavebands (see Section \ref{sec:cov_matrix} for the definition of the covariance matrix).

\subsection{Prior distributions}
In this section we describe the prior distributions assumed for the hierarchical and non-hierarchical method.\\
\\
\textbf{Non-hierarchical:} For the prior distribution of the  parameters $\vec{\theta}$, we assume uniformly distributed (``flat") priors, i.e. $p(\theta)=1$, in the ranges described in Table \ref{tab:para_ranges}.\\
\\
\textbf{Hierarchical:}
For the definition of the prior distributions in the hierarchical framework, we follow \cite{Kelly2012}, \cite{Galliano2018} and the \stan\ manual \citep{Stan_manual}.
\begin{itemize}
\item \textbf{\it parameters}: for the definition of the prior distributions of the parameters given the hyper-parameters, we follow \cite{Kelly2012} and \cite{Galliano2018}. We assume a multivariate Student's \textit{t} distribution with $f=8$ degrees of freedom:
\begin{equation}
p(\vec{\theta_i}|\vec{\mu}, \Sigma) = \text{MultiStudent}  (\vec{\theta_i}|\vec{\mu}, \Sigma, f=8).
\end{equation}
We also tried to vary the number of degrees of freedom and did not see any differences in the results. Assuming a \stud\  allows one to have more galaxies with dust parameters which are `outliers' from the mean of the sample. In this way, we make sure that our assumption that the galaxies belong to the same population is not too stringent. 
We note that the parameters $\vec{\theta_i}$ are not constrained within a certain range but they are allowed to take any value. Their distribution is described by the prior distribution and we set some constraints on the allowed range of the hyper-priors (mean and standard deviation)  that determine the shape of the priors (see next point).
\\

\item \textbf{\it hyper-parameters}: For the mean $\vec{\mu}$ of the parameters, 
 we assume a uniform prior with a large parameter range. 
 In this way we ensure that the prior is proper (i.e $\int p(\theta) d\theta < \infty$), and at the same time we maintain the prior vague enough to not constrain the results  \citep{Tak2018, Gelman2007}. 
 The prior ranges for  $\vec{\mu}$ are shown in Table \ref{tab:mu_prior_range}.
  We note that we set the prior range of $\mu(T_{warm})$ to be > 50 K, because we want the distribution of warm temperatures to be well separated from the distribution of cold temperatures.
For the covariance matrix $\Sigma$, we use the \textit{separation strategy} from \cite{Barnard2000}. This formalism ensures that the prior distributions of the correlations between parameters are uniform over the range $[-1,1]$, meaning that all values of the correlations are equally likely. The separation strategy breaks down the covariance matrix in: 
\begin{equation}
\Sigma=SRS \ ,
\end{equation}
where $S$ is a diagonal matrix with the values of the standard deviation, and $R$ is the correlation matrix. Both $S$ and $R$ have dimension $q\times q$, where $q$ is the number of free parameters in the model. 
  The prior distribution of the hyper-parameters is then:
\begin{equation}
p(\vec{\mu}) \cdot p(\Sigma) \propto p(\vec{\mu}) \cdot p(S) \cdot p(R)\ .
\end{equation}

For the priors on the $S$ and $R$ we follow the recommendations given by the \stan\ manual \citep{Stan_manual}.
For the priors on the diagonal elements of $S$, we use a weakly informative prior, parametrized by a half-Cauchy distribution with a small scale $\sigma$ = 2.5 \citep{Stan_manual}:
\begin{equation}
p(S_{k,k}) = \text{Cauchy}(0, \sigma) =
 \frac{1}{\pi \sigma} \frac{1}{1+\left( \frac{S_{k,k}}{\sigma} \right)^2},
\end{equation}
where $S_{k,k}$ > 0, for $k=1,..,q$. 
 For the priors on the correlation matrix $R$, we use a LKJ correlation distribution with shape $\nu = 2$:
\begin{equation}
p(R) = \text{LKJ\ Corr}(R, \nu) \propto \det(R)^{\nu-1}
\end{equation}
(see \cite{Lewandowski2009} for definitions).
The basic idea of the LKJ correlation distribution is that as $\nu$ increases, the prior increasingly concentrates around the identity matrix.\\

\end{itemize}

\begin{table}
\centering
\caption{Prior parameter ranges  assumed for the Bayesian non-hierarchical SED modelling  using the SMBB function.}
\begin{tabular}{|lc|} 
\hline
Parameter & Range \\ \hline \hline
$\log M_{dust}/M_{\odot}$  & (5, 9) \\ 
$T$ [K] & (5, 50) \\
$\beta$ & ($0.1$, 3) \\
\hline 
\end{tabular}
\label{tab:para_ranges}
\end{table}

\begin{table}
\centering
\caption{Ranges of the priors on the hyper-parameter $\vec{\mu}$ (sample mean) for the Bayesian hierarchical SED modelling  using the SMBB, BMBB and TMBB functions. } 
\begin{tabular}{|lc|}
\hline
\textbf{Hyper-parameter} & Range \\ \hline \hline
\multicolumn{2}{c}{SMBB} \\ \hline
$\mu(\log M_{dust}/M_{\odot})$  & (6, 9) \\ 
$\mu(T)$ [K] & (15, 50) \\ 
$\mu(\beta)$ & (0.5, 3) \\ 
\hline\hline
\multicolumn{2}{c}{BMBB} \\ \hline
$\mu(\log M_{dust}/M_{\odot})$ & (5, 9) \\ 
$\mu(T)$ [K] & (5, 50) \\ 
$\mu(\beta_{1})$ & (0, 5) \\ 
$\mu(\beta_{2})$ & (0, 5) \\ 
$\mu(\lambda_b)$ [\micron ] & (420, 500) \\ 
\hline \hline
\multicolumn{2}{c}{TMBB} \\ \hline
$\mu(\log M_\text{cold}/M_{\odot})$ & (6, 10) \\ 
$\mu(T_\text{cold})$ [K] & (5, 40)\\
$\mu(\beta_\text{cold})$ & (0.5, 5) \\
$\mu(\log M_\text{warm}/M_{\odot})$ & (2, 7) \\
$\mu(T_\text{warm})$ [K] & (50, 90)\\ 
\hline 
\end{tabular}
\label{tab:mu_prior_range}
\end{table}

\subsection{Covariance matrix, beam and filter corrections}

\label{sec:cov_matrix}
In order to perform an accurate fit, it is important to take into account correctly the uncertainties associated with each flux measurement as well as the correlation between these uncertainties.
The covariance matrix $C$ describes the uncertainties associated with the flux densities in the different wave bands, and includes both calibration and measurement uncertainties. Calibration uncertainties can be correlated between bands observed with the same instrument. 
For the definition of the covariance matrix, we follow \cite{Gordon2014}.  The calibration covariance matrix is defined as:
\begin{equation}
C_{j,k}^{cal} = [A_{cor,j,k}+ A_{uncor, j,k}] = [\sigma_{cor,j,k}^2+ \delta_{j,k}\sigma_{uncor, j,k}^2]
\end{equation}
where $A_{cor}$ is the matrix of the noise correlated between bands, $A_{uncor}$ is the diagonal matrix of repeatability that is uncorrelated between bands. $\sigma_{cor, j,k}$ and $\sigma_{uncor, j,k}$ are the percentage of correlated and uncorrelated uncertainties, respectively, between the $j$-th and $k$-th band,  and $\delta_{j,k}$ is one for $j=k$ and zero otherwise.
 The calibration uncertainty values that we use are reported in Table \ref{tab:corr_noise}, given in percentage of the flux.

 The total covariance matrix $C$ is a combination of the calibration and measurement uncertainties:
\begin{equation}
C_{j,k} = C_{j,k}^{cal} \cdot F_{j} \cdot F_{k} + F_{j}^{err} \cdot F_{k}^{err}
\end{equation}
where $F_{j}$ and $F_{k}$ are the fluxes in the $j$-th and $k$-th waveband, and $F_{j}^{err}$ and $F_{k}^{err}$ are the corresponding measurement uncertainties.

The colour and beam corrections applied to our data are described in detail in De Looze et al. (in prep.).\\
\\
\begin{table*}
\centering
\caption{Percentage of correlated and uncorrelated uncertainties  for the different instruments.}
\begin{tabular}{lcccc}
\hline
Instrument & Waveband & Correlated  & Uncorrelated & Reference  \\
 & [\micron] & uncertainty & uncertainty &   \\
  \hline \hline
WISE & 22 & - & 5.7 $\%$ & \cite{Jarrett2011}\\
IRAS & 60 & - & 20 $\%$  & \cite{Sanders2003, Miville-Deschenes2005}  \\
PACS & 100, 160 & 5 $\%$ & 2 $\%$ & \cite{Balog2014}, \cite{Decin2007}\\ 
SPIRE & 250, 350, 500 & 4 $\%$ & 1.5 $\%$ & \cite{Bendo2013} \\ 
SCUBA & 850 & - & 10 $\%$  & \cite{Smith2019}\\ 
\end{tabular}
\label{tab:corr_noise}
\end{table*}
\\
\textbf{Non-hierarchical}: The filter corrections are applied to the model SED by convolving the model flux points with the appropriate filter response curve in each band. 
The \textit{Herschel}/SPIRE fluxes were corrected also for the effective beam area, which depends on the shape of the spectrum due to the absolute SPIRE calibration in units of flux density per beam. The SED shape is described by the dust temperature $T$ and the emissivity index $\beta$. At each step of the Markov chain Monte Carlo (MCMC) algorithm, the \Herschel /SPIRE fluxes are corrected according to the two model parameters, before comparing them to the fluxes of the SED model.
 For the BMBB model, we applied the beam and color corrections using  $\beta_1$ or $\beta_2$ depending on the wavelength position of the break $\lambda_{b}$. For the TMBB model, we calculate which of the two components (warm or cold) contribute the most to the flux in every band. Then we calculate the corrections using the temperature $T$ and $\beta$ values of the dominant component in each band.\\
\\
\textbf{Hierarchical}:
The beam and filter corrections make it more difficult for the code to converge, since in every MCMC step the fluxes are slightly modified. This is more problematic for the hierarchical approach, because it has a larger number of free parameters. Therefore, in order to achieve convergence in a reasonable amount of time, we apply a slightly different approach to implement the beam and filter corrections in the hierarchical case. We first do the hierarchical fit without beam and filter corrections. Then we apply the beam and filter corrections on the fluxes based on the values of $T$ and $\beta$ measured from the fit with no corrections, and finally we repeat the hierarchical fit using the `corrected' fluxes. The beam and filter corrections are generally small compared to the flux uncertainties, therefore this approximation of the corrections does not affect the results significantly.

\subsection{Implementation of the SED fitting}
\label{sec:implementation}

\textbf{Non-hierarchical method:} 
For the implementation of the classical Bayesian SED fitting method, we employ the affine-invariant ensemble sampler for Markov Chain Monte Carlo \citep[MCMC, ][]{Metropolis1953} code {\tt emcee} \citep{Goodman2010,Foreman-Mackey2013}. 
The MCMC algorithm is designed to sample the posterior distribution of the unknown parameters, i.e. the probability of the parameters given the data. The values of the parameters with the corresponding uncertainties can then be inferred from the posterior distribution. We consider as results the median values of the marginalized posterior probability distributions, and we estimate the uncertainties from the values corresponding to the 16th and 84th percentiles.

To monitor the convergence we look at the effective sample size ($N_{eff}$), which is defined as the number of iterations divided by the integrated autocorrelation time $
N_{eff} = N_{iter}/\tau_{int}$. The autocorrelation time $\tau_{int}$ measures  the number of steps after which the drawings are truly independent \citep{Foreman-Mackey2013}. It is recommended to have at least $N_{eff}> 10$, to ensure that the sequence has converged  \citep{Gelman2004}. \\
\\
\textbf{Hierarchical method:}
For the implementation of the hierarchical Bayesian fitting we use \stan\ \citep[][http://mc-stan.org/]{Carpenter2017}, a software for Bayesian inference which employs the No-U-Turn sampler (NUTS), a variant of Hamiltonian Monte Carlo sampler. The Hamiltonian Monte Carlo (HMC) sampling \citep{Duane1987, Neal1994, Neal2011} is a form of MCMC sampling  which uses the gradient of the logarithmic probability function to accelerate the parameter exploration and the convergence to the stationary distribution \citep{Stan_manual}.
 The HMC algorithm is more efficient than other MCMC algorithms (as for example the Metropolis-Hastings algorithm)  in sampling the parameter space and in finding the region of high likelihood, because it samples the probability distribution with fewer samples. Therefore it is particularly well suited for problems with high dimension, as is the case for hierarchical models.
 For example, for the hierarchical fit of 100 galaxies using the SMBB model, which has three free parameters, the dimension is of the order $\sim 300$. 
Another advantage of \stan\  is that it can sample simultaneously the posterior distribution of parameters and hyper-parameters. \stan\ allows to define the model by specifying the probability distribution of each parameter (or hyper-parameter) independently, without the need of computing the full posterior distribution.
For the practical implementation, we used {\tt PyStan}\footnote{http://pystan.readthedocs.io/en/latest/  \\ http://mc-stan.org}, which is the Python interface to \stan\ \citep{pystan}.

The recommended method for monitoring the convergence of the MCMC chains in \stan\ is computing the potential scale reduction statistics $\hat{R}$ \citep{Gelman1992}, which  gives an estimate of the factor by which the scale of the posterior distribution may be reduced as the number of iterations goes to infinity. If $\hat{R}$ is large, it means that increasing the number of iterations is likely to improve the inference.  If $\hat{R}\sim 1$, then we can be confident that the number of iterations that we are using is large enough. Thus we set the requirement that for our runs $\hat{R}< 1.15$. We also check that the effective sample size $N_{eff}$ is always larger than 10.

\subsection{Validation of the method with simulations of mock SEDs}

We test our fitting methods using simulated FIR SEDs. For the mock SEDs, we know the input parameter values, thus we can assess how well our fitting procedure is able to recover them. 
 The simulation code takes as input parameters the dust mass ($\log M_{dust}$), temperature $T$, and emissivity index $\beta$, and it uses these parameters to generate an SED assuming a single modified black-body (SMBB) model. Then it extracts the flux density in the selected wavebands and it adds random noise at each flux point. We assume the noise to be Gaussian distributed around zero, with amplitude equal to the noise level. 
We assume a different noise level in every band. For the wavebands (100, 160, 250, 350, 500, 850) \micron, we use the following noise levels, given as percentages of the flux: (20, 10, 5, 10, 20, 25)\%, respectively. We estimate these values by taking the mean of the error fraction in each band from our data.

\begin{figure*}
\centering
\subfigure{\includegraphics[width=0.48\textwidth]
{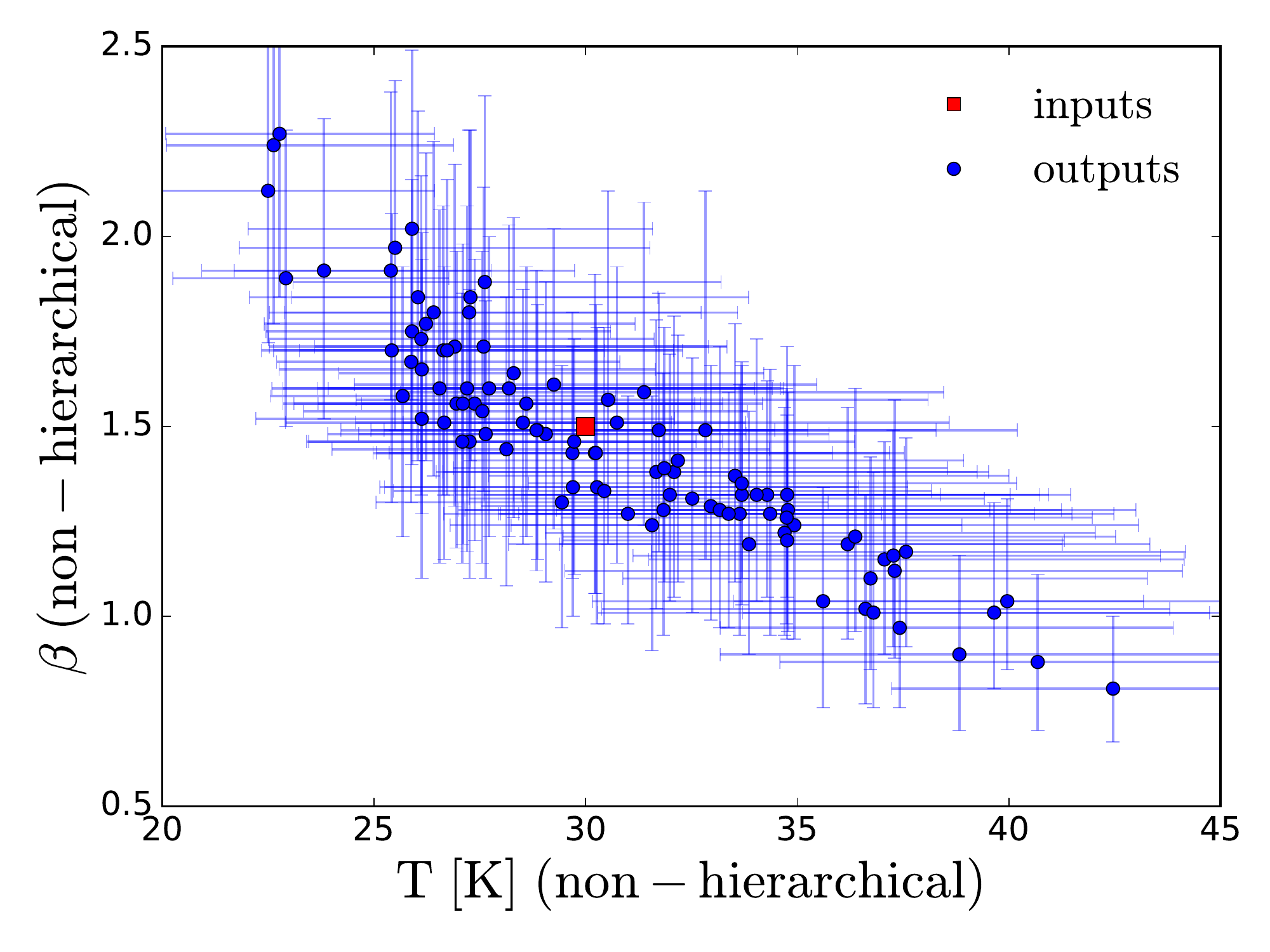}}
\subfigure{\includegraphics[width=0.48\textwidth]{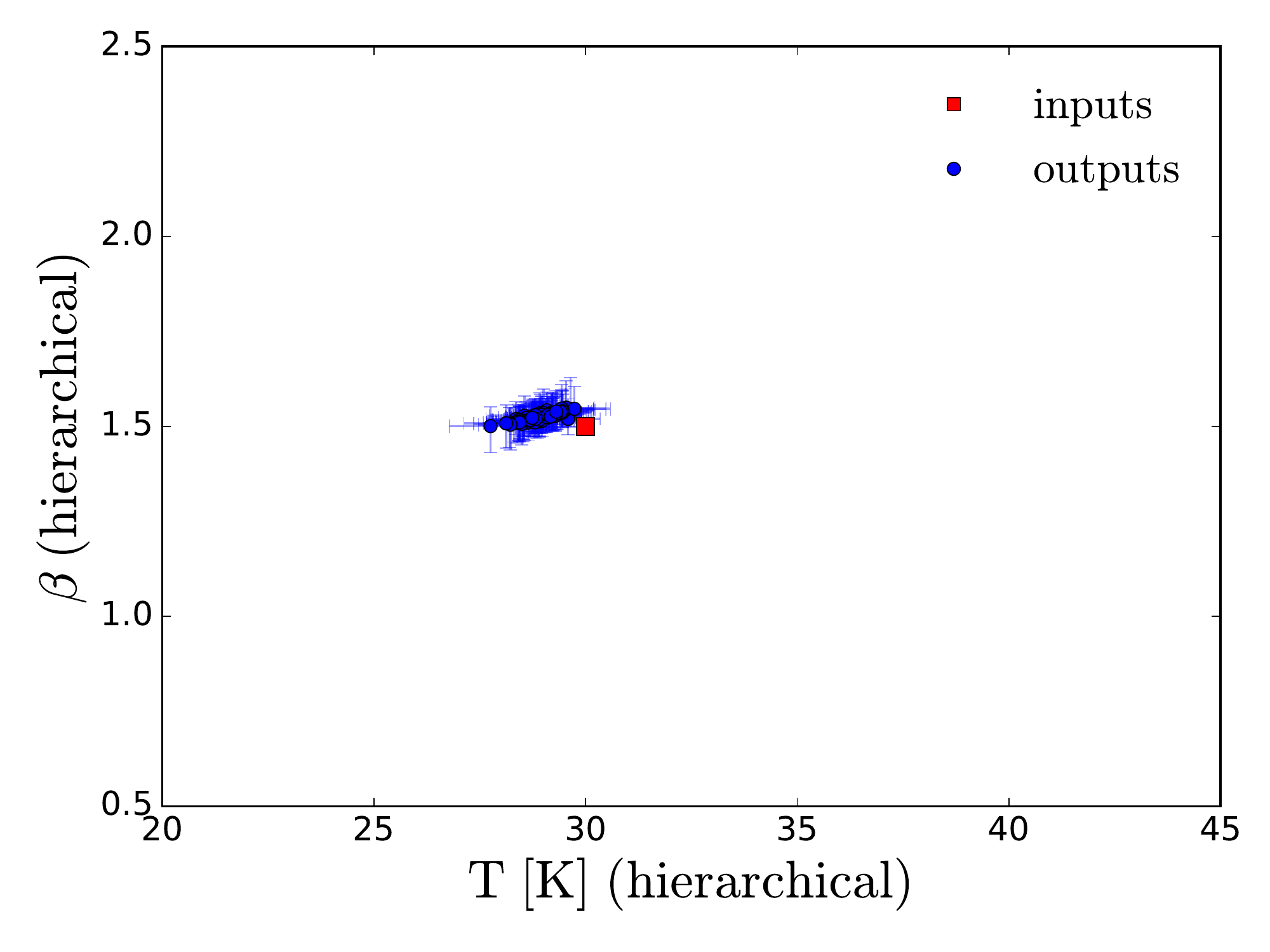}}
\caption{Results of temperature and $\beta$ from the fit of 100 simulated SMBB SEDs with the same input parameters ($\log M_{dust}/M_\odot=8$, $T=30$ K, and $\beta = 1.5$) and 10\% added noise. The output values are derived with the non-hierarchical (left panel) and  hierarchical (right panel) SED fitting method. In red is shown the input value and in blue are the measured values.  
}
\label{fig:T-beta_single_input} 
\end{figure*}

The goal of the test is to assess how well the non-hierarchical Bayesian approach can measure the values of temperature and $\beta$. 
 We simulate 100 SEDs with the same input parameters ($\log M_{dust}=8 $ \Msun, $T= 30 $ K, $ \beta =1.5$), adding to every SED  random noise in every band as explained above. 
 Figure \ref{fig:T-beta_single_input} shows the results in the $T$-$\beta$ plane. As we can see from the figure, an artificial anti-correlation is generated only from the effect of adding noise to the fluxes. This suggests that the non-hierarchical Bayesian approach will always measure a $T$-$\beta$ anti-correlation, even if it is not present in the data. Thus, in order to asses if the $T$-$\beta$ anti-correlation is indeed present in our sample, we need a more sophisticated fitting method.
 
We run the same simulation, but this time we use  the hierarchical code to fit the SEDs. 
The results are in better agreement with the input value, and do not show any artificial correlation or anti-correlation between $T$ and $\beta$. 
 The non-hierarchical method measures a large range of temperatures ($T=22-42$~K) and $\beta$  values ($\beta=0.8-2.3$).
The hierarchical method measures  smaller ranges of $T=27-30$~K and $\beta=1.50-1.55$, which are closer to the input values. 
Consequently, also the dust masses are better measured with the hierarchical method. The dust masses measured with the non-hierarchical method are in the range $\log M_{dust}/M_{\odot}=7.87-8.23$, with typical uncertainties of $\sim0.13$~dex, while the ones measured with the hierarchical method are in the range $\log M_{dust}/M_{\odot}=8.06-8.09$, with typical uncertainties 0.02~dex.

 We also test whether the codes can recover a positive or negative $T$-$\beta$ correlation. In both cases, the hierarchical method perform equally or better than the non-hierarchical code. Details of these simulations can be found in appendix~\ref{sec:simulation}.

\section{Results}
\label{sec:results} 
\subsection{JINGLE sample: non-hierarchical vs. hierarchical results}
In the previous section we have demonstrated,  using simulated SEDs, that the hierarchical method works better than the non-hierarchical approach. Here we apply both methods to the \Ntot\ galaxies of the JINGLE sample and we show the advantages of using the hierarchical method.
 
\begin{figure*}
\centering
\subfigure{\includegraphics[width=1.0\textwidth]
{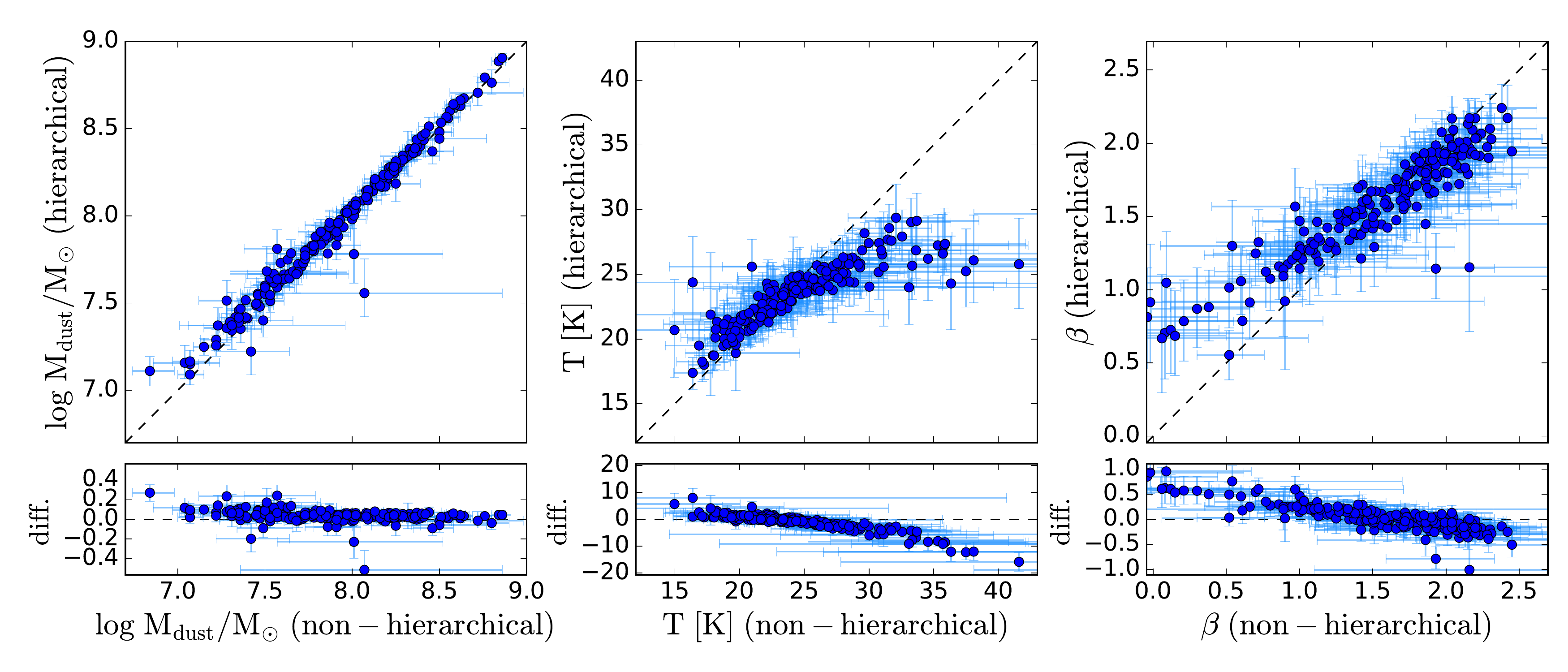}}
\caption{Comparison of dust properties of the JINGLE sample  obtained through the fit of a single modified black-body (SMBB) using the non-hierarchical and hierarchical approaches. The lower panels show the difference between the hierarchical and non-hierarchical fit in each of the derived properties.
}
\label{fig:comp_hier_nonhier} 
\end{figure*}

We start by using the simplest model, the single modified black-body (SMBB). 
Figure \ref{fig:comp_hier_nonhier} shows the comparison of the dust masses, dust temperatures and $\beta$ derived with the two approaches.
In general, dust masses agree quite well between the two methods (median difference = 0.07 dex). The dust masses derived using the hierarchical method are slightly smaller, and this is probably due to the variations in dust temperatures. For a given constant flux, higher dust temperatures correspond to lower dust masses. 
 In the range $15-25$~K the dust temperatures from the hierarchical approach are indeed slightly higher.
 At high temperatures, the differences between the two methods are larger and  the non-hierarchical method measures much higher temperatures ($T>\ 30$ K) than the hierarchical method.
 This is because as the dust temperature increases, the peak of the SED moves to shorter wavelengths. If the SED peaks at wavelengths shorter than 100\micron, it is not sampled by the flux bands considered in the fit, since for the SMBB we are considering the 60\micron\ point as an upper limit. Therefore it is more difficult to constrain the temperature. If we were to include flux points at shorter wavelengths we would need to consider a second MBB component with a warmer temperature, because the assumption of a single temperature MBB does not hold over such a large wavelength range.  Instead, in the hierarchical framework, the code uses the information from the temperature distribution of the galaxy population to constrain $T$, and it will consider more likely for the galaxy to have a temperature close to the population mean temperature than an extreme value. Therefore the hierarchical method can better constrain the dust temperature.

The range of temperatures is smaller in the hierarchical case ($T = 17-30$ K), than in the non-hierarchical case ($T = 15-48$ K).  The same is true for the range of $\beta$: in the hierarchical case $\beta= 0.6-2.2$, while in the non-hierarchical case $\beta= 0.0-2.5$. In the hierarchical approach, we assume that the population follows a common distribution, thus the fitting is less likely to return extreme values of $\beta$. However, the hierarchical code can accommodate some outliers, since we do not define a priori the standard deviation of the prior distribution. Thus if the data require it, the standard deviation can be large, allowing for more `extreme' values of $\beta$.
  But if the extreme objects have large noise on the flux values, then the hierarchical method considers more likely that they are not `true outliers', but that their extreme SED shape is only due to the noise in the data points.  
If we believe that the hierarchical approach gives more accurate results for the cases with high noise level, we conclude that the extreme values found with the non-hierarchical approach are likely not reliable, but only due to the noise in the data.
 The results of the hierarchical fit using the SMBB model are given in Table \ref{tab:results_SMBB}.

\subsection{$T$-$\beta$ relation in the JINGLE sample}

We use the results of the SED fitting using the SMBB model to investigate whether there is a relation between dust temperature and  $\beta$  in our sample of galaxies. An anti-correlation between $T$ and $\beta$  has been observed in many studies \citep[e.g.][]{Dupac2003, Desert2008}, but it has been demonstrated that it can be attributed to the degeneracy between the two parameters and the effect of noise on the data \citep{Shetty2009a, Shetty2009b}.

\begin{figure*}
\centering
\subfigure{\includegraphics[width=0.45\textwidth]
{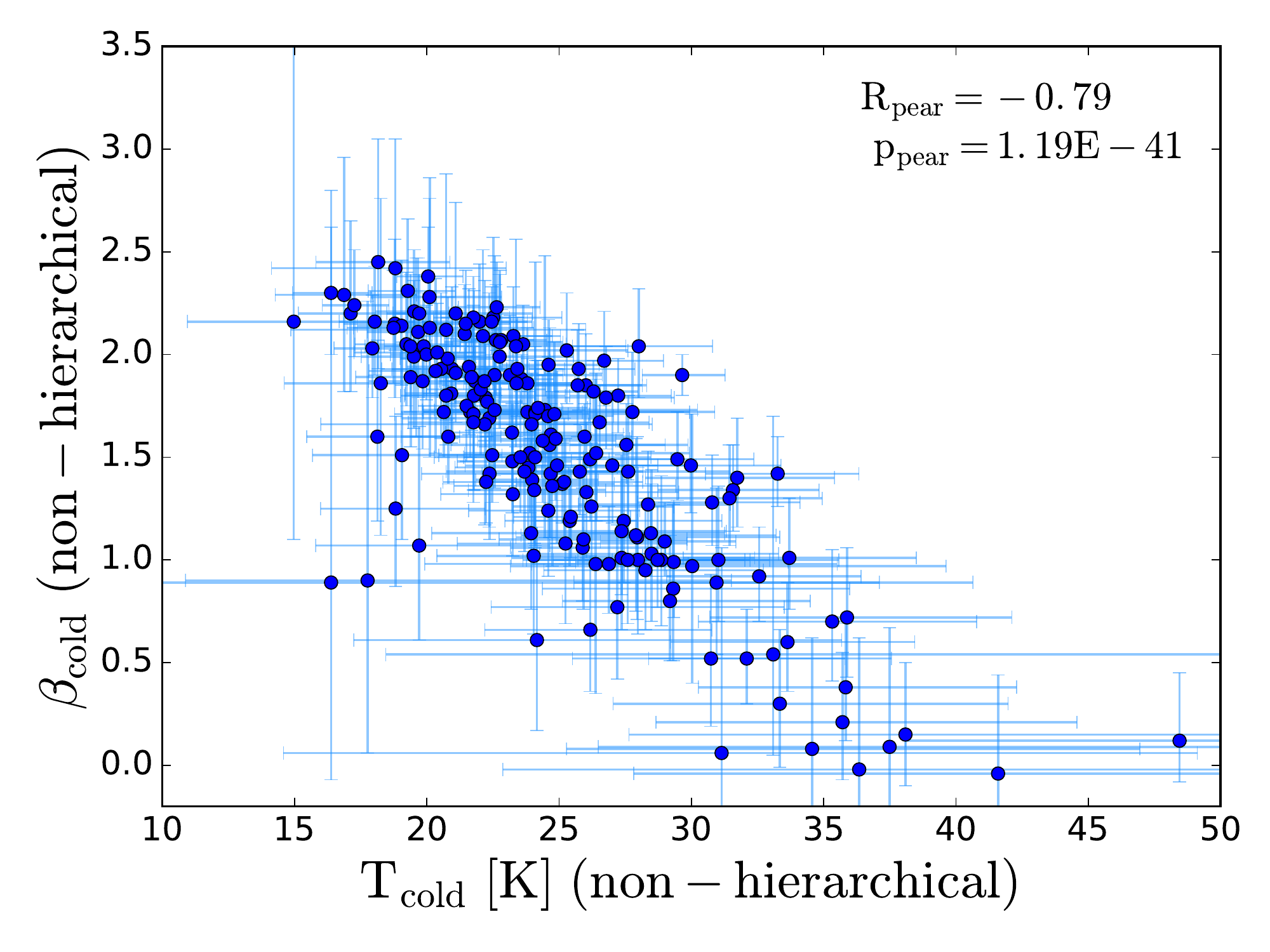}}
\subfigure{\includegraphics[width=0.45\textwidth]
{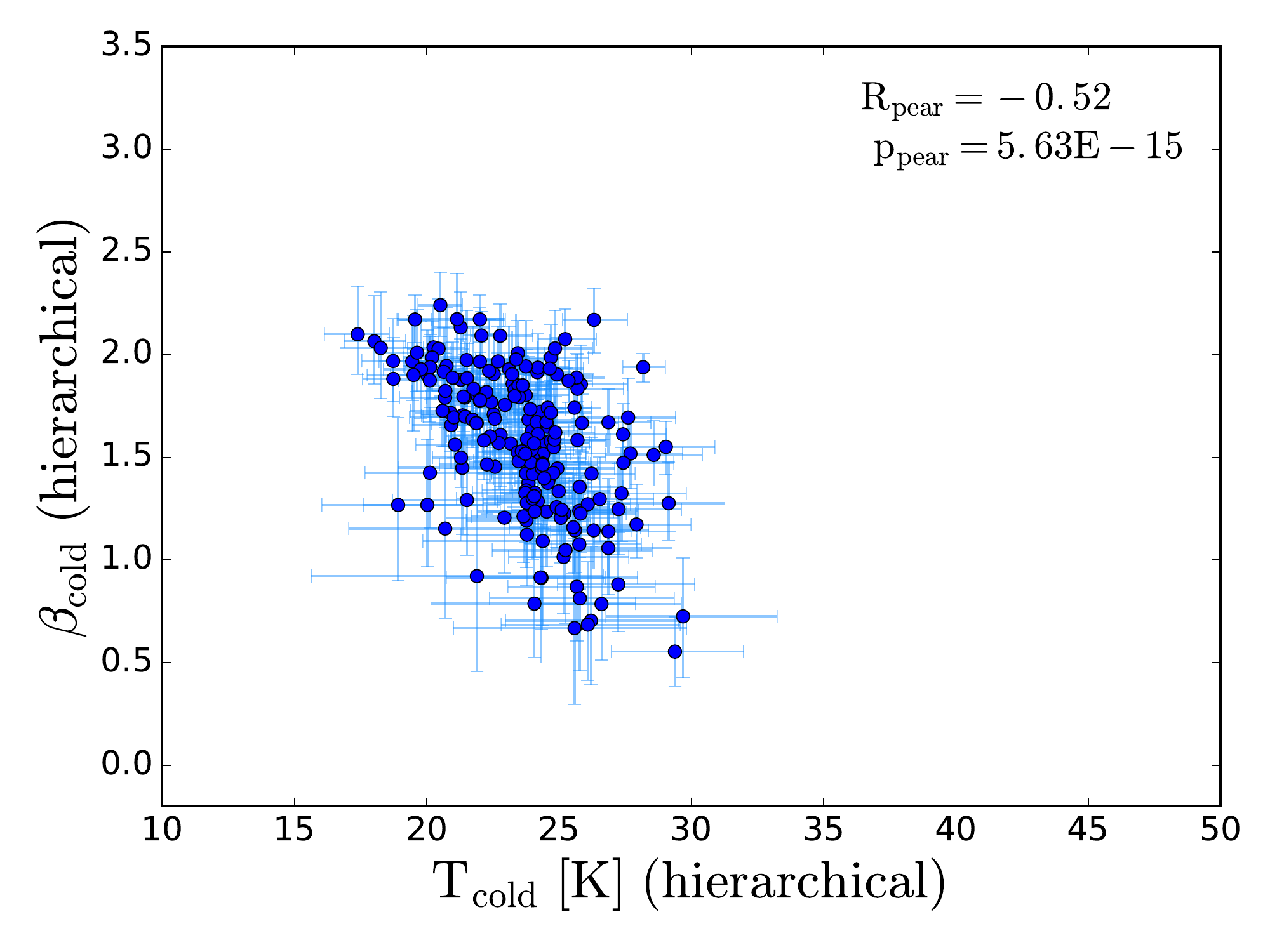}}

\caption{Relation between the dust temperature and dust emissivity index ($T$-$\beta$ relation) for the JINGLE sample derived with non-hierarchical  (left panel) and hierarchical (right panel) Bayesian methods. In both cases, we fit the SED using a single modified black-body (SMBB) model and we include the 850\micron\ flux point in the fit.
}
\label{fig:T-beta} 
\end{figure*}

Figure \ref{fig:T-beta} shows the results from the non-hierarchical and hierarchical approach applied to our sample of \Ntot\ galaxies.
The results from the non-hierarchical method show a significant anti-correlation between $T$ and $\beta$. The Pearson correlation coefficient is $R_{pear}= -0.79$ (p-value $= 1.19\cdot 10^{-41}$). The results from the hierarchical method shows a weaker anti-correlation ($R_{pear}= -0.52$, p-value $= 9.79\cdot 10^{-15}$). 
This shows that the choice of the method used is really important and can deeply influence the results. 
This result confirms previous findings \citep{Shetty2009a,Shetty2009b,Kelly2012, Veneziani2013, Juvela2013} that the observed $T-\beta$ anti-correlation is mainly driven by the fact that they are degenerate parameters, and by the noise on the data.
 There is still an anti-correlation between $T$ and $\beta$ even using the hierarchical approach ($R_{pear}= -0.52$). This could mean that there is indeed a physical relation between these two quantities. However, it is also possible that the hierarchical method is not able to remove completely the $T-\beta$ degeneracy, leaving a residual anti-correlation. With our current data we are not able to distinguish whether the observed relation is a physical effect or whether it is due to a residual degeneracy.

\begin{figure*}
\centering
\subfigure{\includegraphics[width=0.9\textwidth]
{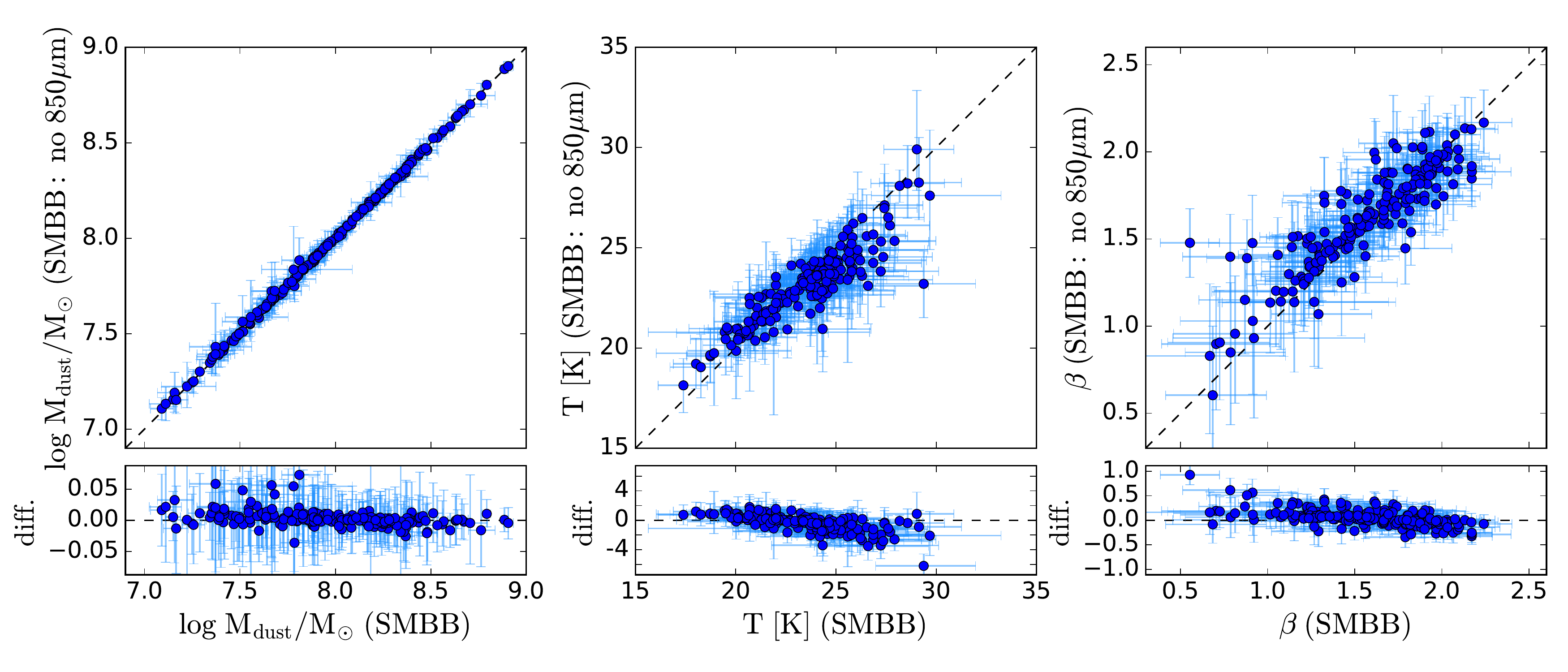}}
\caption{Comparison of the dust masses, temperatures and emissivity indices obtained through the fit of a single modified black-body (SMBB) using the hierarchical approach, with and without the SCUBA-2 flux point at 850\micron.  The lower panels show the difference between fit with and without the SCUBA-2 flux in each of the derived properties.
}
\label{fig:comp_with_without_scuba} 
\end{figure*}

We also compare the results obtained with and without including the 850\micron\ flux point in the fit using the hierarchical approach (see Figure \ref{fig:comp_with_without_scuba}). 
 In general, the emissivity indices $\beta$ measured with the 850\micron\ flux point are equal or lower than the ones measured without the 850\micron\ point. This means that without the SCUBA-2 flux, the fits of the \Herschel\ points alone have  steeper slopes. This suggests that there is indeed a `submm' excess visible at 850\micron, at least in some galaxies. This is visible especially for low values of $\beta < 1$. We note that not all galaxies show this behaviour: for some galaxies the $\beta$ values measured with and without SCUBA-2 flux are in good agreement, or they show a small deficit at 850\micron.
 Consequently, the dust temperatures show the opposite trend: they are in general larger when the 850\micron\ point is included in the fit, because they have to compensate for the lower $\beta$ values.
 The mass measurements are only slightly affected by the presence of the SCUBA-2 flux point (median difference: 0.002 dex). The largest difference in the dust masses measured with and without the SCUBA-2 flux point is 0.07 dex. 
 The fact that the dust masses do not show a larger variation depends on the fact that we assumed a constant absorption coefficient $\kappa_0$. Laboratory studies show that $\kappa$ changes with dust temperature $T$ and $\beta$ \citep{Coupeaud2011, Demyk2017a, Demyk2017b}. 
 Therefore, by keeping $\kappa$ constant we erase the difference in dust masses that would arise from the different temperature and $\beta$ values.  A certain value of $\kappa_0$ will give an accurate dust mass only if the $\beta$ value used for the fit is the same that was used to measure $\kappa_0$  \citep{Bianchi2013}.
 However, a recent  laboratory study by \cite{Demyk2017a} shows that variations in $\kappa_0$ are more prominent for high temperatures ($T > 30$~K) than for low temperatures. For the temperature range considered in this study ($10-30$~K) they do not observe variations in $\kappa_0$.
 A possible approach to account for variations in $\kappa_0$ would be  to change the value of $\kappa_0$ according to the value of $T$ and $\beta$ used for the fitting in an iterative way.
 We plan to investigate this in the future.

\subsection{Comparison of models: SMBB, BMBB, TMBB}

In many cases, the SMBB model is not enough to fit the FIR/submm SED accurately. Especially at long wavelengths, the SED often shows a change in the slope. Therefore we consider also two other models: the broken emissivity law modified black-body (BMBB) and two modified black-bodies  TMBB models, described in Section \ref{sec:models}. In this section we compare the results obtained applying these models to the SED fit of the JINGLE sample. The results of the hierarchical fit using the BMBB and TMBB models are given in Tables \ref{tab:results_BMBB} and \ref{tab:results_TMBB}.

\subsubsection{BMBB}
\begin{figure*}
\centering
\subfigure{\includegraphics[width=0.9\textwidth]
{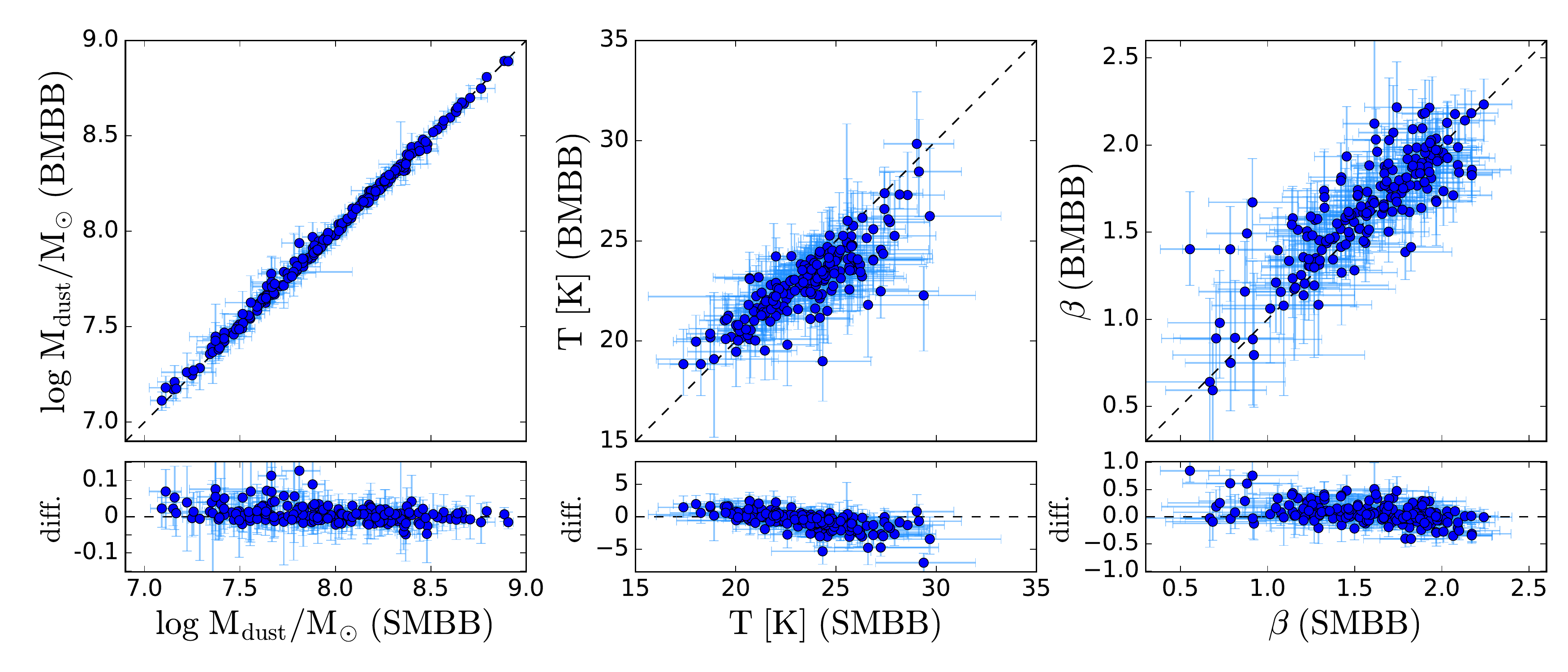}}
\subfigure{\includegraphics[width=0.9\textwidth]
{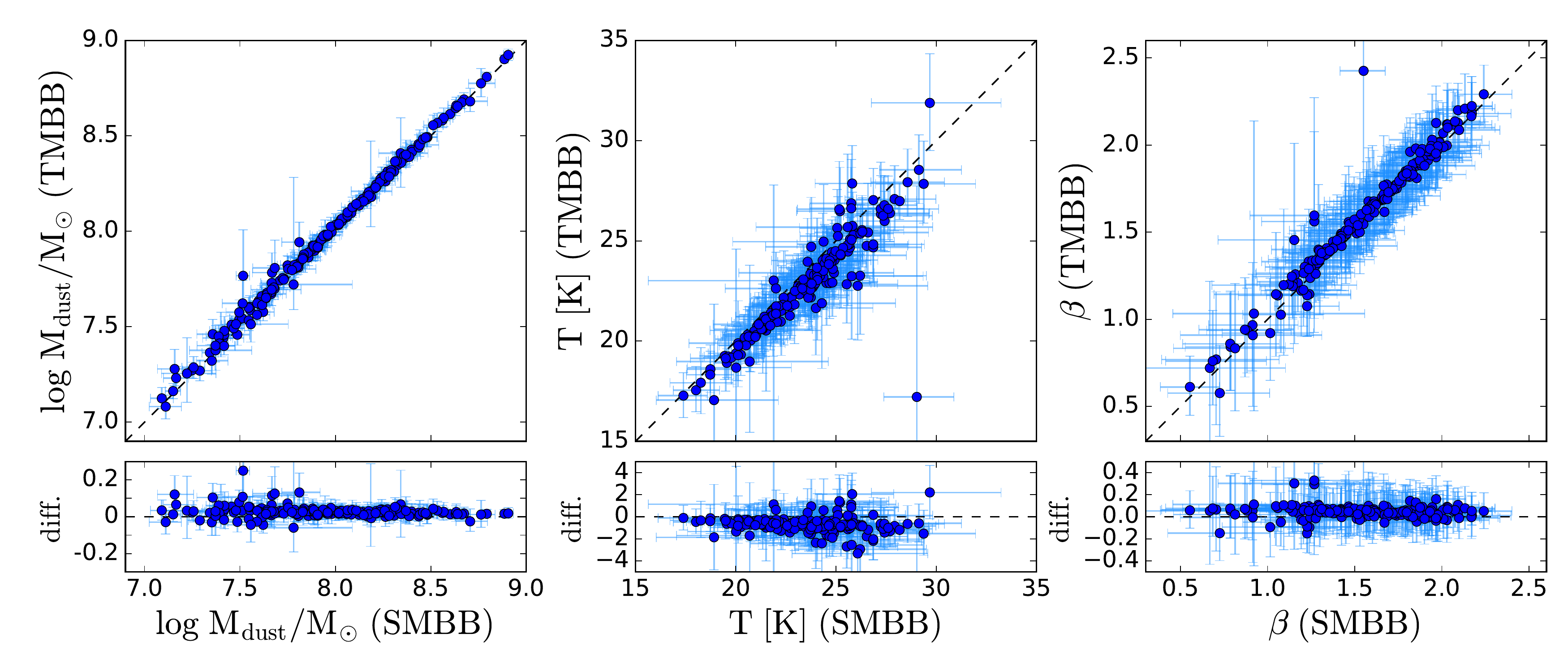}}

\caption{ \textit{Upper panels}: Comparison of the cold dust masses, temperatures and emissivity index obtained through the fit of a single modified black-body (SMBB) and a broken emissivity power law MBB model (BMBB). For the BMBB model, the $\beta$ value shown in the plot is $\beta_1$, i.e. the emissivity index at wavelength < $\lambda_{break}$. The lower sub-panels show the difference between the two models in each of the derived properties. \textit{Bottom panels:} Comparison of the results from the SMBB and  two modified black-bodies (TMBB) model. For the TMBB model, the values shown in the plot are the parameters of the cold component ($\log M_{cold}$, $T_{cold}$, $\beta_{cold}$).}
\label{fig:comp_SMBB_BMBB_TTMBB} 
\end{figure*}

The broken emissivity law modified black-body model \citep[BMBB,][]{Gordon2014} allows for a variation in the wavelength dependence of the dust emissivity law, to account for a submm excess. This is parametrized by using two emissivity indices for shorter and longer wavelengths.  
 The break wavelength is a free parameter in our model.
 For the JINGLE sample we find values in the range $480-488$\micron. The emissivity index at wavelengths shorter than $\lambda_{break}$ ($\beta_1$) is in the range $0.6-2.2$. The range of the second emissivity index at wavelengths $> \lambda_{break}$ ($\beta_2$) is larger ($0.1-3.3$).
 We compared the results obtained using the BMBB model with the results from the SMBB model (Fig.\ref{fig:comp_SMBB_BMBB_TTMBB}). 
The dust masses measured with the BMBB model are in agreement with the ones measured with the SMBB model, with a maximum difference of 0.1~dex.
The BMBB model measures generally slightly lower temperatures than the SMBB model (median difference of 1 K). In the case of a shallower slope of the submm SED, the SMBB model fits it by using a lower value of $\beta$ and a higher $T$. The BMBB can correct using a smaller value of $\beta_2$, without affecting the temperature measurement. Thus $T$ does not depend anymore on the longer wavelength points and can have a lower value. We compare also the emissivity index $\beta$ from the SMBB model, with the parameter $\beta_1$ which describes the slope of the BMBB model before the break. $\beta_1$ tends to be larger than $\beta$ from the SMBB for low values of $\beta$. This is due to the fact that any excess at longer wavelength can be modelled by a second index $\beta_2$, while in the case of the SMBB the excess needs to be taken into account by $\beta$.

\begin{figure*}
\centering
\subfigure{\includegraphics[width=0.44\textwidth]
{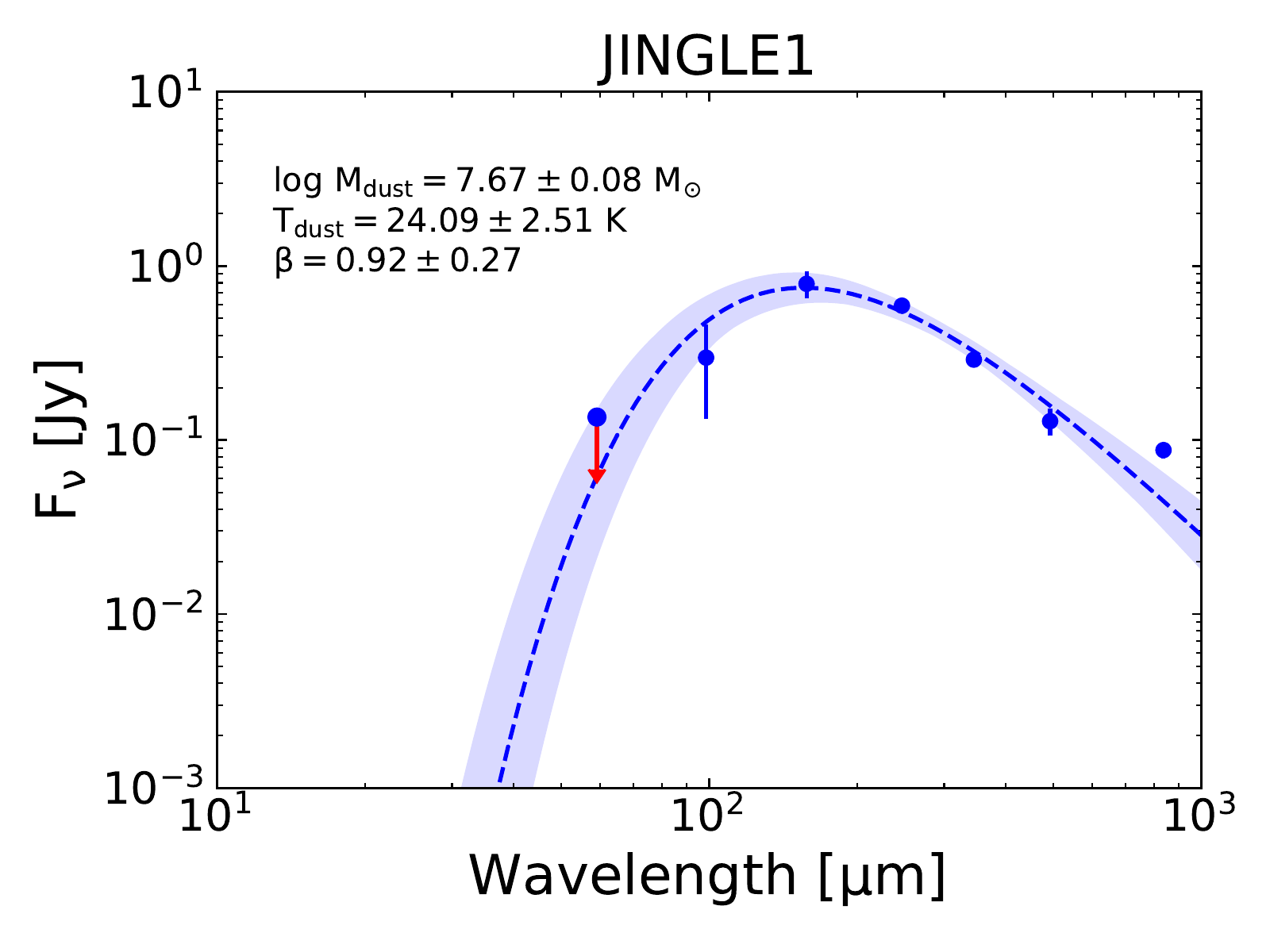}}
\subfigure{\includegraphics[width=0.44\textwidth]
{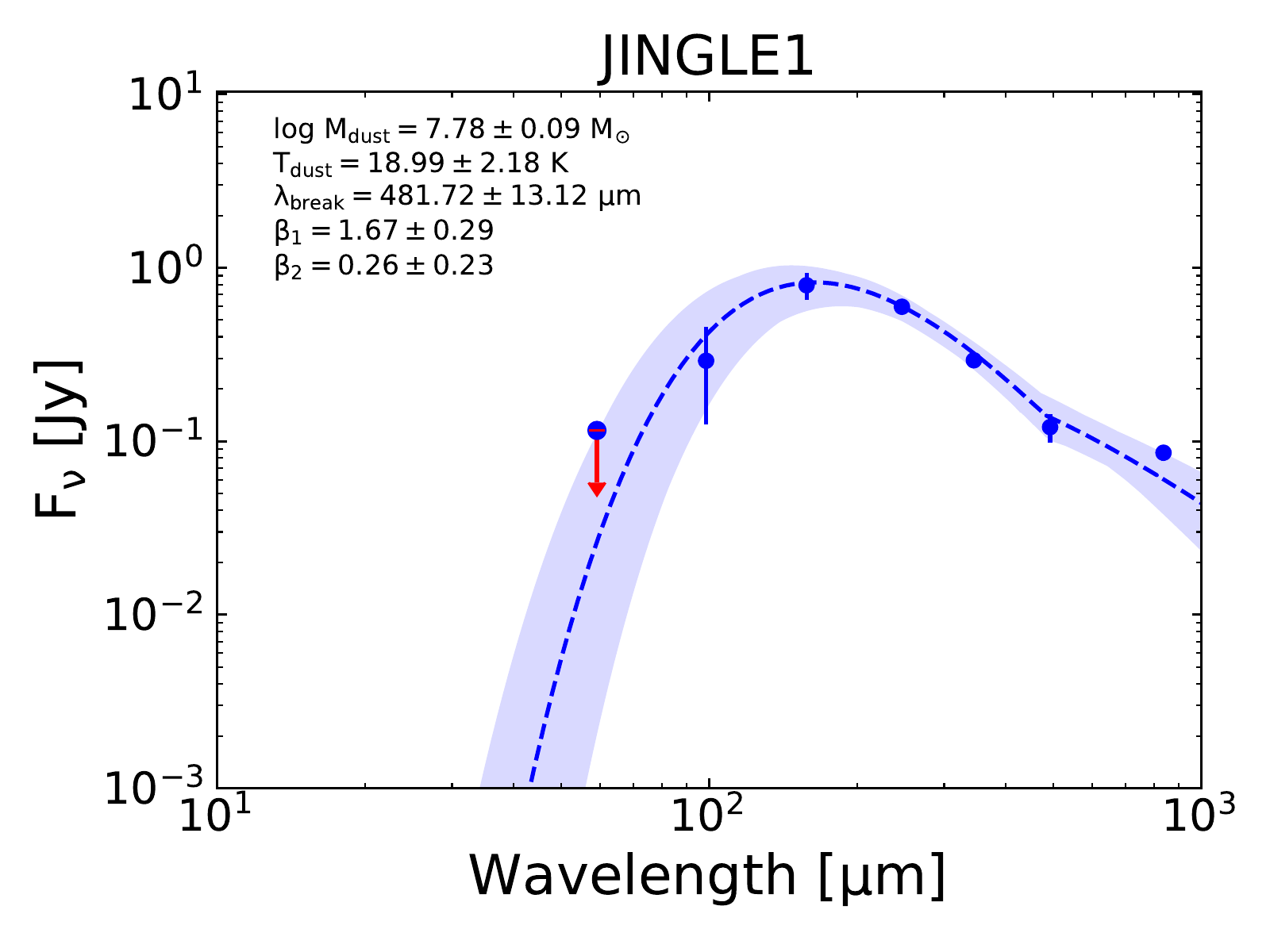}}
\caption{SMBB and BMBB fit for the galaxy JINGLE 1, where there is a clear difference in the dust temperature measured with the two different models.}
\label{fig:outliers_comp_SMBB_BEMBB} 
\end{figure*}

The results from the BMBB model are more similar to the SMBB fit without the 850\micron\ point. This is due to the fact that the BMBB model fits the fluxes at longer wavelengths (500\micron\  and 850\micron\ point) using a second emissivity index $\beta_2$, thus the measurements of $T$ and $\beta_1$ are not sensitive to the flux measurement at 500\micron\ and 850\micron. Figure \ref{fig:outliers_comp_SMBB_BEMBB} shows an example of the SMBB and BMBB fit of one  galaxy for which the difference in temperature is more evident (JINGLE 1). 
This model is especially useful to quantify the possible sub-mm excess, given by the difference between the two emissivity indices $\beta_1$ and $\beta_2$. Further discussion on the submm excess is presented in Section \ref{sec:submm_excess}.

\subsubsection{TMBB}
The bottom panels of Figure \ref{fig:comp_SMBB_BMBB_TTMBB} show the comparison of the SMBB and the two MBB model (TMBB). The dust masses are in good agreement, with the cold dust masses derived from the TMBB being slightly higher (median offset: 0.03 dex). 

\begin{figure*}
\centering
\subfigure{\includegraphics[width=0.44\textwidth]{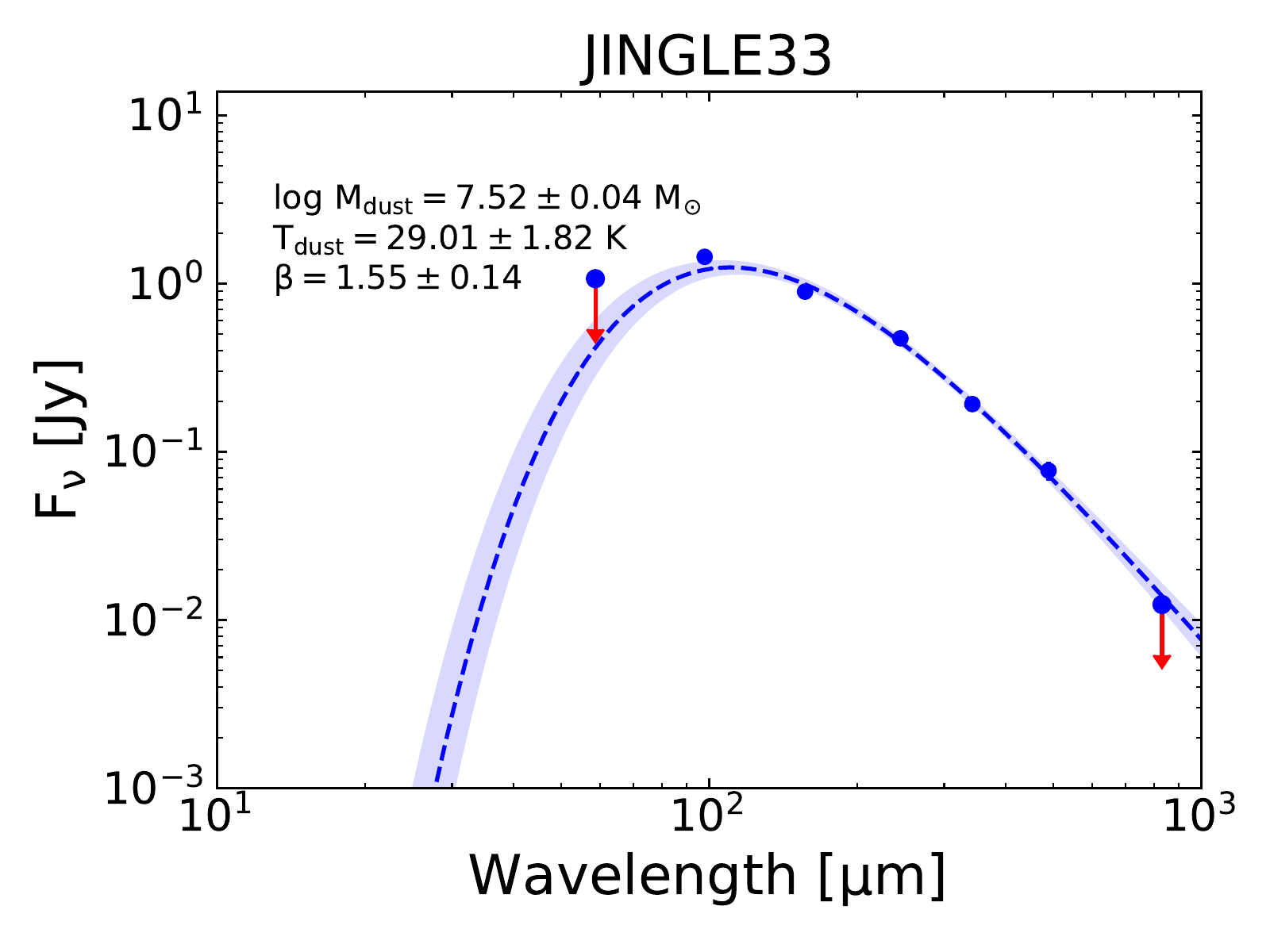}}
\subfigure{\includegraphics[width=0.44\textwidth]
{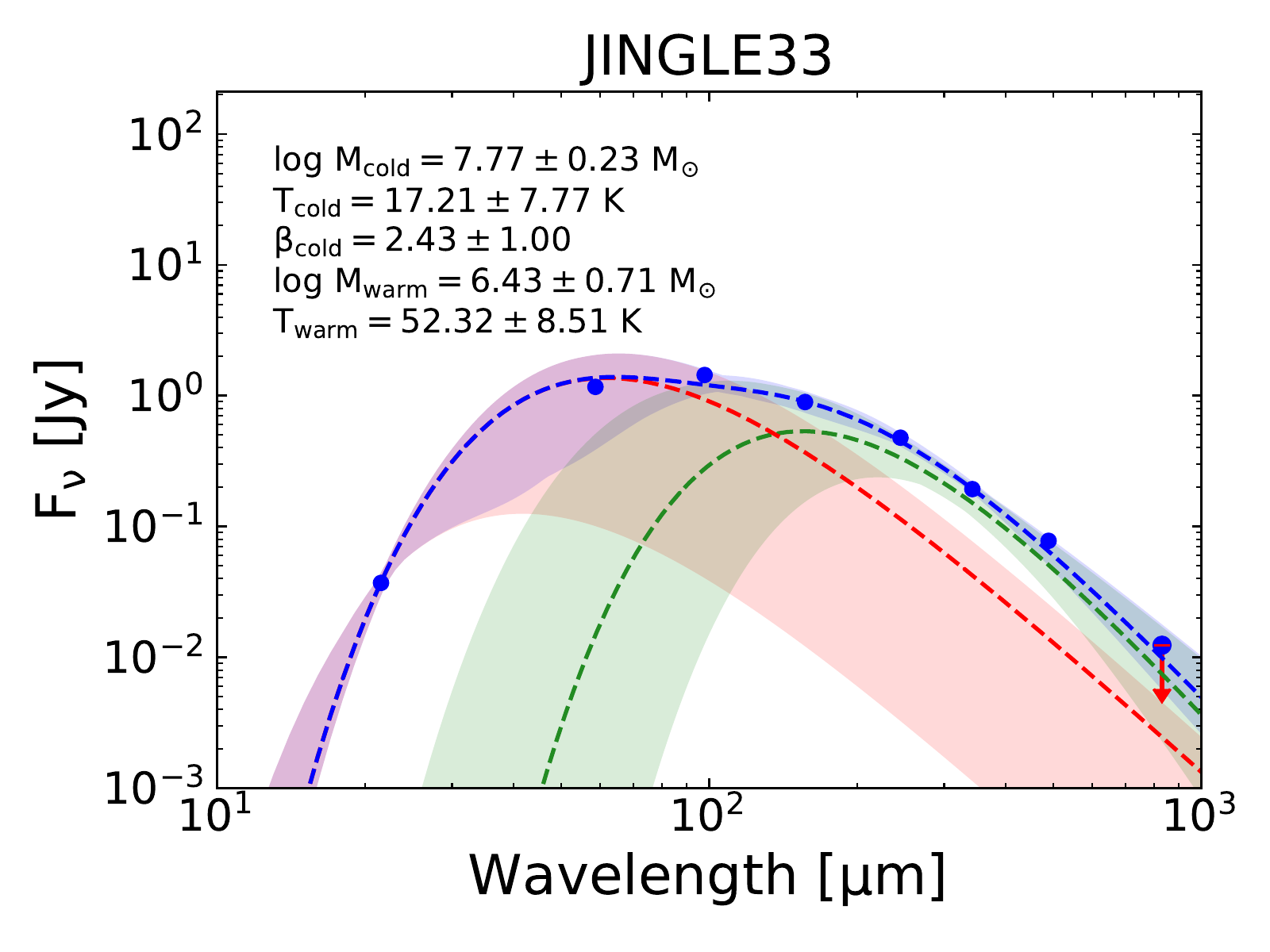}}
\caption{SMBB and TMBB fit for the galaxy JINGLE 33, which shows a  clear difference in the cold dust mass measured with the two different methods. The warm component has a large contribution to the total dust emission  in this galaxy.}
\label{fig:outliers_comp_SMBB_TMBB} 
\end{figure*}

The dust temperatures of the cold component obtained with the TMBB model tend to be  lower than the ones measured from the SMBB model by about 3\% (or 0.8 K). This is expected, since the warm component is contributing to the fit of the 100\micron\ flux, allowing the cold component to shift to longer wavelengths, corresponding to colder temperatures. Consequently, the $\beta_{cold}$ values from the TMBB are also slightly higher (median offset: 0.05).
 The outlier is JINGLE 33 (Fig.~\ref{fig:outliers_comp_SMBB_TMBB}). This galaxy has a high 60\micron\ flux, compared to the 100\micron\ flux, which results in the warm dust component (with $T_{warm}=52.3$~K) reproducing most of the emission, and skewing the cold dust component to a lower temperature ($T_{cold}=17.2$~K) and a higher dust mass.

The warm dust component does not contribute much to the entire dust mass. Warm dust masses are in the range $10^{3.4}-10^{6.6}$ M$_\odot$, which correspond to only 0.01-4.4\% of the total dust mass of the galaxies. 
  Nevertheless, it is important to take into account this component because, as we have shown, it can affect the measurement of the temperature and emissivity $\beta$ of the cold component. The temperatures of the warm component are in the range $66-76$~K, with the exception of JINGLE 33 which has a lower temperature (52.3~K).

If we compare the total dust masses ($M_{dust, tot} = M_{cold}+M_{warm}$) from the TMBB with the cold dust masses $M_{cold}$ from the SMBB, the latter are smaller by 10\% ($\sim0.08$~dex) on average. 
Other studies found that fitting the SED using the TMBB model will result in higher cold dust masses. For example \cite{Gordon2014} found that the dust masses of the Small and Large Magellanic Clouds are 6-15 times larger when estimated using a TMBB model instead of the SMBB model.
\cite{Clark2015} found that the warm dust mass can contribute up to 38\% of the total dust mass of galaxies in the \textit{Herschel}-ATLAS survey.
  The disagreement with our findings is probably due the fact that these studies do not include the 22\micron\ flux point in their fit. Consequently, their warm component is shifted to longer wavelength and has lower temperature than ours, thus contributing more to the total dust mass. The cold dust temperature of the TMBB will also be smaller than in the SMBB case, thus resulting in higher cold dust masses.

\subsection{Model comparison with information criterion}
In order to decide which of the models provides a better fit to the data, we applied a criterion based on the comparison of the likelihoods. We consider the Bayesian Information Criterion (BIC) \citep{Schwarz1978} which takes into account not only the likelihood of the fit, but also the number of free parameters of the models. 
 The latter point is important, since increasing the number of free parameters would generally lead to better fits. 
The Bayesian Information Criterion (BIC) \citep{Schwarz1978} is defined as:
\begin{equation}
BIC = -2\cdot \ln(L) + q\cdot \ln(m)\,
\end{equation}
where $L$ is the likelihood (i.e. the probability of the data given the parameter $p(\vec{F}|\vec{\theta})$), 
 $q$ is the number of free parameters of the model, and $m$ is the number of data points (wavebands).
The model with the lowest BIC value is the preferred model according to this criterion. To calculate the likelihood $L_i$ for the $i$-th galaxy we consider the product of the likelihood $p(F_{i,j}^{obs}|\vec{\theta_i}, F_{i,j}^{err}, \delta_j)$ in all wavebands $j=1,...,m$.
\begin{equation}
L_i = \prod_{j=1}^m p(F_{i,j}^{obs}|\vec{\theta_i}, F_{i,j}^{err}, \delta_j)
\end{equation}

\begin{figure}
\centering
\subfigure{\includegraphics[width=0.44\textwidth]
{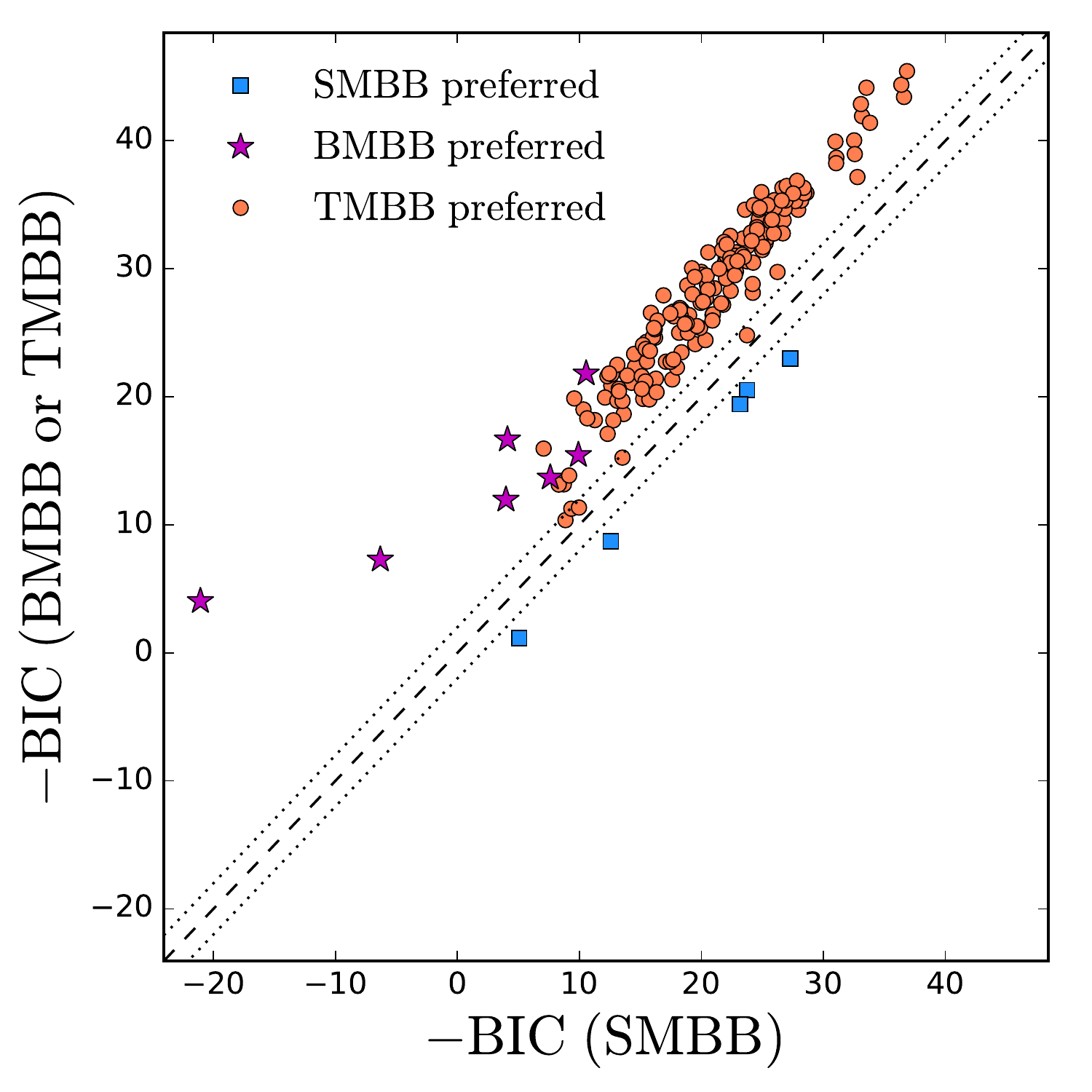}}
\caption{Comparison of the negative Bayesian Information Criterion (-BIC) for the fit using the three models: SMBB, BMBB, TMBB. The model with the largest value of -BIC is the preferred model. If the difference between the BICs is smaller than two (shown by the dotted lines) there is little evidence to prefer one model over an other.
}
\label{fig:comp_info_crit} 
\end{figure}

Figure \ref{fig:comp_info_crit} shows the BIC values for the BMBB and TMBB models compared to the SMBB model. For most of the galaxies (180/\Ntot, 94\%), the TMBB model is preferred. This is probably due to the fact that the additional warm component can help to improve the fit at 100\micron, without affecting the fit of the points at longer wavelengths.

For seven galaxies the preferred model is the BMBB model (JINGLE ID: 35, 56, 77, 101, 118, 133, and 147). In all these galaxies there is a clear submm excess at 850\micron. 
  The BIC criterion does not identify all galaxies for which the 850\micron\ flux is enhanced with respect to the SMBB model, but selects the ones for which the discrepancy can not be attributed to flux uncertainties or uncertainties in the model.
 
There are five galaxies which are best modelled with the SMBB model (JINGLE ID 83, 110, 142, 159, and  186).  
 The TMBB model is not able to fit well the 60\micron\ and 100\micron\ flux points of these galaxies. For JINGLE 83 and JINGLE 159 the 60\micron\ flux is too low and is not well fitted by the TMBB model. For JINGLE 110 the 60\micron\ flux is instead too high compared to the 100\micron\ flux. 
 For JINGLE 186, the uncertainty on the 60\micron\ flux is very small, and therefore even a small deviation from the perfect fit of that data point results in a low likelihood. In JINGLE 142, the 500$\micron$ point is enhanced with respect to the 350\micron\ flux point and the 850\micron\ upper limit. In general neither the SMBB and TMBB models are able to produce a good fit for this galaxy. 
 The SED fits with the BMBB and TMBB models for all galaxies are shown in Figure \ref{fig:three_SED_models}.

We conclude that the TMBB model produces the best fit of the FIR SED for most of the galaxies. Additionally, the comparison of the BIC of the SMBB and BMBB model can be used to identify galaxies which show a strong submm excess or deficit.

\section{Relation between dust properties and galaxy properties}
\label{sec:dust_scal_relation}

In this section, we investigate how dust properties correlate with global galaxy properties.
We use the results obtained using the SMBB model, even though the TMBB model is preferred according to the Bayesian information criterion. We decide to use the SMBB model because one of the goals of this analysis is to provide prescriptions to estimate $T$ and $\beta$ from other galaxy quantities. These prescriptions can be useful in those cases where only a few photometric data points are available and in such cases it is preferred to use the model with the smallest number of free parameter (i.e. the SMBB model). Additionally, as we have shown in the previous section, the differences in $T$ and $\beta$ derived from the SMBB and the TMBB models are not very large and they are mainly systematic shifts, that can be accounted for.

 We include in this analysis also the galaxies from the \Herschel\ Reference Survey \citep[HRS,][]{Boselli2010b}, which allow us to extend the parameter range to lower SFR and specific SFR, since a large fraction of the HRS sample are galaxies which lie below the star-formation main-sequence (see Fig.~\ref{fig:SFR_Mstar_JINGLE_HRS}). 
  In this case, the total sample of galaxies consists of two populations: star-forming galaxies (main-sequence galaxies) and passive galaxies (below main-sequence). Therefore the basic assumption for the use of the hierarchical method that all galaxies belong to the same population does not hold any more. We therefore divide the `total' sample (JINGLE+HRS) into two sub-samples according to their position in the SFR-$M_*$ plane and fit each separately.  
  In this way, the assumption that the galaxies in one sub-sample belong to the same population is still valid.
We define the two sub-samples as follows:
\begin{itemize}
\item \textit{main-sequence galaxies/ star-forming sample}: galaxies belonging to the SF main-sequence or laying above it. 
This sample consists of all galaxies which fall above the lower limit of the SF main-sequence, defined as 0.4 dex below the SF main-sequence from \cite{Saintonge2016}.

\item \textit{below main-sequence sample/passive sample}: galaxies laying below the SF main sequence. These are the galaxies which lie more than 0.4 dex below the SF main-sequence defined by \cite{Saintonge2016}.
\end{itemize}

The star-forming sample consists of 313 galaxies (177 from JINGLE and 136 from HRS) and the passive sample of 159 galaxies (15 from JINGLE and 144 from HRS). We did a test fitting galaxies belonging to the two sub-samples together. This test confirms that it is necessary to separate the sample in two, to avoid to force the two sub-samples to move toward a common mean, introducing systematic biases in the results.

\begin{figure*}
\centering
\subfigure{\includegraphics[width=0.44\textwidth]
{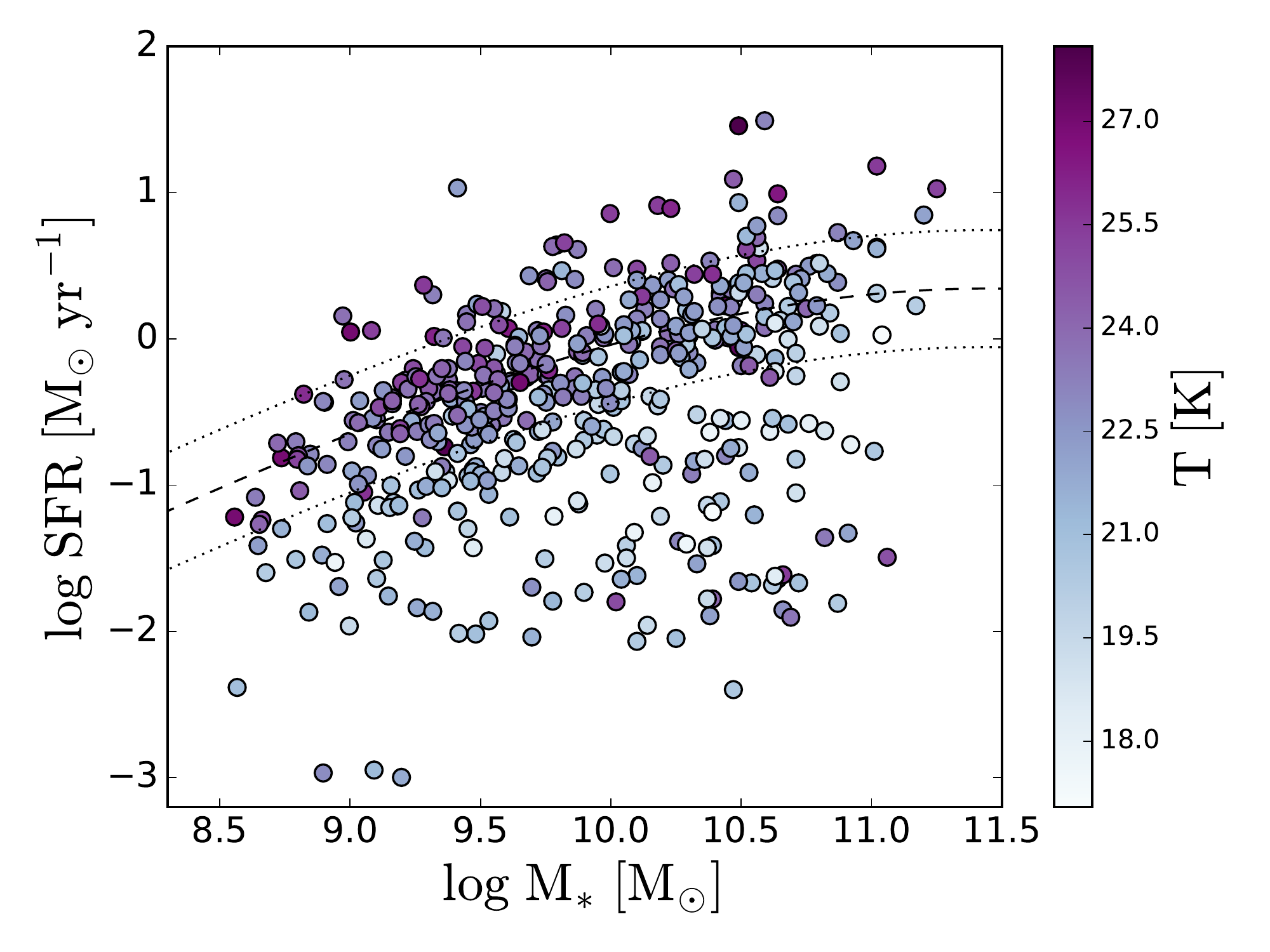}}
\subfigure{\includegraphics[width=0.44\textwidth]
{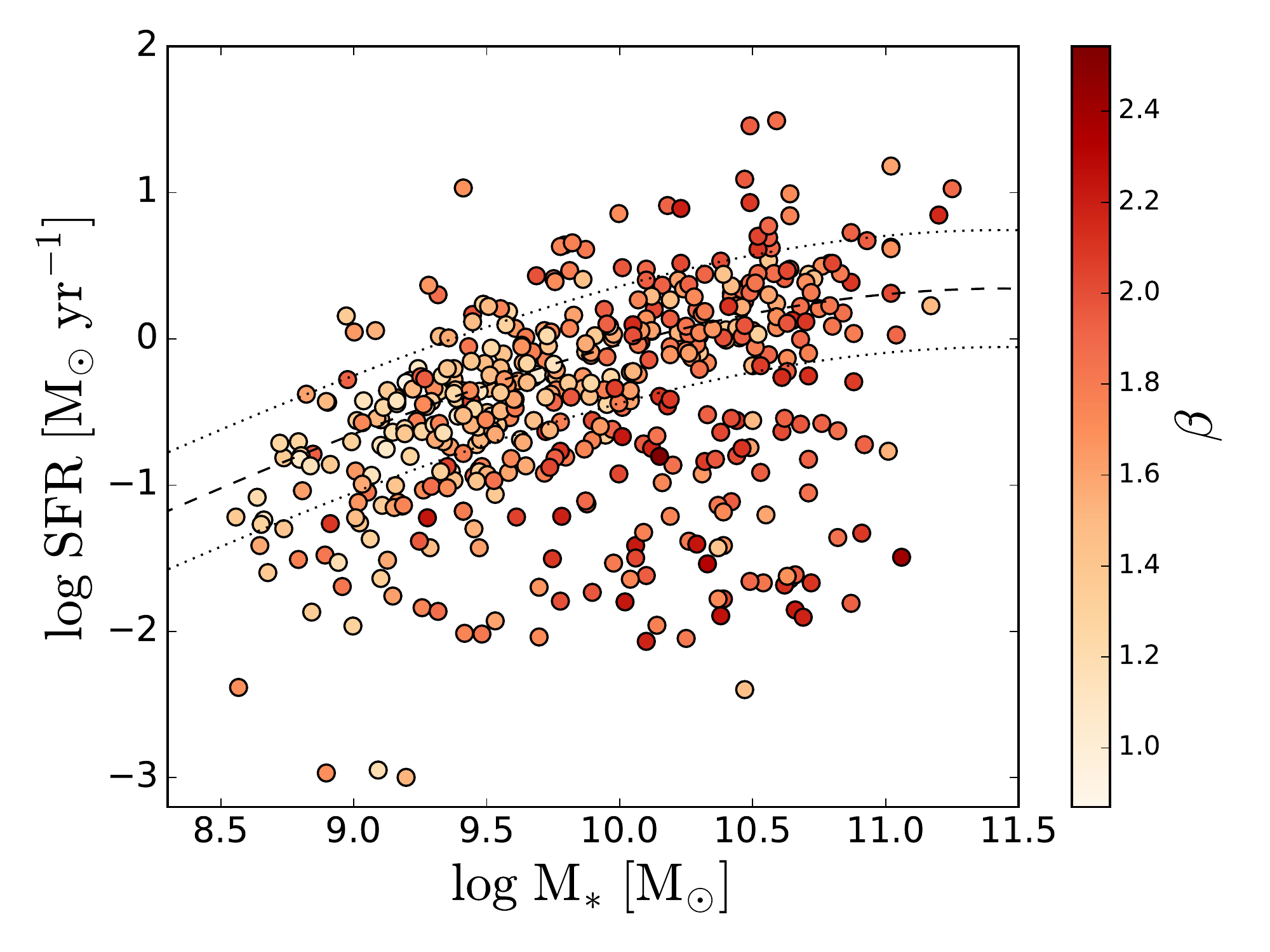}}
\caption{Distribution of the JINGLE and HRS sample in the SFR-$M_*$ plane, color coded by dust temperature (left) and emissivity index $\beta$ (right). Dust temperatures and $\beta$ are measured using the SMBB model and the hierarchical SED fitting approach. The position of the star formation main sequence \citep{Saintonge2016} is shown as a dashed lines, the 0.4 dex dispersion is shown by dotted lines.}
\label{fig:SFR_Mstar_temp_beta}
\end{figure*}

Figure \ref{fig:SFR_Mstar_temp_beta} shows the galaxies on the SFR-$M_*$ plane, color-coded by dust temperature $T$ and emissivity index $\beta$. The dust temperature increases when moving from the bottom-right corner (high $M_*$, low SFR) to the upper-left corner (low $M_*$, high SFR). The emissivity $\beta$ instead tends to increase with $M_*$. From this figure we can already see that $T$ and $\beta$ are related to different galaxy properties, with $T$ varying depending on the SSFR and $\beta$ on the stellar mass.

We quantify the strength of these relations by calculating the correlation coefficients between $T$, $\beta$ and the following quantities:
 stellar mass, stellar mass surface density ($\mu_*=M_*/(2\pi R_{50}^2)$, where $R_{50}$ is the optical half-light radius in the $i$ band from SDSS), metallicity \citep[12+log(O/H), using the O3N2 calibration of][]{Pettini2004}, HI mass fraction ($M_\text{HI}/M_*$), star-formation rate (SFR), specific SFR (SSFR), SFR surface density ($\Sigma_\text{SFR}$), and SFR divided by dust mass. We consider all quantities in log space.
 
We calculate the Pearson correlation coefficient $R$ and perform a linear fit  when the absolute value of the correlation coefficient is higher than 0.4, both for the total sample and for the JINGLE and HRS samples separately. We did the fit also for the two samples separately to see whether there are differences in the correlations derived using JINGLE or HRS.
 We apply a correction to account for the fact that the stellar mass distribution of our sample does not exactly represent the stellar mass distribution in the local Universe, using the method developed for the xCOLD GASS survey \citep{Saintonge2017}.  
 We compare the mass distribution of our sample, in bins of 0.1 dex in $\log M_{*}$, to the expected mass distribution of a volume-limited sample based on the stellar mass function from \cite{Baldry2012}. For each mass bin, we calculate the ratio between the normalized number of galaxies in our sample and in the mass distribution from \cite{Baldry2012}. We apply this ratio as a statistical weight when we fit the dust scaling relations.
 The correlation coefficients and parameters of the linear fits are summarized in Table \ref{tab:corr_coeff}.

\begin{table*}
\centering
\caption{The table shows the Pearson correlation coefficient $R$ between dust properties (dust emissivity index $\beta$ and dust temperature $T$) and global galaxy properties. If |$R$|> 0.4 we provide the best fit relation (slope and intercept)  between the selected galaxy property ($p$) and $T$ (or $\beta$).}
\label{tab:corr_coeff} 
\begin{tabular}{|lccccccc|}
\hline
Properties $p$ & \multicolumn{3}{c}{correlation with $\beta$} & \multicolumn{3}{c}{correlation with $T$}   \\
 & $R$  & slope & intercept & $R$ &  slope & intercept \\ 
 \hline
 \hline
$\log M_*$ & $0.58$ & $0.23\pm0.02$ & $-0.60\pm0.22$ &$-0.29$ & & \\
$\log \mu_*$ & $0.62$ & $0.30\pm0.03$ & $-0.84\pm0.27$ &$-0.19$ & & \\
12+log(O/H) & $0.58$ & $0.95\pm0.13$ & $-6.64\pm1.16$ &$-0.19$ & & \\
$\log M_{HI}/M_*$ & $-0.65$ & $-0.25\pm0.04$ & $1.56\pm0.02$ &$0.41$ & $0.38\pm0.23$ & $23.07\pm0.15$ \\

log SFR & $0.20$ & & &$0.21$ & & \\
log SSFR & $-0.40$ & & &$0.54$ & $1.83\pm0.19$ & $41.02\pm1.90$ \\
log $\Sigma_\text{SFR}$ & $0.13$ & & &$0.49$ & $2.49\pm0.23$ & $26.74\pm0.38$ \\
log SFR/$M_{dust}$ & $-0.15$ & & &$0.73$ & $3.40\pm0.29$ & $49.52\pm2.32$ \\
\hline

\end{tabular}
\end{table*}

\begin{figure*}
\centering
\subfigure{\includegraphics[width=0.9\textwidth]
{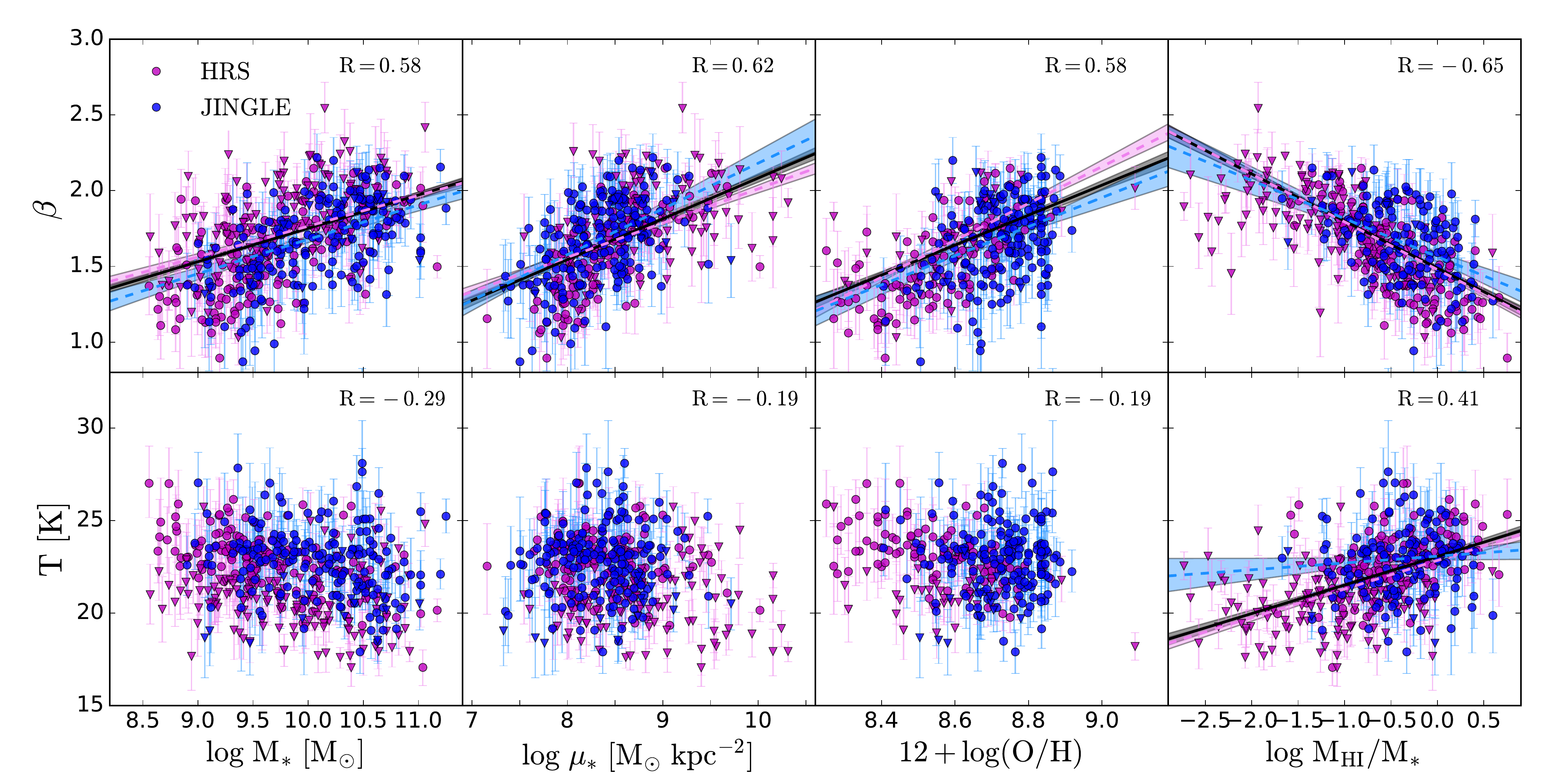}}
\subfigure{\includegraphics[width=0.9\textwidth]
{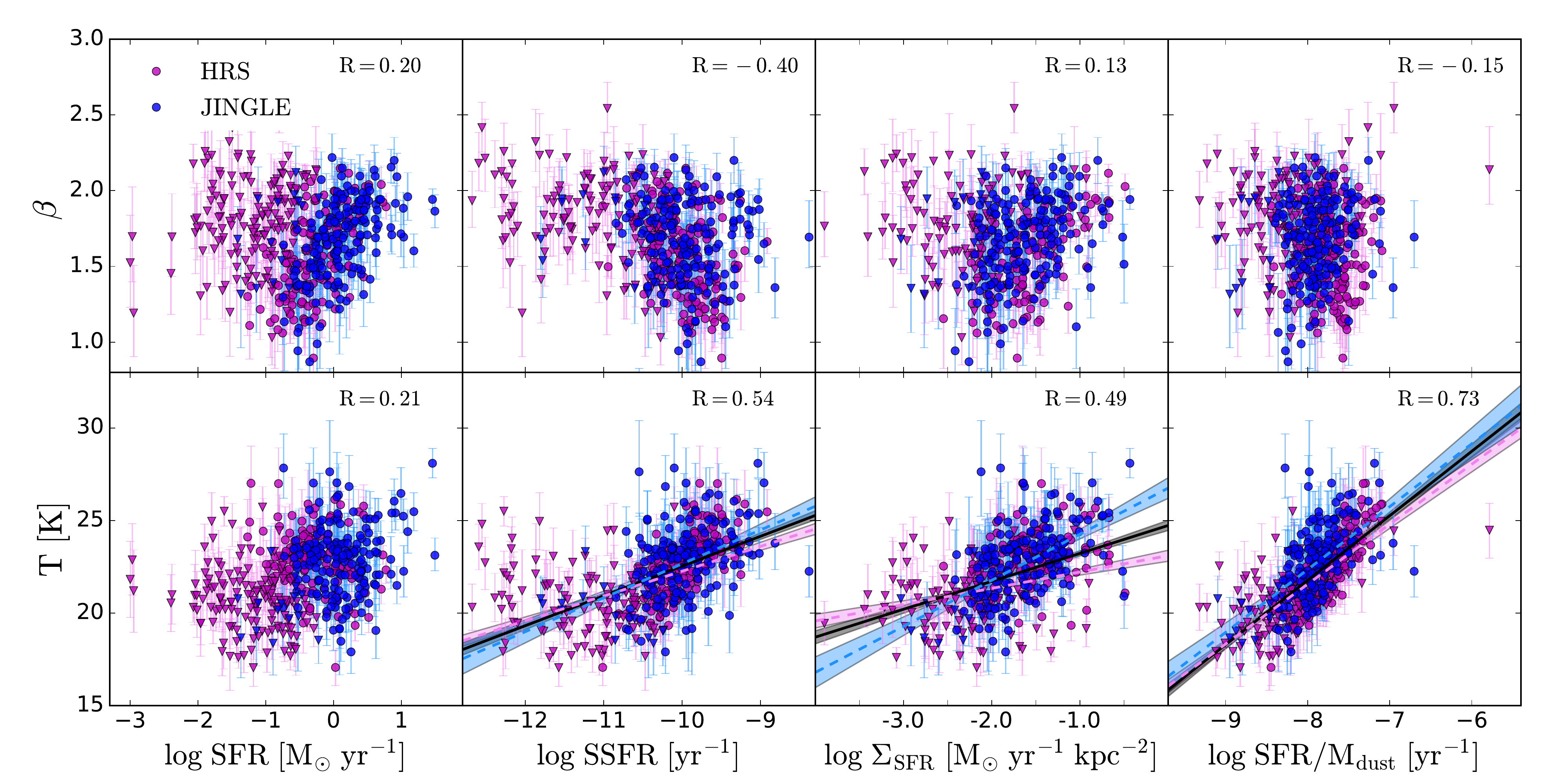}}
\caption{Dust scaling relations: correlation of dust temperature $T$ and effective $\beta$ with other global galaxy properties: stellar mass ($M_*$), stellar mass surface density ($\mu_*=M_*/(2\pi R_{50}^2)$, where $R_{50}^2$ is the optical half-light radius in the $i$ band from SDSS in kpc), metallicity  \citep[12+log(O/H), O3N2 calibration of][]{Pettini2004}, HI mass fraction ($M_{HI}/M_*$), star-formation rate (SFR), specific SFR (SSFR), SFR surface density ($\Sigma_\text{SFR}$), and SFR over dust mass (SFR/M$_\text{dust}$).
Dust temperatures and $\beta$ are measured using the SMBB model and the hierarchical SED fitting approach.
 The JINGLE sample is shown in blue and the HRS sample in magenta. Galaxies of the `main-sequence' sample are shown with circles and galaxies of the `below main-sequence' sample are shown with triangles. In every panel we show the Pearson correlation coefficient $R$. For the cases where $R>0.4$, the plot shows the linear fit to the JINGLE sample (in blue), to the HRS sample (in magenta), and to the two samples together (in black).}
\label{fig:comp_T_beta_prop}
\end{figure*}

We find that the emissivity $\beta$ shows a positive correlation with log $M_*$ (Pearson correlation coefficient $R=0.58$), log $\mu_*$ ($R=0.62$), and metallicity ($R=0.58$). 
  Since these galaxy properties are all correlated with each other, it is not surprising that they all correlate with $\beta$. 
  These trends were already observed by \cite{Cortese2014} in the HRS sample. They also observed negative correlations of these quantities  with dust temperature $T$, due to the fact that they used a non-hierarchical method for the fitting and therefore they could not break the degeneracy between  $T$ and $\beta$. 
Thus, they were not able to distinguish whether the fundamental physical correlations were driven by the temperature or by the emissivity index. In our analysis, these three quantities do not show a strong anti-correlation with temperature ($-0.29\leq R\leq -0.19$).
 We note that for the JINGLE galaxies the metallicities are measured from the SDSS fibre spectra and therefore represent only the metallicities in the central 3 arcsec of the galaxies. For the HRS sample, metallicities are measured from long-slit integrated optical spectra \citep{Boselli2013, Hughes2013}, and thus represent better the global metallicities of the galaxies. Indeed we find that the correlation between $\beta$ and metallicity is higher ($R=0.67$ if we consider only the HRS sample. 
We also find an anti-correlation between $\beta$ and the HI mass fraction ($R=-0.65$), that was already observed in \cite{Cortese2014}. In this case, the HI mass fraction shows  a weaker correlation with dust temperature ($R= 0.41$). The HI mass fraction is known to correlate with the inverse of the stellar mass surface density and with SSFR \citep{Catinella2013}. Thus it is expected to see an anti-correlation with $\beta$ and a positive correlation with $T$, due to the correlation of SSFR with $T$.

The dust temperature correlates with log SSFR ($R=0.54$), log $\Sigma_{SFR}$ ($R=0.49$), and log SFR/$M_{dust}$  ($R=0.73$).
These correlations have already been observed by \cite{Clemens2013} and \cite{Cortese2014}. 
As stated in \cite{Clemens2013}, the fact the cold dust temperature correlates with SFR surface density but not with stellar mass surface density suggests that the cold dust is heated more by ongoing star-formation or by young stars.  Also \cite{Kirkpatrick2014} observed a correlation between cold dust temperature and SFR normalized by the 500\micron\ luminosity, that is a proxy for the dust mass, on spatially resolved scales in galaxies from the KINGFISH sample \citep{Kennicutt2011}. According to their work, this correlation suggests that the number of photons from young stars relative to the amount of dust has an important heating effect on the diffuse cold dust component. Moreover, \cite{Galametz2012} studied a sub-sample of galaxies from the KINGFISH sample and observed that the higher dust temperatures coincide with the center of star-forming regions, showing a connection between dust temperature and star-formation.
 
 The temperature of the dust is regulated by the radiation from star-formation, weighted by the amount of dust present in the galaxy.
The relation between $T$ and SSFR shows more scatter at low SSFR. This may be related in part to the fact that SFR measurements are less accurate for low SSFR \citep[log~SSFR< -10.6, ][]{Hunt2019}. 
Also it is likely that the contribution of the older stellar population to the dust heating is higher in low SSFR galaxies, since the star-formation is weak and the contribution from old stars can be more significant.

\subsection{Primary correlation analysis}
In this section we investigate which are the primary parameters driving the correlation with dust properties. This analysis has two goals:  1) to provide prescriptions to estimate the temperature $T$ and the emissivity index $\beta$ of the dust from other galaxy properties, 2) to understand which are the physical quantities that influence and set $T$ and $\beta$ in a galaxy.  

 We perform a Bayesian inference analysis to find the best combination of parameters that can be used to estimate the dust properties.
  We consider the galaxy parameters which, alone or combined, show some correlation with $\beta$ and $T$: stellar mass $M_*$,  star-formation rate, dust mass, metallicity, and surface area ($A=2\pi R_{50}^2$, where $R_{50}^2$ is the optical half-light radius in the $i$ band from SDSS in kpc). The surface area is used to calculate for example the SFR and stellar mass `surface density'.
  We fit first-order polynomial models with a different number of parameters, exploring all possible combinations of parameters. The number of possible combination of $k$ parameters selected from a total sample of $n$ parameters is $C_{n,k} =\frac{n!}{k!(n-k)!}$. We use a first-order polynomial model in log space:
\begin{equation}
Q_{model}(x_1, ..., x_k) = \sum_{j=1}^{k} a_j \log (x_j) +b,
\end{equation}
where $k$ is the number of galaxy properties $x_j$ considered, and $Q_{model}$ is the value of the dust quantity ($T$ or $\beta$) approximated by the model.
 We use a Bayesian inference method to determine the optimal number of parameters needed to fit the data and the best fitting relations. We model the probability of observing our data, given the model and the uncertainties, as a normal distribution:
\begin{multline}
p(Q_i |Q_{model,i}(x_{1,i}, ..., x_{k,i}), Q_{err,i}) = \\ w_i \cdot \text{Normal}(Q_{model,i}, Q_{err,i}),
\end{multline}
 for each galaxy $i$ in our sample, where $w_i$ is the weight correcting for the flat $M_*$ distribution (see Sec.~\ref{sec:dust_scal_relation}).
  We consider only the uncertainties on the dust quantity $Q_i$, but not on the galaxy properties $x_{j,i}$. We make this choice because we want to minimise the difference between  $Q_{i}$ and $Q_{model,i}$, given the quantities $x_{j,i}$. 
 We perform a MCMC fit using \stan\ to find the best fitting parameters and measure the likelihood of the different models. Then we apply the Bayesian Information Criterion (BIC) to find the optimal number of parameters and the best model. 

We consider first the models to estimate $\beta$. According to the BIC, the preferred model has five parameters: stellar mass, surface area, metallicity, star-formation rate, and HI mass. The best fit relation is given by:
\begin{multline}
\beta_{model} = 0.26_{-0.03}^{+0.03}\cdot \log M_{*} 
- 0.27_{-0.03}^{+0.03}\cdot \log Area \\
+ 0.60_{-0.09}^{+0.09}\cdot \left[12+\log(O/H)\right]
+ 0.18_{-0.03}^{+0.03}\cdot log SFR \\
- 0.23_{-0.03}^{+0.03}\cdot  log M_{HI}
-3.54_{-0.84}^{+0.82}. 
\end{multline}

 This model includes five parameters, several of which are known to be correlated, therefore it is difficult to know which one is more fundamentally related to $\beta$.
 To assess this, we measure the increase in $R^2$ that each parameter produces when it is added to a model that contains already all other parameters. This change represents the amount of variance that can be explained by each parameter and that is not explained by the other variables. We measure $R^2$ ($0<R^2<1$) as the squared Pearson correlation coefficient between the dust parameter ($\beta$ or $T$) and the `modelled' parameter ($\beta_{model}$, $T_{model}$), i.e. the parameter estimated by the linear combination of galaxy properties.
Table \ref{tab:beta_R2} shows the results. 
 From the analysis of the increase of the $R^2$, we can see that the most fundamental parameter determining $\beta$ is the stellar mass (increase in $R^2$: $\Delta R^2=11.2\%$). The second one is the surface area ($\Delta R^2= 8.0\%$).  Since they have opposite coefficients in the fit with almost the same magnitude ($0.26\pm0.03$ for $M_*$ and $-0.27\pm0.03$ for the surface area), this can be interpreted as the stellar mass surface density correlating with $\beta$. If we consider  stellar mass and surface area combined as a single parameter  in the analysis, the increase in $R^2$ due to stellar mass surface density is  $\Delta R^2= 17.9\%$. The following parameter in order of importance is the metallicity ($\Delta R^2= 7.1\%$). SFR and HI mass cause  a smaller increase in $R^2$ ($\Delta R^2=5.0\%$ and $5.7\%$ respectively), and the dust mass has a negligible contribution ($\Delta R^2 = 0.5\%$).
 
 \cite{Smith2012b} studied the variation of $\beta$ in M31 (Andromeda). They found that $\beta$ decreases with galactocentric radius. Since also the stellar mass surface density, $\mu_*$, in M31 decreases with radius \citep{Tamm2012}, their result is consistent with a correlation between $\beta$ and $\mu_*$.
 \cite{Koehler2015} found that the emissivity index of grains evolve from lower to higher $\beta$ values when transitioning from diffuse to denser inter-stellar medium (ISM) due to grain coagulations.  If the stellar mass density is  related to the density of the ISM, this could explain the relation between $\beta$ and the stellar mass surface density.
  
 As we have seen in the previous section, $\beta$ correlates also with metallicity and with the inverse of the HI mass fraction. This indicates a relation between $\beta$ and the state of evolution of a galaxy: more evolved galaxies tend to have higher metallicity and lower HI fraction.
A possible interpretation of the variation of $\beta$ with metallicity and HI mass fraction  is related to the structure and  composition of dust grains. Crystalline or carbonaceous dust is characterized by a lower $\beta$ with respect to  amorphous or silicate dust \citep{Desert1990,Jones2013}.
We expect less evolved (metal-poor) galaxies undergoing an elevated period of star formation activity to produce a lot of dust in stars  \citep{Zhukovska2014},  and this dust has a more crystalline structure at the beginning \citep{Waters1996, Waelkens1996, deVries2010}  and tends to become more `amorphous' with time \citep[e.g.][]{Demyk2001}. Therefore more evolved galaxies can be expected to have more amorphous dust and higher $\beta$. Additionally, silicate dust is thought to survive for a longer time compared to carbon dust \citep[e.g.][]{Jones2011}. Thus we expect dust in a more evolved galaxy to have a larger fraction of silicate grains that are associated with higher values of $\beta$.
 Another possible explanation for the relation between $\beta$ and metallicity is the observation that the abundance of carbon stars, which produce carbon dust, decreases at high metallicities \citep{Boyer2019}. Thus we can expect high-metallicity galaxies to have less carbonaceous dust and consequently a higher $\beta$.
 
Another possibility is that the low $\beta$ values are due to temperature mixing. In our analysis we are not measuring directly the emissivity of dust grains but we are measuring an `effective $\beta$', which includes both the actual emissivity of the dust and the effect of temperature mixing  \citep[e.g.][]{Hunt2015}.
It has been shown that variations of the dust temperatures along the line-of-sight can broaden the SED and mimic the effect of a low $\beta$ value \citep{Shetty2009a}.
  \citet{Remy-Ruyer2015} find the SED of low-metallicity dwarf galaxies to be broader than the one of higher metallicity galaxies, consistent with our finding of lower $\beta$ in low-metallicity galaxies.
  They explain this effect with the fact that dwarf galaxies have a clumpier ISM that produces a wider distribution of dust temperatures.

Since the preferred relation to approximate $\beta$ needs a large number of parameters, we also provide the 
best relation with two parameters (stellar mass and surface area) and with three parameters (stellar mass, surface area, and metallicity), that are more practical to use: 

\begin{multline}
\beta_{model} = 0.42_{-0.02}^{+0.02}\cdot \log M_{*} 
- 0.37_{-0.03}^{+0.03}\cdot \log Area
-1.97_{-0.18}^{+0.18}. 
\end{multline}
\begin{multline}
\beta_{model} = 0.28_{-0.03}^{+0.03}\cdot \log M_{*} 
- 0.38_{-0.03}^{+0.03}\cdot \log Area \\
+ 0.80_{-0.09}^{+0.09}\cdot \left[12+\log \text{(O/H)}\right]
-7.48_{-0.67}^{+0.64}. 
\end{multline}

A summary with the  best relations for every number $k$ of parameters can be found in Table \ref{tab:corr_coeff_best}.

We perform a similar analysis to investigate which combination of parameters gives the better approximation of the dust temperature $T$. 
 According to the BIC, the preferred model has three parameters: SFR, dust mass, and metallicity (BIC$=848.8$). Also the two-parameter model with SFR and dust mass has a similar BIC (BIC$=849.6$), meaning that adding the metallicity parameter has only a small effect on improving the correlation. 
This confirms our previous finding that dust temperature correlates strongly with SFR per unit dust mass.
The $R^2$ analysis gives the same result: the most important parameter is clearly the SFR ($\Delta R^2= 87.9\%$), with a secondary dependence on the dust mass ($\Delta R^2= 16.6\%$). The other four parameters have a very small effect ($\Delta R^2< 3\%$).

This relation is however of limited practical interest since it requires prior knowledge of the dust mass. 
  Therefore we consider also the two-parameter model with the best BIC that do not include $\log M_{dust}$ as a parameter. The two-parameter model uses SFR and stellar mass ($R=0.50$):
  \begin{equation}
T_{model} = 2.50_{-0.22}^{+0.22}\cdot \log SFR  -2.14_{-0.19}^{+0.20}\cdot \log M_{*} +44.24_{-2.02}^{+1.93}.
\end{equation}

Tables for $T$ and $\beta$ with all the relations with two or three parameters are in the appendix (Tables \ref{tab:corr_beta} and \ref{tab:corr_T}).

\begin{table}
\centering
\caption{Increase in $R^2$ when the parameter is added to a model that already contains the other parameters.}
\begin{tabular}{|lcc|}
\hline
& $\beta$ & $T$  \\
Parameter & increase in $\Delta R^2$ ($\%$) & increase in $\Delta R^2$ ($\%$)  \\ 
  \hline \hline
$\log M_*$  & 11.2 & 0.5\\
log SFR  & 5.0 & 80.0  \\
log Area & 8.0 & 2.4\\
12+log(O/H) & 7.1 & 1.5 \\
$ \log M_{dust}$ & 0.5 & 13.6\\
$ \log M_{HI}$ & 5.7 & 0.5 \\
\hline 
\end{tabular}
\label{tab:beta_R2}
\end{table}

\begin{table*}
\centering
\caption{Results of the correlation analysis to derive an expression to approximate the emissivity $\beta$ and the dust temperature using global galaxy properties. The table shows the coefficients $a_j$ of the best polynomial expression ($Q_{model}(x_1, ..., x_k) = \sum_{j=1}^{k} a_j \log (x_j) +b$) to estimate $\beta$ and $T$ using a different number  of parameters $k$. The table also shows the Baysian Information Criterion (BIC) and the Pearson correlation coefficient $R$ between the dust parameter ($\beta$ or $T$) and the `modelled' parameter ($\beta_{model}$, $T_{model}$), i.e. the parameter estimated by the linear combination of galaxy properties.}
\label{tab:corr_coeff_best} 
\begin{tabular}{|lccccccccc|}
\hline
\multicolumn{9}{c}{\bf emissivity index $\beta$}\\
Parameters & $\log M_*$ & log SFR & log Area & 12+log(O/H) &  $\log M_{dust}$ & $ \log M_{HI}$ & intercept & BIC & $R$\\ 
 & [M$_\odot$] &[M$_\odot$ yr$^{-1}$]  & [kpc$^2$] &  & [M$_\odot$]  &[M$_\odot$]   &  & & \\
  \hline

$k$ = 1 & &  &  & 0.98 $\pm$ 0.06 &  &  & -6.77 $\pm$ 0.59 & 170.56 & 0.61 \\
$k$ = 2 &0.42 $\pm$ 0.02 &  & -0.37 $\pm$ 0.03 &  &  &  & -1.97 $\pm$ 0.18 & 53.19 & 0.64 \\
$k$ = 3 &0.28 $\pm$ 0.03 &  & -0.38 $\pm$ 0.03 & 0.80 $\pm$ 0.09 &  &  & -7.48 $\pm$ 0.64 & -14.37 & 0.70 \\
$k$ = 4 &0.33 $\pm$ 0.03 &  & -0.29 $\pm$ 0.03 & 0.69 $\pm$ 0.10 &  & -0.13 $\pm$ 0.03 & -5.92 $\pm$ 0.69 & -27.87 & 0.71 \\
$k$ = 5 &0.26 $\pm$ 0.03 & 0.18 $\pm$ 0.03 & -0.27 $\pm$ 0.03 & 0.60 $\pm$ 0.09 &  & -0.23 $\pm$ 0.03 & -3.54 $\pm$ 0.82 & -54.40 & 0.73 \\
$k$ = 6 &0.31 $\pm$ 0.04 & 0.23 $\pm$ 0.04 & -0.25 $\pm$ 0.04 & 0.66 $\pm$ 0.10 & -0.13 $\pm$ 0.08 & -0.20 $\pm$ 0.04 & -3.84 $\pm$ 0.89 & -51.19 & 0.73 \\

\hline \hline

\multicolumn{10}{c}{\bf Temperature}  \\ 

Parameters & $\log M_*$ & log SFR & log Area & 12+log(O/H) &  $\log M_{dust}$ & $ \log M_{HI}$ & intercept & BIC & $R$\\ 
 & [M$_\odot$] &[M$_\odot$ yr$^{-1}$]  & [kpc$^2$] &  & [M$_\odot$]  &[M$_\odot$]   &  &  &  \\
  \hline 

$k$ = 1 & & 0.65 $\pm$ 0.13 &  &  &  &  & 22.93 $\pm$ 0.08 & 1024.78 & 0.15 \\
$k$ = 2 & & 4.19 $\pm$ 0.29 &  &  & -3.73 $\pm$ 0.30 &  & 51.88 $\pm$ 2.20 & 849.60 & 0.68 \\
$k$ = 3 & & 4.06 $\pm$ 0.29 &  & -1.85 $\pm$ 0.75 & -3.31 $\pm$ 0.31 &  & 64.7 $\pm$ 5.44 & 848.76 & 0.68 \\
$k$ = 4 & & 3.93 $\pm$ 0.31 & -0.66 $\pm$ 0.28 & -2.24 $\pm$ 0.79 & -2.71 $\pm$ 0.41 &  & 64.13 $\pm$ 5.67 & 849.08 & 0.69 \\
$k$ = 5 &0.36 $\pm$ 0.35 & 3.99 $\pm$ 0.32 & -0.63 $\pm$ 0.29 & -2.36 $\pm$ 0.76 & -3.08 $\pm$ 0.57 &  & 64.86 $\pm$ 5.82 & 853.77 & 0.69 \\
$k$ = 6 &0.29 $\pm$ 0.39 & 4.01 $\pm$ 0.33 & -0.58 $\pm$ 0.30 & -2.59 $\pm$ 0.81 & -2.86 $\pm$ 0.64 & -0.23 $\pm$ 0.29 & 67.87 $\pm$ 7.28 & 859.06 & 0.70 \\

\hline 
\end{tabular}
\end{table*}

\subsection{Submm excess}
\label{sec:submm_excess}

In this section we discuss the behaviour of the SED at long wavelengths ($\lambda > 500\micron$). In particular, we are interested in galaxies which show a so-called `submm excess'.
An excess at submm wavelength has been observed in dwarf galaxies \citep{Lisenfeld2002, Galliano2003}, in late-type galaxies \citep{Dumke2004, Bendo2006, Galametz2009}, and in the Magellanic Clouds \citep{Israel2010, Bot2010b}. The most significant excesses can not be explained by contribution from synchrotron, free-free or molecular line emission \citep[e.g.][]{Galliano2003}. Different explanations  proposed to explain this phenomenon are for example the presence of a very cold dust component, a temperature-dependent emissivity \citep{Meny2007}, and the presence of rotating or magnetic  grains \citep{Draine2012}.

\begin{figure}
\centering
\subfigure{\includegraphics[width=0.44\textwidth]
{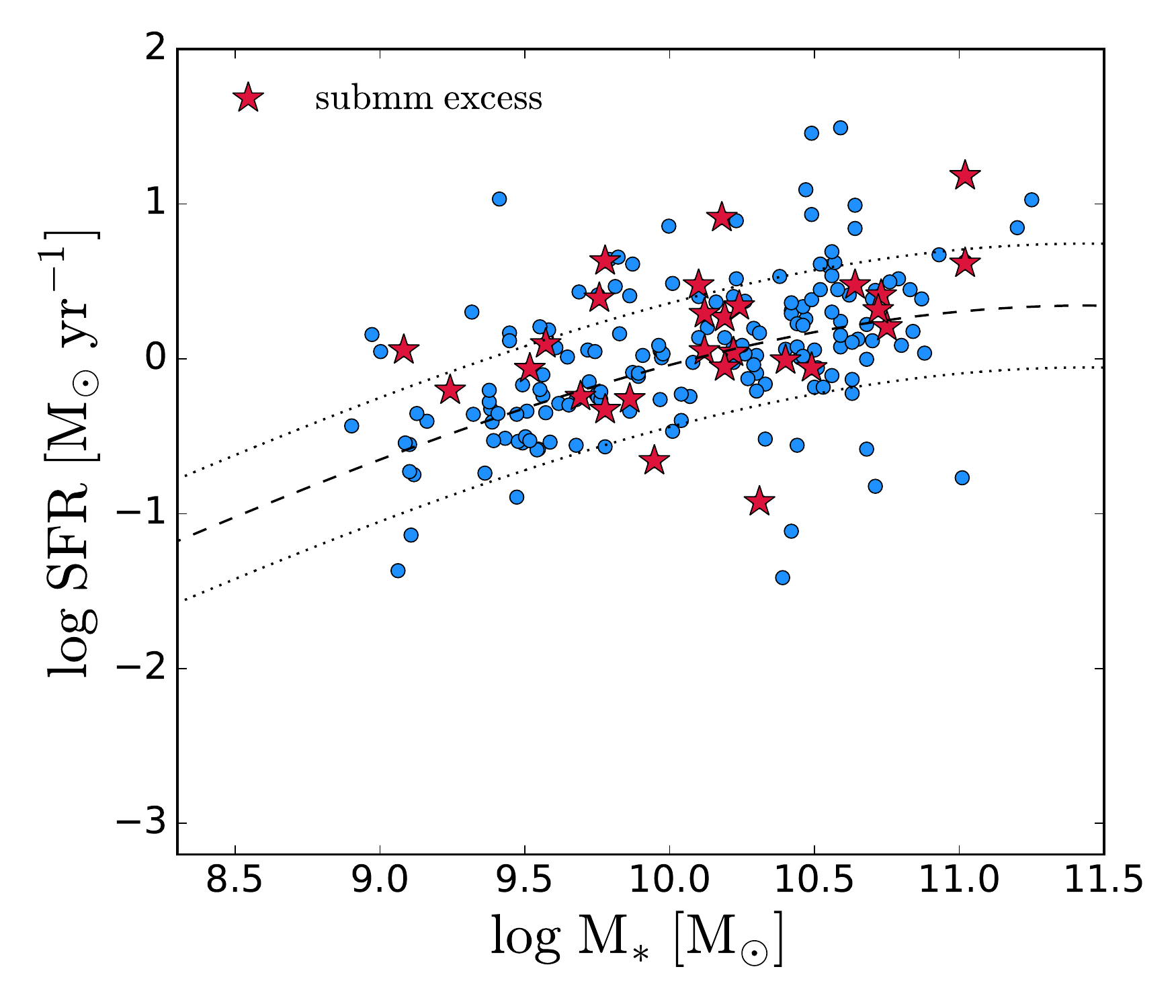}}
\caption{JINGLE galaxies with a submm excess are shown by red stars symbols, while the JINGLE sample is shown in light blue. The position of the star formation main sequence \citep{Saintonge2016} is shown as a dashed lines, the 0.4 dex dispersion is shown by dotted lines.}
\label{fig:SFR_Mstar_submm_excess}
\end{figure}

We identify the galaxies with an excess at 850\micron\ with respect to the SMBB model, taking into account uncertainties on the SCUBA-2 fluxes and on the SMBB model:
\begin{equation}
F^{obs} -F^{model} > F^{obs}_{err} +F^{model}_{err}.
\end{equation}
 There are 27/\Ntot (14\%) galaxies which satisfy this criterion. If we adopt a more stringent criterion, requiring the galaxy to have an excess above 2$\sigma$ (i.e. $(F^{obs} -F^{model}) > 2\cdot F^{obs}_{err}$), we find that 24 galaxies (12\%) satisfy this criterion. From a normal distribution, we would expect to find only 2.5\% of the galaxies with an excess above 2$\sigma$, thus we think that it is a statistically significant result.
 The galaxies with submm excess do not appear to be in a particular region of the SFR-$M_*$ plane (see Fig. \ref{fig:SFR_Mstar_submm_excess}).
There also some galaxies which show a deficit at 850\micron. 

A weak point of this analysis is that the submm excess is determined only by a single point, the 850\micron\ SCUBA-2 flux. Therefore the presence of an excess can also be due to a number of factors including  measurement errors, uncertainties on the apertures, contamination by other sources, and uncertainties on the CO(3-2) contribution. In order to better characterise and quantify the submm excess, additional flux points at longer wavelengths are needed. We plan to investigate this in the future.
We have an accepted proposal to observe 18 JINGLE targets at 1mm and 2mm with NIKA-2 on the IRAM-30m telescope. 
With two additional flux points we will be able to characterize better the submm excess and to test different models proposed to account for it. 

\section{Conclusions}
\label{sec:conclusions}
In this paper we  analyse a sample of \Ntot\ star-forming galaxies from the JINGLE survey. We also include in the analysis 323 galaxies from the Herschel Reference Survey (HRS) to expand our analysis to galaxies with lower specific star-formation rate.
We fit their far-infrared/submm SED  with modified black-body (MBB) models using a hierarchical Bayesian approach that allows to reduce the degeneracy between parameters, especially between dust temperature and emissivity index $\beta$. We consider three models: single modified black-body (SMBB), two modified black-bodies (TMBB), and MBB with a broken emissivity law (BMBB).

The main results of our study are:

\begin{itemize}
\item \textit{Dust masses}: the choice of the model (SMBB, BMBB or TMBB) has only a small effect on the dust mass estimates. The cold dust masses measured with the TMBB are larger than the ones measured by the SMBB by only 0.04 dex on average, and the dust masses measured with the BMBB model agree very well with the SMBB results.\\ 

\item $T$-$\beta$ relation: the use of the hierarchical Bayesian approach to fit the FIR SED is crucial to infer the intrinsic relation between dust temperature and dust emissivity index $\beta$.
 In the JINGLE sample, the anti-correlation between $T$ and $\beta$ is reduced when we use the hierarchical approach ($R=-0.52$) with respect to the non-hierarchical result ($R=-0.79$). Using the hierarchical approach, both $T$ and $\beta$ span  smaller ranges ($17\ \text{K} < T < 30$ K, $0.6 < \beta < 2.2$) with fewer outliers.
\\

\item \textit{Dust scaling relations:} the hierarchical approach is able to reduce the degeneracy between $T$ and $\beta$ and to separate their relations with other galaxy properties. We find that the dust emissivity index $\beta$ correlates with stellar mass surface density, metallicity and anti-correlates with HI mass fraction ($M_{HI}/M_*$). The strongest relation is with stellar mass surface density. 
 The dust temperature correlates with HI mass fraction, SSFR, SFR surface density and  SFR per unit dust mass. The strongest relation is with SFR per unit dust mass.
 These relations can be used to estimate the dust temperature or emissivity index in galaxies where insufficient data prevents determining them directly through SED fitting.\\

\item \textit{Submm excess}:  we observe an excess at 850\micron\ with respect to the flux predicted from the SMBB fit in 26/\Ntot\ (14\%) galaxies, but we do not find these galaxies to lie in a particular region in the stellar mass-SFR plane. 
Additional flux points at longer wavelengths are needed to better characterize the submm excess and to  investigate its origin.
\end{itemize}

The dust scaling relations derived in this work based on low-redshift galaxies show that dust properties correlate with global galaxy properties.  After calibrating these relations with data at higher redshift, they could be applied to the study of high-redshift galaxies.
 Thanks to ALMA it is now possible to detect dust emission in galaxies at redshifts as high as $z>7$ \citep[e.g.][]{Watson2015, Laporte2017}, but the measurement of dust masses in these objects is difficult due to the scarcity of photometric points. 
 The possibility to use scaling relations to predict what dust properties to apply in the SED modelling  will  increase the precision of the dust mass measurements in the early Universe, and consequently will help our understanding of dust evolution over cosmic time.

\section*{Acknowledgements}

We thank Boris Leistedt, Fr{\'e}d{\'e}ric Galliano, Luca Cortese, and Lorne Whiteway for useful discussions. 
A.S.~acknowledges support from the Royal Society through the award of a University Research Fellowship. 
I.D.L.~gratefully acknowledges the support of the Research 
Foundation Flanders (FWO).
C.D.W. acknowledges support from the Natural Science and Engineering Research Council of Canada and the Canada Research Chairs program. 
E.B.~acknowledges support from the UK Science and Technology Facilities Council [grant number ST/M001008/1]. 
M.J.M.~acknowledges the support of the National Science Centre, Poland, through the SONATA BIS grant 2018/30/E/ST9/00208.
 The James Clerk Maxwell Telescope is operated by the East Asian Observatory on behalf of The National Astronomical Observatory of Japan; Academia Sinica Institute of Astronomy and Astrophysics; the Korea Astronomy and Space Science Institute; the Operation, Maintenance and Upgrading Fund for Astronomical Telescopes and Facility Instruments, budgeted from the Ministry of Finance (MOF) of China and administrated by the Chinese Academy of Sciences (CAS), as well as the National Key R$\&$D Program of China (No. 2017YFA0402700). Additional funding support is provided by the Science and Technology Facilities Council of the United Kingdom and participating universities in the United Kingdom and Canada. The authors wish to recognize and acknowledge the very significant cultural role and reverence that the summit of Mauna Kea has always had within the indigenous Hawaiian community. We are most fortunate to have the opportunity to conduct observations from this mountain.

This research has made use of data from HRS project. HRS is a Herschel Key Programme utilising Guaranteed Time from the SPIRE instrument team, ESAC scientists and a mission scientist.
The HRS data was accessed through the Herschel Database in Marseille (HeDaM - http://hedam.lam.fr) operated by CeSAM and hosted by the Laboratoire d'Astrophysique de Marseille.

This research made use of Astropy, a community-developed core Python package for Astronomy \citep{astropy}, {\tt Matplotlib} \citep{Hunter2007} and {\tt NumPy} \citep{VanDerWalt2011}. This research used the {\tt TOPCAT} tool for catalogue cross-matching \citep{Taylor2005}.
This research used the \stan\ interface for Python {\tt PyStan} \citep{pystan}. 
This research used the {\tt CORNER} Python package \citep{corner}.




\bibliographystyle{mn2e}
\bibliography{Biblio/JINGLE_V.bib} 

\begin{thebibliography}{117}
\expandafter\ifx\csname natexlab\endcsname\relax\def\natexlab#1{#1}\fi

\bibitem[{Armus {et~al}\mbox{.}(2009)Armus, Mazzarella, Evans, Surace, Sanders,
  Iwasawa, Frayer, Howell, Chan, Petric, Vavilkin, Kim, Haan, Inami, Murphy,
  Appleton, Barnes, Bothun, Bridge, Charmandaris, Jensen, Kewley, Lord, Madore,
  Marshall, Melbourne, Rich, Satyapal, Schulz, Spoon, Sturm, U, Veilleux, \&
  Xu}]{Armus2009}
Armus L. {et~al.}, 2009, Publications of the Astronomical Society of the
  Pacific, 121, 559

\bibitem[{Baldry {et~al}\mbox{.}(2012)Baldry, Driver, Loveday, Taylor, Kelvin,
  Liske, Norberg, Robotham, Brough, Hopkins, Bamford, Peacock, Bland-Hawthorn,
  Conselice, Croom, Jones, Parkinson, Popescu, Prescott, Sharp, \&
  Tuffs}]{Baldry2012}
Baldry I.~K. {et~al.}, 2012, Monthly Notices of the Royal Astronomical Society,
  421, 621

\bibitem[{Balog {et~al}\mbox{.}(2014)Balog, M{\"u}ller, Nielbock, Altieri,
  Klaas, Blommaert, Linz, Lutz, Mo{\'o}r, Billot, Sauvage, \&
  Okumura}]{Balog2014}
Balog Z. {et~al.}, 2014, Experimental Astronomy, 37, 129

\bibitem[{Barlow(1978)}]{Barlow1978c}
Barlow M.~J., 1978, Monthly Notices of the Royal Astronomical Society, 183, 417

\bibitem[{Barnard, McCulloch \& Meng(2000)Barnard, McCulloch, \&
  Meng}]{Barnard2000}
Barnard J., McCulloch R., Meng X.-L., 2000, Statistica Sinica, 10

\bibitem[{Bendo {et~al}\mbox{.}(2006)Bendo, Dale, Draine, Engelbracht,
  Kennicutt, Calzetti, Gordon, Helou, Hollenbach, Li, Murphy, Prescott, \&
  Smith}]{Bendo2006}
Bendo G.~J. {et~al.}, 2006, The Astrophysical Journal, 652, 283

\bibitem[{Bendo {et~al}\mbox{.}(2013)Bendo, Griffin, Bock, Conversi, Dowell,
  Lim, Lu, North, Papageorgiou, Pearson, Pohlen, Polehampton, Schulz, Shupe,
  Sibthorpe, Spencer, Swinyard, Valtchanov, \& Xu}]{Bendo2013}
Bendo G.~J. {et~al.}, 2013, Monthly Notices of the Royal Astronomical Society,
  433, 3062

\bibitem[{Bertoldi {et~al}\mbox{.}(2003)Bertoldi, Carilli, Cox, Fan, Strauss,
  Beelen, Omont, \& Zylka}]{Bertoldi2003}
Bertoldi F., Carilli C.~L., Cox P., Fan X., Strauss M.~A., Beelen A., Omont A.,
  Zylka R., 2003, A{\&}A, 406, L55

\bibitem[{Bianchi(2013)}]{Bianchi2013}
Bianchi S., 2013, A{\&}A, 552, A89

\bibitem[{Boselli {et~al}\mbox{.}(2012)Boselli, Ciesla, Cortese, Buat, Boquien,
  Bendo, Boissier, Eales, Gavazzi, Hughes, Pohlen, Smith, Baes, Bianchi,
  Clements, Cooray, Davies, Gear, Madden, Magrini, Panuzzo, Remy, Spinoglio, \&
  Zibetti}]{Boselli2012}
Boselli A. {et~al.}, 2012, A{\&}A, 540, A54

\bibitem[{Boselli {et~al}\mbox{.}(2010)Boselli, Eales, Cortese, Bendo, Chanial,
  Buat, Davies, Auld, Rigby, Baes, Barlow, Bock, Bradford, Castro-Rodriguez,
  Charlot, Clements, Cormier, Dwek, Elbaz, Galametz, Galliano, Gear, Glenn,
  Gomez, Griffin, Hony, Isaak, Levenson, Lu, Madden, O'Halloran, Okamura,
  Oliver, Page, Panuzzo, Papageorgiou, Parkin, Perez-Fournon, Pohlen, Rangwala,
  Roussel, Rykala, Sacchi, Sauvage, Schulz, Schirm, Smith, Spinoglio, Stevens,
  Symeonidis, Vaccari, Vigroux, Wilson, Wozniak, Wright, \&
  Zeilinger}]{Boselli2010b}
Boselli A. {et~al.}, 2010, Publications of the Astronomical Society of Pacific,
  122, 261

\bibitem[{Boselli {et~al}\mbox{.}(2013)Boselli, Hughes, Cortese, Gavazzi, \&
  Buat}]{Boselli2013}
Boselli A., Hughes T.~M., Cortese L., Gavazzi G., Buat V., 2013, A{\&}A, 550,
  A114

\bibitem[{Bot {et~al}\mbox{.}(2010)Bot, Ysard, Paradis, Bernard, Lagache,
  Israel, \& Wall}]{Bot2010b}
Bot C., Ysard N., Paradis D., Bernard J.~P., Lagache G., Israel F.~P., Wall
  W.~F., 2010, A{\&}A, 523, A20

\bibitem[{Boyer {et~al}\mbox{.}(2019)Boyer, Williams, Aringer, Chen, Dalcanton,
  Girardi, Guhathakurta, Marigo, Olsen, Rosenfield, \& Weisz}]{Boyer2019}
Boyer M.~L. {et~al.}, 2019, arXiv, arXiv:1904.02172

\bibitem[{Carpenter {et~al}\mbox{.}(2017)Carpenter, , Gelman, Hoffman, Lee,
  Goodrich, Betancourt, Brubaker, Guo, Li, \& Riddell}]{Carpenter2017}
Carpenter B. {et~al.}, 2017, Journal of Statistical Software 76(1)

\bibitem[{Casey(2012)}]{Casey2012}
Casey C.~M., 2012, Monthly Notices of the Royal Astronomical Society, 425, 3094

\bibitem[{Catinella {et~al}\mbox{.}(2013)Catinella, Schiminovich, Cortese,
  Fabello, Hummels, Moran, Lemonias, Cooper, Wu, Heckman, \&
  Wang}]{Catinella2013}
Catinella B. {et~al.}, 2013, Monthly Notices of the Royal Astronomical Society,
  436, 34

\bibitem[{Ceccarelli {et~al}\mbox{.}(2018)Ceccarelli, Viti, Balucani, \&
  Taquet}]{Ceccarelli2018}
Ceccarelli C., Viti S., Balucani N., Taquet V., 2018, Monthly Notices of the
  Royal Astronomical Society, 476, 1371

\bibitem[{Ciesla {et~al}\mbox{.}(2012)Ciesla, Boselli, Smith, Bendo, Cortese,
  Eales, Bianchi, Boquien, Buat, Davies, Pohlen, Zibetti, Baes, Cooray,
  De~Looze, di~Serego~Alighieri, Galametz, Gomez, Lebouteiller, Madden,
  Pappalardo, Remy, Spinoglio, Vaccari, Auld, \& Clements}]{Ciesla2012}
Ciesla L. {et~al.}, 2012, A{\&}A, 543, A161

\bibitem[{Clark {et~al}\mbox{.}(2015)Clark, Dunne, Gomez, Maddox, De~Vis,
  Smith, Eales, Baes, Bendo, Bourne, Driver, Dye, Furlanetto, Grootes, Ivison,
  Schofield, Robotham, Rowlands, Valiante, Vlahakis, van~der Werf, Wright, \&
  De~Zotti}]{Clark2015}
Clark C. J.~R. {et~al.}, 2015, Monthly Notices of the Royal Astronomical
  Society, 452, 397

\bibitem[{Clark {et~al}\mbox{.}(2016)Clark, Schofield, Gomez, \&
  Davies}]{Clark2016}
Clark C. J.~R., Schofield S.~P., Gomez H.~L., Davies J.~I., 2016, Monthly
  Notices of the Royal Astronomical Society, 459, 1646

\bibitem[{Clemens {et~al}\mbox{.}(2013)Clemens, Negrello, De~Zotti,
  Gonzalez-Nuevo, Bonavera, Cosco, Guarese, Boaretto, Salucci, Baccigalupi,
  Clements, Danese, Lapi, Mandolesi, Partridge, Perrotta, Serjeant, Scott, \&
  Toffolatti}]{Clemens2013}
Clemens M.~S. {et~al.}, 2013, Monthly Notices of the Royal Astronomical
  Society, 433, 695

\bibitem[{Clements {et~al}\mbox{.}(1996)Clements, Sutherland, McMahon, \&
  Saunders}]{Clements1996}
Clements D.~L., Sutherland W.~J., McMahon R.~G., Saunders W., 1996, Monthly
  Notices of the Royal Astronomical Society, 279, 477

\bibitem[{Cortese {et~al}\mbox{.}(2014)Cortese, Fritz, Bianchi, Boselli,
  Ciesla, Bendo, Boquien, Roussel, Baes, Buat, Clemens, Cooray, Cormier,
  Davies, De~Looze, Eales, Fuller, Hunt, Madden, Munoz-Mateos, Pappalardo,
  Pierini, R{\'e}my-Ruyer, Sauvage, di~Serego~Alighieri, Smith, Spinoglio,
  Vaccari, \& Vlahakis}]{Cortese2014}
Cortese L. {et~al.}, 2014, Monthly Notices of the Royal Astronomical Society,
  440, 942

\bibitem[{Coupeaud {et~al}\mbox{.}(2011)Coupeaud, Demyk, Meny, Nayral, Delpech,
  Leroux, Depecker, Creff, Brubach, \& Roy}]{Coupeaud2011}
Coupeaud A. {et~al.}, 2011, A{\&}A, 535, A124

\bibitem[{da~Cunha, Charlot \& Elbaz(2008)da~Cunha, Charlot, \&
  Elbaz}]{daCunha2008}
da~Cunha E., Charlot S., Elbaz D., 2008, Monthly Notices of the Royal
  Astronomical Society, 388, 1595

\bibitem[{De~Vis {et~al}\mbox{.}(2017)De~Vis, Dunne, Maddox, Gomez, Clark,
  Bauer, Viaene, Schofield, Baes, Baker, Bourne, Driver, Dye, Eales,
  Furlanetto, Ivison, Robotham, Rowlands, Smith, Smith, Valiante, \&
  Wright}]{DeVis2017}
De~Vis P. {et~al.}, 2017, Monthly Notices of the Royal Astronomical Society,
  464, 4680

\bibitem[{de~Vries {et~al}\mbox{.}(2010)de~Vries, Min, Waters, Blommaert, \&
  Kemper}]{deVries2010}
de~Vries B.~L., Min M., Waters L. B. F.~M., Blommaert J. A. D.~L., Kemper F.,
  2010, A{\&}A, 516, A86

\bibitem[{Decin \& Eriksson(2007)}]{Decin2007}
Decin L., Eriksson K., 2007, A{\&}A, 472, 1041

\bibitem[{Demyk {et~al}\mbox{.}(2001)Demyk, Carrez, Leroux, Cordier, Jones,
  Borg, Quirico, Raynal, \& d'Hendecourt}]{Demyk2001}
Demyk K. {et~al.}, 2001, A{\&}A, 368, L38

\bibitem[{Demyk {et~al}\mbox{.}(2017{\natexlab{a}})Demyk, Meny, Leroux,
  Depecker, Brubach, Roy, Nayral, Ojo, \& Delpech}]{Demyk2017b}
Demyk K. {et~al.}, 2017{\natexlab{a}}, A{\&}A, 606, A50

\bibitem[{Demyk {et~al}\mbox{.}(2017{\natexlab{b}})Demyk, Meny, Lu,
  Papatheodorou, Toplis, Leroux, Depecker, Brubach, Roy, Nayral, Ojo, Delpech,
  Paradis, \& Gromov}]{Demyk2017a}
Demyk K. {et~al.}, 2017{\natexlab{b}}, A{\&}A, 600, A123

\bibitem[{D{\'e}sert, Boulanger \& Puget(1990)D{\'e}sert, Boulanger, \&
  Puget}]{Desert1990}
D{\'e}sert F.~X., Boulanger F., Puget J.~L., 1990, Astronomy and Astrophysics
  (ISSN 0004-6361), 237, 215

\bibitem[{D{\'e}sert {et~al}\mbox{.}(2008)D{\'e}sert, Mac{\'\i}as-P{\'e}rez,
  Mayet, Giardino, Renault, Aumont, Beno{\^\i}t, Bernard, Ponthieu, \&
  Tristram}]{Desert2008}
D{\'e}sert F.~X. {et~al.}, 2008, A{\&}A, 481, 411

\bibitem[{Draine \& Hensley(2012)}]{Draine2012}
Draine B.~T., Hensley B., 2012, The Astrophysical Journal, 757, 103

\bibitem[{Draine \& Li(2007)}]{Draine2007}
Draine B.~T., Li A., 2007, The Astrophysical Journal, 657, 810

\bibitem[{Duane {et~al}\mbox{.}(1987)Duane, Kennedy, Pendleton, \&
  Roweth}]{Duane1987}
Duane S., Kennedy A.~D., Pendleton B.~J., Roweth D., 1987, Physics Letters B,
  195, 216

\bibitem[{Dumke, Krause \& Wielebinski(2004)Dumke, Krause, \&
  Wielebinski}]{Dumke2004}
Dumke M., Krause M., Wielebinski R., 2004, A{\&}A, 414, 475

\bibitem[{Dupac {et~al}\mbox{.}(2003)Dupac, Bernard, Boudet, Giard, Lamarre,
  Meny, Pajot, Ristorcelli, Serra, Stepnik, \& Torre}]{Dupac2003}
Dupac X. {et~al.}, 2003, A{\&}A, 404, L11

\bibitem[{Eales {et~al}\mbox{.}(2017)Eales, de~Vis, Smith, Appah, Ciesla,
  Duffield, \& Schofield}]{Eales2017}
Eales S., de~Vis P., Smith M. W.~L., Appah K., Ciesla L., Duffield C.,
  Schofield S., 2017, Monthly Notices of the Royal Astronomical Society, 465,
  3125

\bibitem[{Eales {et~al}\mbox{.}(2010)Eales, Dunne, Clements, Cooray, De~Zotti,
  Dye, Ivison, Jarvis, Lagache, Maddox, Negrello, Serjeant, Thompson,
  Van~Kampen, Amblard, Andreani, Baes, Beelen, Bendo, Benford, Bertoldi, Bock,
  Bonfield, Boselli, Bridge, Buat, Burgarella, Carlberg, Cava, Chanial,
  Charlot, Christopher, Coles, Cortese, Dariush, da~Cunha, Dalton, Danese,
  Dannerbauer, Driver, Dunlop, Fan, Farrah, Frayer, Frenk, Geach, Gardner,
  Gomez, Gonzalez-Nuevo, Gonz{\'a}lez-Solares, Griffin, Hardcastle,
  Hatziminaoglou, Herranz, Hughes, Ibar, Jeong, Lacey, Lapi, Lawrence, Lee,
  Leeuw, Liske, L{\'o}pez-Caniego, M{\"u}ller, Nandra, Panuzzo, Papageorgiou,
  Patanchon, Peacock, Pearson, Phillipps, Pohlen, Popescu, Rawlings, Rigby,
  Rigopoulou, Robotham, Rodighiero, Sansom, Schulz, Scott, Smith, Sibthorpe,
  Smail, Stevens, Sutherland, Takeuchi, Tedds, Temi, Tuffs, Trichas, Vaccari,
  Valtchanov, van~der Werf, Verma, Vieria, Vlahakis, \& White}]{Eales2010}
Eales S. {et~al.}, 2010, Publications of the Astronomical Society of the
  Pacific, 122, 499

\bibitem[{Eales {et~al}\mbox{.}(2012)Eales, Smith, Auld, Baes, Bendo, Bianchi,
  Boselli, Ciesla, Clements, Cooray, Cortese, Davies, De~Looze, Galametz, Gear,
  Gentile, Gomez, Fritz, Hughes, Madden, Magrini, Pohlen, Spinoglio,
  Verstappen, Vlahakis, \& Wilson}]{Eales2012}
Eales S. {et~al.}, 2012, The Astrophysical Journal, 761, 168

\bibitem[{Ferrara, Viti \& Ceccarelli(2016)Ferrara, Viti, \&
  Ceccarelli}]{Ferrara2016}
Ferrara A., Viti S., Ceccarelli C., 2016, Monthly Notices of the Royal
  Astronomical Society: Letters, 463, L112

\bibitem[{Foreman-Mackey(2016)}]{corner}
Foreman-Mackey D., 2016, The Journal of Open Source Software, 24

\bibitem[{Foreman-Mackey {et~al}\mbox{.}(2013)Foreman-Mackey, Hogg, Lang, \&
  Goodman}]{Foreman-Mackey2013}
Foreman-Mackey D., Hogg D.~W., Lang D., Goodman J., 2013, Publications of the
  Astronomical Society of Pacific, 125, 306

\bibitem[{Galametz {et~al}\mbox{.}(2012)Galametz, Kennicutt, Albrecht, Aniano,
  Armus, Bertoldi, Calzetti, Crocker, Croxall, Dale, Donovan~Meyer, Draine,
  Engelbracht, Hinz, Roussel, Skibba, Tabatabaei, Walter, Wei{\ss}, Wilson, \&
  Wolfire}]{Galametz2012}
Galametz M. {et~al.}, 2012, Monthly Notices of the Royal Astronomical Society,
  425, 763

\bibitem[{Galametz {et~al}\mbox{.}(2009)Galametz, Madden, Galliano, Hony,
  Schuller, Beelen, Bendo, Sauvage, Lundgren, \& Billot}]{Galametz2009}
Galametz M. {et~al.}, 2009, A{\&}A, 508, 645

\bibitem[{Galametz {et~al}\mbox{.}(2011)Galametz, Madden, Galliano, Hony,
  Bendo, \& Sauvage}]{Galametz2011}
Galametz M., Madden S.~C., Galliano F., Hony S., Bendo G.~J., Sauvage M., 2011,
  A{\&}A, 532, A56

\bibitem[{Galliano(2018)}]{Galliano2018}
Galliano F., 2018, Monthly Notices of the Royal Astronomical Society

\bibitem[{Galliano, Galametz \& Jones(2018)Galliano, Galametz, \&
  Jones}]{Galliano2018b}
Galliano F., Galametz M., Jones A.~P., 2018, Annual Review of Astronomy and
  Astrophysics, 56, 673

\bibitem[{Galliano {et~al}\mbox{.}(2003)Galliano, Madden, Jones, Wilson,
  Bernard, \& Le~Peintre}]{Galliano2003}
Galliano F., Madden S.~C., Jones A.~P., Wilson C.~D., Bernard J.~P., Le~Peintre
  F., 2003, A{\&}A, 407, 159

\bibitem[{Gelman {et~al}\mbox{.}(2004)Gelman, Carlin, Ster, Dunson, Vehtari, \&
  Rubin}]{Gelman2004}
Gelman A., Carlin J.~B., Ster H.~S., Dunson D.~B., Vehtari A., Rubin D.~B.,
  2004, Chapman and Hall/CRC

\bibitem[{Gelman \& Hill(2007)}]{Gelman2007}
Gelman A., Hill J., 2007, Cambridge University Press

\bibitem[{Gelman \& Rubin(1992)}]{Gelman1992}
Gelman A., Rubin D.~B., 1992, Statistical Science, 7, 457

\bibitem[{Goodman \& Weare(2010)}]{Goodman2010}
Goodman J., Weare J., 2010, Communications in Applied Mathematics and
  Computational Science, 5, 65

\bibitem[{Gordon {et~al}\mbox{.}(2014)Gordon, Roman-Duval, Bot, Meixner,
  Babler, Bernard, Bolatto, Boyer, Clayton, Engelbracht, Fukui, Galametz,
  Galliano, Hony, Hughes, Indebetouw, Israel, Jameson, Kawamura, Lebouteiller,
  Li, Madden, Matsuura, Misselt, Montiel, Okumura, Onishi, Panuzzo, Paradis,
  Rubio, Sandstrom, Sauvage, Seale, Sewi{\l}o, Tchernyshyov, \&
  Skibba}]{Gordon2014}
Gordon K.~D. {et~al.}, 2014, The Astrophysical Journal, 797, 85

\bibitem[{Groves {et~al}\mbox{.}(2015)Groves, Schinnerer, Leroy, Galametz,
  Walter, Bolatto, Hunt, Dale, Calzetti, Croxall, \& Kennicutt}]{Groves2015}
Groves B.~A. {et~al.}, 2015, The Astrophysical Journal, 799, 96

\bibitem[{Hermelo {et~al}\mbox{.}(2016)Hermelo, Rela{\~n}o, Lisenfeld, Verley,
  Kramer, Ruiz-Lara, Boquien, Xilouris, \& Albrecht}]{Hermelo2016}
Hermelo I. {et~al.}, 2016, A{\&}A, 590, A56

\bibitem[{Hildebrand(1983)}]{Hildebrand1983}
Hildebrand R.~H., 1983, Quarterly Journal of the Royal Astronomical Society,
  24, 267

\bibitem[{Hughes {et~al}\mbox{.}(2013)Hughes, Cortese, Boselli, Gavazzi, \&
  Davies}]{Hughes2013}
Hughes T.~M., Cortese L., Boselli A., Gavazzi G., Davies J.~I., 2013, A{\&}A,
  550, A115

\bibitem[{Hunt {et~al}\mbox{.}(2019)Hunt, De~Looze, Boquien, Nikutta, Rossi,
  Bianchi, Dale, Granato, Kennicutt, Silva, Ciesla, Rela{\~n}o, Viaene, Brandl,
  Calzetti, Croxall, Draine, Galametz, Gordon, Groves, Helou, Herrera-Camus,
  Hinz, Koda, Salim, Sandstrom, Smith, Wilson, \& Zibetti}]{Hunt2019}
Hunt L.~K. {et~al.}, 2019, A{\&}A, 621, A51

\bibitem[{Hunt {et~al}\mbox{.}(2015)Hunt, Draine, Bianchi, Gordon, Aniano,
  Calzetti, Dale, Helou, Hinz, Kennicutt, Roussel, Wilson, Bolatto, Boquien,
  Croxall, Galametz, Gil~de Paz, Koda, Mu{\~n}oz-Mateos, Sandstrom, Sauvage,
  Vigroux, \& Zibetti}]{Hunt2015}
Hunt L.~K. {et~al.}, 2015, A{\&}A, 576, A33

\bibitem[{Hunter(2007)}]{Hunter2007}
Hunter J.~D., 2007, Computing in Science and Engineering, 9, 90

\bibitem[{Israel {et~al}\mbox{.}(2010)Israel, Wall, Raban, Reach, Bot, Oonk,
  Ysard, \& Bernard}]{Israel2010}
Israel F.~P., Wall W.~F., Raban D., Reach W.~T., Bot C., Oonk J. B.~R., Ysard
  N., Bernard J.~P., 2010, A{\&}A, 519, A67

\bibitem[{Jarrett {et~al}\mbox{.}(2011)Jarrett, Cohen, Masci, Wright, Stern,
  Benford, Blain, Carey, Cutri, Eisenhardt, Lonsdale, Mainzer, Marsh, Padgett,
  Petty, Ressler, Skrutskie, Stanford, Surace, Tsai, Wheelock, \&
  Yan}]{Jarrett2011}
Jarrett T.~H. {et~al.}, 2011, The Astrophysical Journal, 735, 112

\bibitem[{Jones {et~al}\mbox{.}(2013)Jones, Fanciullo, K{\"o}hler, Verstraete,
  Guillet, Bocchio, \& Ysard}]{Jones2013}
Jones A.~P., Fanciullo L., K{\"o}hler M., Verstraete L., Guillet V., Bocchio
  M., Ysard N., 2013, A{\&}A, 558, A62

\bibitem[{Jones \& Nuth(2011)}]{Jones2011}
Jones A.~P., Nuth J.~A., 2011, A{\&}A, 530, A44

\bibitem[{Juvela {et~al}\mbox{.}(2013)Juvela, Montillaud, Ysard, \&
  Lunttila}]{Juvela2013}
Juvela M., Montillaud J., Ysard N., Lunttila T., 2013, A{\&}A, 556, A63

\bibitem[{Kelly {et~al}\mbox{.}(2012)Kelly, Shetty, Stutz, Kauffmann, Goodman,
  \& Launhardt}]{Kelly2012}
Kelly B.~C., Shetty R., Stutz A.~M., Kauffmann J., Goodman A.~A., Launhardt R.,
  2012, The Astrophysical Journal, 752, 55

\bibitem[{Kennicutt {et~al}\mbox{.}(2011)Kennicutt, Calzetti, Aniano, Appleton,
  Armus, Beir{\~a}o, Bolatto, Brandl, Crocker, Croxall, Dale, Donovan~Meyer,
  Draine, Engelbracht, Galametz, Gordon, Groves, Hao, Helou, Hinz, Hunt,
  Johnson, Koda, Krause, Leroy, Li, Meidt, Montiel, Murphy, Rahman, Rix,
  Roussel, Sandstrom, Sauvage, Schinnerer, Skibba, Smith, Srinivasan, Vigroux,
  Walter, Wilson, Wolfire, \& Zibetti}]{Kennicutt2011}
Kennicutt R.~C. {et~al.}, 2011, Publications of the Astronomical Society of the
  Pacific, 123, 1347

\bibitem[{Kirkpatrick {et~al}\mbox{.}(2014)Kirkpatrick, Calzetti, Kennicutt,
  Galametz, Gordon, Groves, Hunt, Dale, Hinz, \& Tabatabaei}]{Kirkpatrick2014}
Kirkpatrick A. {et~al.}, 2014, The Astrophysical Journal, 789, 130

\bibitem[{K{\"o}hler, Ysard \& Jones(2015)K{\"o}hler, Ysard, \&
  Jones}]{Koehler2015}
K{\"o}hler M., Ysard N., Jones A.~P., 2015, A{\&}A, 579, A15

\bibitem[{Laporte {et~al}\mbox{.}(2017)Laporte, Ellis, Boone, Bauer,
  Qu{\'e}nard, Roberts-Borsani, Pell{\'o}, Perez-Fournon, \&
  Streblyanska}]{Laporte2017}
Laporte N. {et~al.}, 2017, The Astrophysical Journal Letters, 837, L21

\bibitem[{Lewandowski, Kurowicka \& Joe(2009)Lewandowski, Kurowicka, \&
  Joe}]{Lewandowski2009}
Lewandowski D., Kurowicka D., Joe H., 2009, Journal of Multivariate Analysis,
  100:1989–2001, 556

\bibitem[{Lisenfeld {et~al}\mbox{.}(2002)Lisenfeld, Israel, Stil, \&
  Sievers}]{Lisenfeld2002}
Lisenfeld U., Israel F.~P., Stil J.~M., Sievers A., 2002, A{\&}A, 382, 860

\bibitem[{Maddox {et~al}\mbox{.}(2018)Maddox, Valiante, Cigan, Dunne, Eales,
  Smith, Dye, Furlanetto, Ibar, De~Zotti, Millard, Bourne, Gomez, Ivison,
  Scott, \& Valtchanov}]{Maddox2018}
Maddox S.~J. {et~al.}, 2018, The Astrophysical Journal Supplement Series, 236,
  30

\bibitem[{Magdis {et~al}\mbox{.}(2012)Magdis, Daddi, B{\'e}thermin, Sargent,
  Elbaz, Pannella, Dickinson, Dannerbauer, da~Cunha, Walter, Rigopoulou,
  Charmandaris, Hwang, \& Kartaltepe}]{Magdis2012}
Magdis G.~E. {et~al.}, 2012, The Astrophysical Journal, 760, 6

\bibitem[{Meny {et~al}\mbox{.}(2007)Meny, Gromov, Boudet, Bernard, Paradis, \&
  Nayral}]{Meny2007}
Meny C., Gromov V., Boudet N., Bernard J.~P., Paradis D., Nayral C., 2007,
  A{\&}A, 468, 171

\bibitem[{Metropolis {et~al}\mbox{.}(1953)Metropolis, Rosenbluth, Rosenbluth,
  Teller, \& Teller}]{Metropolis1953}
Metropolis N., Rosenbluth A.~W., Rosenbluth M.~N., Teller A.~H., Teller E.,
  1953, The Journal of Chemical Physics, 21, 1087

\bibitem[{Micha{\l}owski(2015)}]{Michalowski2015}
Micha{\l}owski M.~J., 2015, A{\&}A, 577, A80

\bibitem[{Miville-Desch{\^e}nes \& Lagache(2005)}]{Miville-Deschenes2005}
Miville-Desch{\^e}nes M.-A., Lagache G., 2005, The Astrophysical Journal
  Supplement Series, 157, 302

\bibitem[{Neal(2011)}]{Neal2011}
Neal R., 2011, MCMC using Hamiltonian dynamics. In Brooks, S., AGelman, A.,
  Jones, G.L. and Meng, X.L., editors, \textit{Handbook of Markov Chain Monte
  Carlo}, Chapman and Hall/CRC., 116

\bibitem[{Neal(1994)}]{Neal1994}
Neal R.~M., 1994, Journal of Computational Physics (ISSN 0021-9991), 111, 194

\bibitem[{Pettini \& Pagel(2004)}]{Pettini2004}
Pettini M., Pagel B. E.~J., 2004, Monthly Notices of the Royal Astronomical
  Society, 348, L59

\bibitem[{Pilbratt {et~al}\mbox{.}(2010)Pilbratt, Riedinger, Passvogel, Crone,
  Doyle, Gageur, Heras, Jewell, Metcalfe, Ott, \& Schmidt}]{Pilbratt2010}
Pilbratt G.~L. {et~al.}, 2010, A{\&}A, 518, L1

\bibitem[{Priddey {et~al}\mbox{.}(2003)Priddey, Isaak, McMahon, Robson, \&
  Pearson}]{Priddey2003}
Priddey R.~S., Isaak K.~G., McMahon R.~G., Robson E.~I., Pearson C.~P., 2003,
  Monthly Notices of the Royal Astronomical Society, 344, L74

\bibitem[{Rela{\~n}o {et~al}\mbox{.}(2018)Rela{\~n}o, De~Looze, Kennicutt,
  Lisenfeld, Dariush, Verley, Braine, Tabatabaei, Kramer, Boquien, Xilouris, \&
  Gratier}]{Relano2018}
Rela{\~n}o M. {et~al.}, 2018, A{\&}A, 613, A43

\bibitem[{R{\'e}my-Ruyer {et~al}\mbox{.}(2013)R{\'e}my-Ruyer, Madden, Galliano,
  Hony, Sauvage, Bendo, Roussel, Pohlen, Smith, Galametz, Cormier,
  Lebouteiller, Wu, Baes, Barlow, Boquien, Boselli, Ciesla, De~Looze,
  Karczewski, Panuzzo, Spinoglio, Vaccari, \& Wilson}]{Remy-Ruyer2013}
R{\'e}my-Ruyer A. {et~al.}, 2013, A{\&}A, 557, A95

\bibitem[{R{\'e}my-Ruyer {et~al}\mbox{.}(2015)R{\'e}my-Ruyer, Madden, Galliano,
  Lebouteiller, Baes, Bendo, Boselli, Ciesla, Cormier, Cooray, Cortese,
  De~Looze, Doublier-Pritchard, Galametz, Jones, Karczewski, Lu, \&
  Spinoglio}]{Remy-Ruyer2015}
R{\'e}my-Ruyer A. {et~al.}, 2015, A{\&}A, 582, A121

\bibitem[{Rowlands {et~al}\mbox{.}(2014)Rowlands, Gomez, Dunne,
  Arag{\'o}n-Salamanca, Dye, Maddox, da~Cunha, \& van~der Werf}]{Rowlands2014}
Rowlands K., Gomez H.~L., Dunne L., Arag{\'o}n-Salamanca A., Dye S., Maddox S.,
  da~Cunha E., van~der Werf P., 2014, Monthly Notices of the Royal Astronomical
  Society, 441, 1040

\bibitem[{Saintonge {et~al}\mbox{.}(2016)Saintonge, Catinella, Cortese, Genzel,
  Giovanelli, Haynes, Janowiecki, Kramer, Lutz, Schiminovich, Tacconi, Wuyts,
  \& Accurso}]{Saintonge2016}
Saintonge A. {et~al.}, 2016, Monthly Notices of the Royal Astronomical Society,
  462, 1749

\bibitem[{Saintonge {et~al}\mbox{.}(2017)Saintonge, Catinella, Tacconi,
  Kauffmann, Genzel, Cortese, Dav{\'e}, Fletcher, Graci{\'a}-Carpio, Kramer,
  Heckman, Janowiecki, Lutz, Rosario, Schiminovich, Schuster, Wang, Wuyts,
  Borthakur, Lamperti, \& Roberts-Borsani}]{Saintonge2017}
Saintonge A. {et~al.}, 2017, The Astrophysical Journal Supplement Series, 233,
  22

\bibitem[{Saintonge {et~al}\mbox{.}(2018)Saintonge, Wilson, Xiao, Lin, Hwang,
  Tosaki, Bureau, Cigan, Clark, Clements, De~Looze, Dharmawardena, Gao, Gear,
  Greenslade, Lamperti, Lee, Li, Micha{\l}owski, Mok, Pan, Sansom, Sargent,
  Smith, Williams, Yang, Zhu, Accurso, Barmby, Brinks, Bourne, Brown, Chung,
  Chung, Cibinel, Coppin, Davies, Davis, Eales, Fanciullo, Fang, Gao, Glass,
  Gomez, Greve, He, Ho, Huang, Jeong, Jiang, Jiao, Kemper, Kim, Kim, Kim, Ko,
  Kong, Lacaille, Lacey, Lee, Lee, Lee, Masters, Oh, Papadopoulos, Park, Park,
  Parsons, Rowlands, Scicluna, Scudder, Sethuram, Serjeant, Shao, Sheen, Shi,
  Shim, Smith, Spekkens, Tsai, Verma, Urquhart, Violino, Viti, Wake, Wang,
  Wouterloot, Yang, Yim, Yuan, \& Zheng}]{Saintonge2018}
Saintonge A. {et~al.}, 2018, Monthly Notices of the Royal Astronomical Society,
  481, 3497

\bibitem[{Sanders {et~al}\mbox{.}(2003)Sanders, Mazzarella, Kim, Surace, \&
  Soifer}]{Sanders2003}
Sanders D.~B., Mazzarella J.~M., Kim D.~C., Surace J.~A., Soifer B.~T., 2003,
  The Astronomical Journal, 126, 1607

\bibitem[{Sawicki(2012)}]{Sawicki2012}
Sawicki A., 2012, Journal of Physics A: Mathematical and Theoretical, 45,
  505202

\bibitem[{Schwarz(1978)}]{Schwarz1978}
Schwarz G., 1978, Ann. Statist., 6(2), 461

\bibitem[{Scoville {et~al}\mbox{.}(2014)Scoville, Aussel, Sheth, Scott,
  Sanders, Ivison, Pope, Capak, Vanden~Bout, Manohar, Kartaltepe, Robertson, \&
  Lilly}]{Scoville2014}
Scoville N. {et~al.}, 2014, The Astrophysical Journal, 783, 84

\bibitem[{Shetty {et~al}\mbox{.}(2009{\natexlab{a}})Shetty, Kauffmann, Schnee,
  \& Goodman}]{Shetty2009a}
Shetty R., Kauffmann J., Schnee S., Goodman A.~A., 2009{\natexlab{a}}, The
  Astrophysical Journal, 696, 676

\bibitem[{Shetty {et~al}\mbox{.}(2009{\natexlab{b}})Shetty, Kauffmann, Schnee,
  Goodman, \& Ercolano}]{Shetty2009b}
Shetty R., Kauffmann J., Schnee S., Goodman A.~A., Ercolano B.,
  2009{\natexlab{b}}, The Astrophysical Journal, 696, 2234

\bibitem[{Smith {et~al}\mbox{.}(2019)Smith, Clark, De~Looze, Lamperti,
  Saintonge, Wilson, Accurso, Brinks, Bureau, Chung, Cigan, Clements,
  Dharmawardena, Fanciullo, Gao, Gao, Gear, Gomez, Greenslade, Hwang, Kemper,
  Lee, Li, Lin, Liu, Moln{\'a}r, Mok, Pan, Sargent, Scicluna, Smith, Urquhart,
  Williams, Xiao, Yang, \& Zhu}]{Smith2019}
Smith M. W.~L. {et~al.}, 2019, Monthly Notices of the Royal Astronomical
  Society

\bibitem[{Smith {et~al}\mbox{.}(2012{\natexlab{a}})Smith, Eales, Gomez,
  Roman-Duval, Fritz, Braun, Baes, Bendo, Blommaert, Boquien, Boselli,
  Clements, Cooray, Cortese, De~Looze, Ford, Gear, Gentile, Gordon, Kirk,
  Lebouteiller, Madden, Mentuch, O'Halloran, Page, Schulz, Spinoglio,
  Verstappen, Wilson, \& Thilker}]{Smith2012b}
Smith M. W.~L. {et~al.}, 2012{\natexlab{a}}, The Astrophysical Journal, 756, 40

\bibitem[{Smith {et~al}\mbox{.}(2012{\natexlab{b}})Smith, Gomez, Eales, Ciesla,
  Boselli, Cortese, Bendo, Baes, Bianchi, Clemens, Clements, Cooray, Davies,
  De~Looze, di~Serego~Alighieri, Fritz, Gavazzi, Gear, Madden, Mentuch,
  Panuzzo, Pohlen, Spinoglio, Verstappen, Vlahakis, Wilson, \&
  Xilouris}]{Smith2012}
Smith M. W.~L. {et~al.}, 2012{\natexlab{b}}, The Astrophysical Journal, 748,
  123

\bibitem[{{Stan Development Team}(2017)}]{Stan_manual}
{Stan Development Team}, 2017, Stan Modeling Language: User's Guide and
  Reference Manual. Version 2.17.0. (http://mc-stan.org)

\bibitem[{{Stan Development Team}(2018)}]{pystan}
{Stan Development Team}, 2018, PyStan: the Python interface to Stan, Version
  2.17.1.0

\bibitem[{Tak, Ghosh \& Ellis(2018)Tak, Ghosh, \& Ellis}]{Tak2018}
Tak H., Ghosh S.~K., Ellis J.~A., 2018, Monthly Notices of the Royal
  Astronomical Society, 481, 277

\bibitem[{Tamm {et~al}\mbox{.}(2012)Tamm, Tempel, Tenjes, Tihhonova, \&
  Tuvikene}]{Tamm2012}
Tamm A., Tempel E., Tenjes P., Tihhonova O., Tuvikene T., 2012, A{\&}A, 546, A4

\bibitem[{Taylor(2005)}]{Taylor2005}
Taylor M.~B., 2005, Astronomical Data Analysis Software and Systems XIV ASP
  Conference Series, 347, 29

\bibitem[{{The Astropy Collaboration} {et~al}\mbox{.}(2013){The Astropy
  Collaboration}, {Robitaille, Thomas P.}, {Tollerud, Erik J.}, {Greenfield,
  Perry}, {Droettboom, Michael}, {Bray, Erik}, {Aldcroft, Tom}, {Davis, Matt},
  {Ginsburg, Adam}, {Price-Whelan, Adrian M.}, {Kerzendorf, Wolfgang E.},
  {Conley, Alexander}, {Crighton, Neil}, {Barbary, Kyle}, {Muna, Demitri},
  {Ferguson, Henry}, {Grollier, Fr\'ed\'eric}, {Parikh, Madhura M.}, {Nair,
  Prasanth H.}, {G\"unther, Hans M.}, {Deil, Christoph}, {Woillez, Julien},
  {Conseil, Simon}, {Kramer, Roban}, {Turner, James E. H.}, {Singer, Leo},
  {Fox, Ryan}, {Weaver, Benjamin A.}, {Zabalza, Victor}, {Edwards, Zachary I.},
  {Azalee Bostroem, K.}, {Burke, D. J.}, {Casey, Andrew R.}, {Crawford, Steven
  M.}, {Dencheva, Nadia}, {Ely, Justin}, {Jenness, Tim}, {Labrie, Kathleen},
  {Lim, Pey Lian}, {Pierfederici, Francesco}, {Pontzen, Andrew}, {Ptak, Andy},
  {Refsdal, Brian}, {Servillat, Mathieu}, \& {Streicher, Ole}}]{astropy}
{The Astropy Collaboration} {et~al.}, 2013, A\&A, 558, A33

\bibitem[{Utomo {et~al}\mbox{.}(2019)Utomo, Chiang, Leroy, Sandstrom, \&
  Chastenet}]{Utomo2019}
Utomo D., Chiang I.~D., Leroy A.~K., Sandstrom K.~M., Chastenet J., 2019, The
  Astrophysical Journal, 874, 141

\bibitem[{Van Der~Walt, Colbert \& Varoquaux(2011)Van Der~Walt, Colbert, \&
  Varoquaux}]{VanDerWalt2011}
Van Der~Walt S., Colbert S.~C., Varoquaux G., 2011, Computing in Science and
  Engineering, 13, 22

\bibitem[{Veneziani {et~al}\mbox{.}(2013)Veneziani, Piacentini, Noriega-Crespo,
  Carey, Paladini, \& Paradis}]{Veneziani2013}
Veneziani M., Piacentini F., Noriega-Crespo A., Carey S., Paladini R., Paradis
  D., 2013, The Astrophysical Journal, 772, 56

\bibitem[{Waelkens {et~al}\mbox{.}(1996)Waelkens, Waters, de~Graauw, Huygen,
  Malfait, Plets, Vandenbussche, Beintema, Boxhoorn, Habing, Heras, Kester,
  Lahuis, Morris, Roelfsema, Salama, Siebenmorgen, Trams, van~der Bliek,
  Valentijn, \& Wesselius}]{Waelkens1996}
Waelkens C. {et~al.}, 1996, A{\&}A, 315, L245

\bibitem[{Waters {et~al}\mbox{.}(1996)Waters, Molster, de~Jong, Beintema,
  Waelkens, Boogert, Boxhoorn, de~Graauw, Drapatz, Feuchtgruber, Genzel,
  Helmich, Heras, Huygen, Izumiura, Justtanont, Kester, Kunze, Lahuis, Lamers,
  Leech, Loup, Lutz, Morris, Price, Roelfsema, Salama, Schaeidt, Tielens,
  Trams, Valentijn, Vandenbussche, van~den Ancker, van Dishoeck, Van~Winckel,
  Wesselius, \& Young}]{Waters1996}
Waters L. B. F.~M. {et~al.}, 1996, A{\&}A, 315, L361

\bibitem[{Watson {et~al}\mbox{.}(2015)Watson, Christensen, Knudsen, Richard,
  Gallazzi, \& Micha{\l}owski}]{Watson2015}
Watson D., Christensen L., Knudsen K.~K., Richard J., Gallazzi A.,
  Micha{\l}owski M.~J., 2015, Nature, 519, 327

\bibitem[{Yang \& Phillips(2007)}]{Yang2007}
Yang M., Phillips T., 2007, The Astrophysical Journal, 662, 284

\bibitem[{Ysard {et~al}\mbox{.}(2018)Ysard, Jones, Demyk, Bout{\'e}raon, \&
  Koehler}]{Ysard2018}
Ysard N., Jones A.~P., Demyk K., Bout{\'e}raon T., Koehler M., 2018, A{\&}A,
  617, A124

\bibitem[{Zhukovska(2014)}]{Zhukovska2014}
Zhukovska S., 2014, A{\&}A, 562, A76

\end{thebibliography}




\appendix

\section{Normal distribution vs. Student's $t$-distribution }
\label{sec:normal_vs_student}

In this section we investigate how the choice of the prior distributions affect the results. In particular, the distribution of the parameter population $p(\vec{\theta}|\vec{\mu}, \Sigma)$ and the noise distribution $p(F^{obs}_j|\vec{\theta}, F^err_j, \delta_j)$.

The \stud\ is appropriate for robust statistical models \citep{Kelly2012,Gelman2004} and it is recommended when the measurement errors are assumed to be Gaussian, but their standard deviation is not known but only estimated.  If we assume that the true variance $\sigma^2$ follows a Scaled Inverse-$\chi^2$ distribution with scale parameter $\sigma$'$^2$ (the estimated variance), then  modelling the noise as a \stud\ with standard deviation $\sigma$'  is equivalent to the assumption that the noise is normal distributed with a standard deviation $\sigma$  \citep{Gelman2004}. For the choice of the degrees of freedom we follow \cite{Kelly2012} and used $f=3$, since it is the smallest value for which mean and variance of the distribution are finite. The results do not depends strongly on change on $f$ which are less than an order of magnitude and  $f$<10 is a typical choice for robust models \citep[e.g.][]{Gelman2004}. For a \stud\ with $f=3$, (61.5\%, 86.5\%, 94.6\%) of the distribution lie within (1$\sigma$, 2$\sigma$, 3$\sigma$) from the mean, respectively. In comparison, for a Gaussian distribution the percentages are (68.3\%, 95.4\%, 99.7\%).

\begin{figure*}
\centering
\subfigure{\includegraphics[width=0.9\textwidth]
{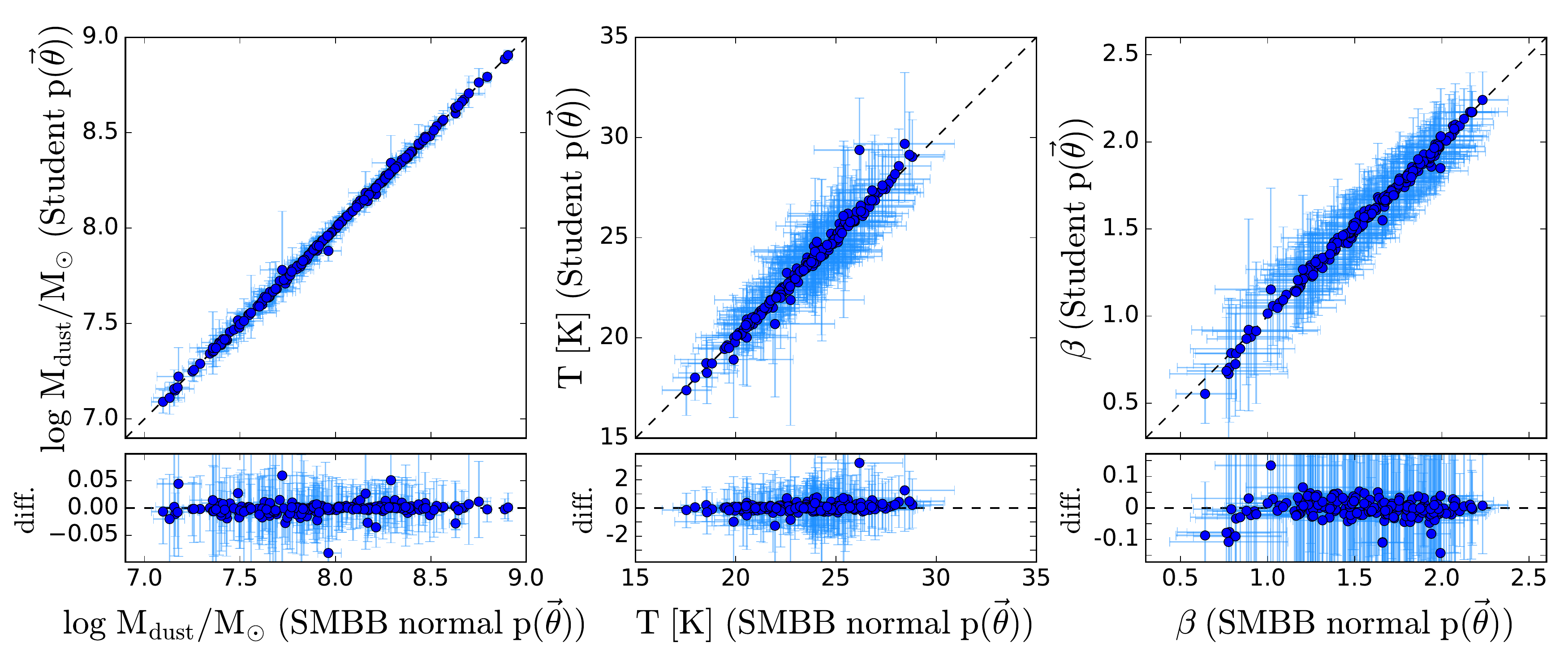}}
\caption{Comparison of the dust masses, temperatures and emissivity index obtained through the fit of a single modified black-body (SMBB) using the hierarchical approach, assuming a normal or a \stud\ with 8 degrees of freedom for the distribution of the parameters given the hyper-parameters $p(\vec{\theta}|\vec{\mu}, \Sigma)$.}
\label{fig:comp_normal_stud} 
\end{figure*}

First, we focus on the population distribution of the parameters given the hyper-parameters (mean and standard deviation). 
 We consider two distributions: normal and   \stud, which compared to the normal distribution allows for more outliers in the tail of the distribution. For the \stud\ we use $f=8$ degrees of freedom, which is the value we use for the analysis in this paper.
As we can see from Fig. \ref{fig:comp_normal_stud}, the results do not change much. The dust masses do not vary depending on the choice of the sample distribution. Temperature and $\beta$ show small differences, within the uncertainties, and no systematic offset.
We conclude that the choice of the distribution does not affect the results critically.

\begin{figure*}
\centering
\subfigure{\includegraphics[width=0.9\textwidth]
{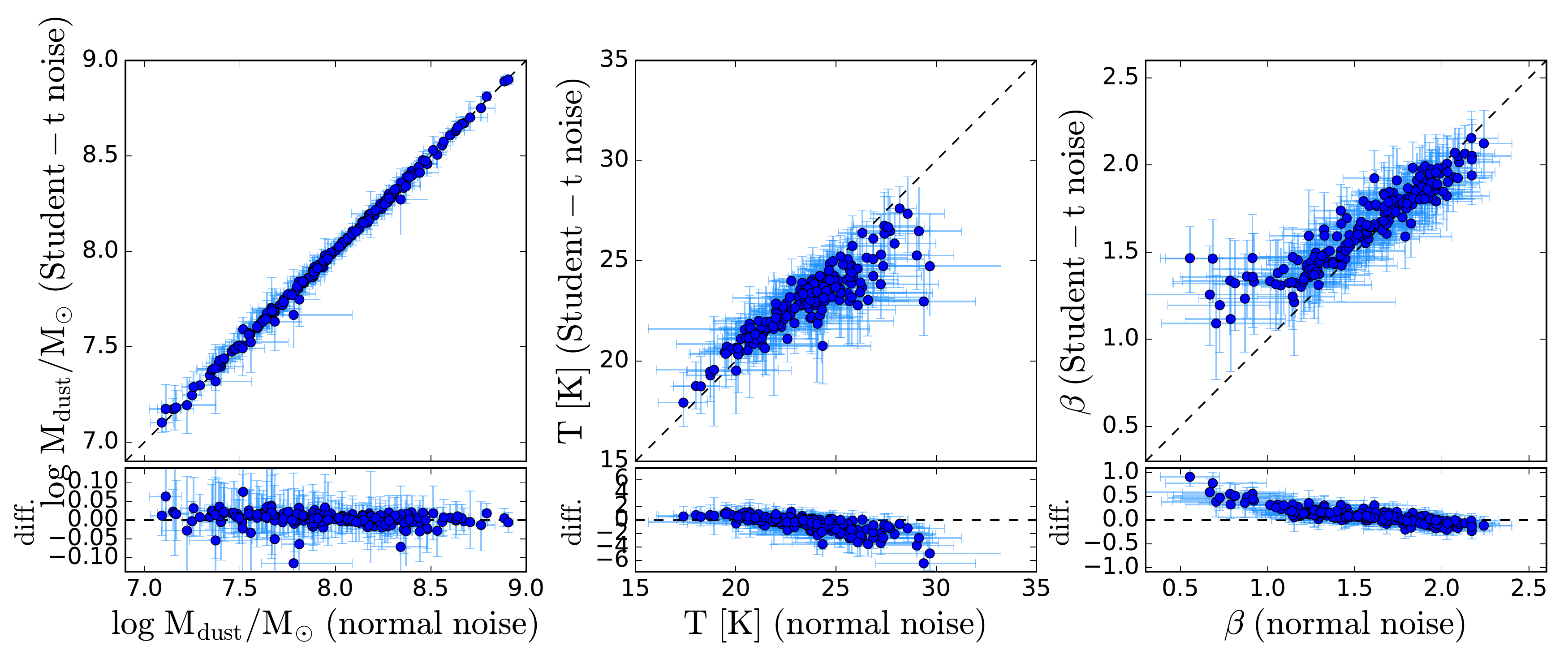}}
\caption{Comparison of the dust masses, temperatures and emissivity index obtained through the fit of a single modified black-body (SMBB) using the hierarchical approach, assuming a normal distribution or a \stud\ with three degrees of freedom for the noise.}
\label{fig:comp_normal_stud_noise} 
\end{figure*}

The second assumption on the priors is about the noise  distribution. We consider also in this case a normal and a \stud\ with three degrees of freedom (see description in Sec.\ref{sec:noise_distr}). The \stud\ is less sensitive to flux points which may be outliers, due to large uncertainties or to the noise being underestimate. 
Figure \ref{fig:comp_normal_stud_noise} shows the comparison plots. Again, the dust masses are robust with respect to the choice of the noise distribution.
Temperature and $\beta$, on the other hand, show some variations, with the results obtained using the \stud\ covering a smaller range of $T$ and $\beta$ values with respect to the results from the normal distribution ($T =18-27$~K, $\beta=1.1-2.1$ for the \stud, $T =17-30$~K, $\beta=0.6-2.2$ for the normal distribution).
The values of $T$ or $\beta$ which differ more from the mean values are determined mostly by the flux points at long (850\micron) or short wavelengths (100\micron). These are also the flux points which on average have the largest measurement uncertainties. With the assumption of a \stud, we imply that the deviation of the SED shape from a SMBB with mean parameter values is not due to a change in $T$ or $\beta$, but is more likely due to uncertainties in the flux measurements, which lead to 'outlier' flux points. Therefore the measured $T$ and $\beta$ will cover a smaller range of values.

\begin{figure}
\centering
\subfigure{\includegraphics[width=0.49\textwidth]
{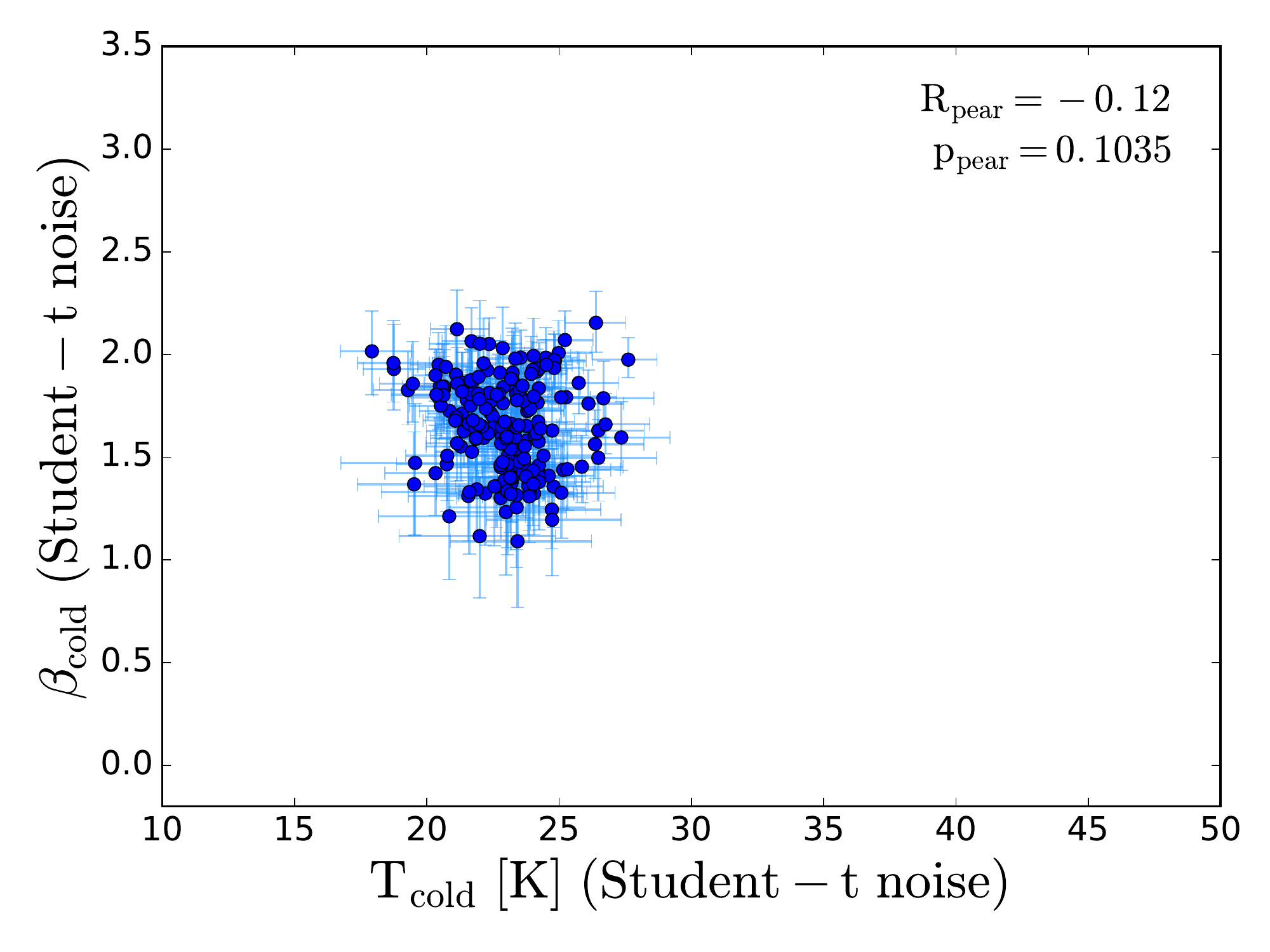}}
\caption{ Relation between the dust temperature and dust emissivity index ($T$-$\beta$ relation) from the SMBB hierarchical fit of the JINGLE sample assuming a \stud\ with three degrees of freedom for the noise.}
\label{fig:T-beta_stud_noise} 
\end{figure}

The choice of the noise distribution affects consequently also the derived relation between $T$ and $\beta$, shown in Fig. \ref{fig:T-beta_stud_noise}. Assuming a \stud\ for the noise, the results  show a weak anti-correlation ($R=-0.12$) between $T$ and $\beta$. This is similar to the result obtained from the fit without the 850\micron\ point. 
The fit with Student noise assumes that the variations at 850\micron\ are due to larger uncertainties on the estimate of the 850\micron\ uncertainties, rather than to a real variations in the sub-mm slope of the SED. Therefore the fit tends to `ignore' the extreme 850\micron\ flux points.  
In some cases the 850\micron\ point does not follow the same SED slope as the other points, but it shows an excess or a deficit. We have two ways to model this type of SED.  One possibility is to  assume that the true uncertainties on the 850$\micron$ point are larger than the estimated ones, and therefore model the SED  assuming a \stud. If instead we believe that the different behaviour of the SED at wavelengths longer than 500\micron\ is real, we can model the noise using a normal distribution. In this paper, we decide to model the noise using a normal distribution.

\section{Upper limits formalism}
In the case of a non-detection in one of the bands,  the likelihood needs to be modified to include an upper limit for the non-detection. 
 Following the formalism described in \cite{Sawicki2012}, the upper limit of an observation provides a limit on the evaluation of a definite integral.  For a measured flux, $F_j$, which is clearly detected,  the probability of observing our data, given the true value of the observables $F_j^{true}$ and the measurement uncertainties $F_{j}^{err}$, is:
\begin{equation}
p(F_j|F_j^{true}, F_j^{err}).
\end{equation}
 In the case of a single non-detection, we consider the upper limit $F_{lim,j}$, and the probability is:
\begin{equation}
p(F_{lim,j}|F_j^{true}, F_j^{err}) = \int_{-\infty}^{F_{lim,j}} p(F_j|F_j^{true}, F_j^{err}) dF_j.
\end{equation}

\textbf{\textit{Non-hierarchical}}: in the non-hierarchical approach, the likelihood in case of a non detection on the $j$-th flux measurement is given by:
\begin{multline}
p(\vec{F}^{obs}|\vec{\theta}) = \int_{-\infty}^{F_{lim,j}} \text{MultiNormal}(\vec{F}^{obs}|\vec{F}^{mod}(\vec{\theta}), C) dF_j^{obs}.
\end{multline}

Since the likelihood evaluation in case of upper limits includes  the computation of integrals, the use of upper limit is computationally expensive. Thus we allow our code to perform the SED fit with one flux point as upper limit at most, to avoid that the code has to calculate too many integrals. If more than one band has an upper limit, we consider only the upper limit in one band and we neglect the other flux point. We prefer to keep the 850\micron\ point, if it is an upper limit, since it is the longest wavelength point and it is the one that places more constraints on the SED slope.\\ 
\\
\textbf{\textit{Hierarchical}}: similarly, for the hierarchical method the likelihood for the $i$-th galaxy in case of a non-detection in the $j$-th band is:
\small
\begin{multline}
p(\vec{F_i}^{obs}|\vec{\theta_i}) =\\ =\int_{-\infty}^{F_{lim,i,j}} \text{MultiNormal}(\vec{F_i}^{obs}|\vec{F_i}^{mod}(\vec{\theta_i}), C_i) dF_{i, j}^{obs}.
\label{UL_hier}
\end{multline}
\normalsize

If the upper limit is in a band whose uncertainties are not correlated with other bands (i.e. the SCUBA-2 850\micron\ band or the IRAS 60\micron\ band), the expression for the upper limit can be divided in two parts, and the part that does not depend on $F_{i,j}$ can be taken out of the integral:

\small
\begin{multline}
p(\vec{F_i}^{obs}|\vec{\theta_i}) = \\
 = \text{MultiNormal}(\vec{F_i^{'}}^{obs}|\vec{F_i^{'}}^{mod}(\vec{\theta_i}), C_i^{'})\\ \cdot \int_{-\infty}^{F_{lim,i,j}} p(F_{i, j}^{obs}|F_{i, j}^{mod}(\vec{\theta_i}), F_{i, j}^{err})  dF_{i, j}^{obs} \\
= \text{MultiNormal}(\vec{F_i^{'}}^{obs}|\vec{F_i^{'}}^{mod}(\vec{\theta_i}), C_i^{'})\\ \cdot \int_{-\infty}^{F_{lim,i,j}} \text{Normal}\left(F_{i, j}^{obs}|F_{i, j}^{mod}(\vec{\theta_i}), \sqrt{C_{i, jj}}\right)  dF_{i, j}^{obs},
\label{UL_hier}
\end{multline}
\normalsize

where $\vec{F_i^{'}}$ is the ($m-1$)-dimensional vector equal to the vector $\vec{F_i}$ but without the $j$-th component. Similarly, $C_i^{'}$ is equal to the covariance matrix $C_i$, but without the $j$-th component. $C_{i, jj}$ is the $jj$ component of the covariance matrix $C_{i}$ for the $i$-th galaxy.

The integral of the univariate normal distribution can then be computed analytically:
\small
\begin{multline}
\int_{-\infty}^{y_{lim}} \text{Normal}(y|\mu, \sigma)  dy 
= \frac{1}{\sqrt{2\pi}\sigma} \cdot \int_{-\infty}^{y_{lim}} \exp \left(- \frac{1}{2}\left(\frac{y-\mu}{\sigma} \right)^2 \right) dy  \\
 = \frac{1}{2} \left[\text{erf}\left( \frac{y_{lim}-\mu}{\sqrt{2}\sigma}\right) +1 \right],
\label{UL_hier_normal}
\end{multline}
\normalsize
where `erf' is the error function. If the upper limit is in one of the \Herschel\ bands, the integral can also be computed analytically, but it requires more computations and it slows down code. Therefore we decide to ignore the points with non-detections in the \Herschel\ bands.

\section{Additional simulations}
\label{sec:simulation}

\textbf{$T-\beta$ anti-correlated:}

The second test we did was to see whether the hierarchical code can recover a $T$-$\beta$ anti-correlation. 
We simulated 100 SEDs with temperatures uniformly distributed in the range 20 - 30 K and the corresponding $\beta$ given by the relation:
\begin{equation}
\beta = -0.121 \cdot T + 4.595.
\end{equation} 
The slope and intercept of this relation are derived from the results of the non-hierarchical fit to the real data. We also added some scatter to the $T-\beta$ anti-correlation.
As before, we kept the dust mass constant ($\log M=8$ \Msun). 

\begin{figure*}
\centering
\subfigure{\includegraphics[width=0.48\textwidth]
{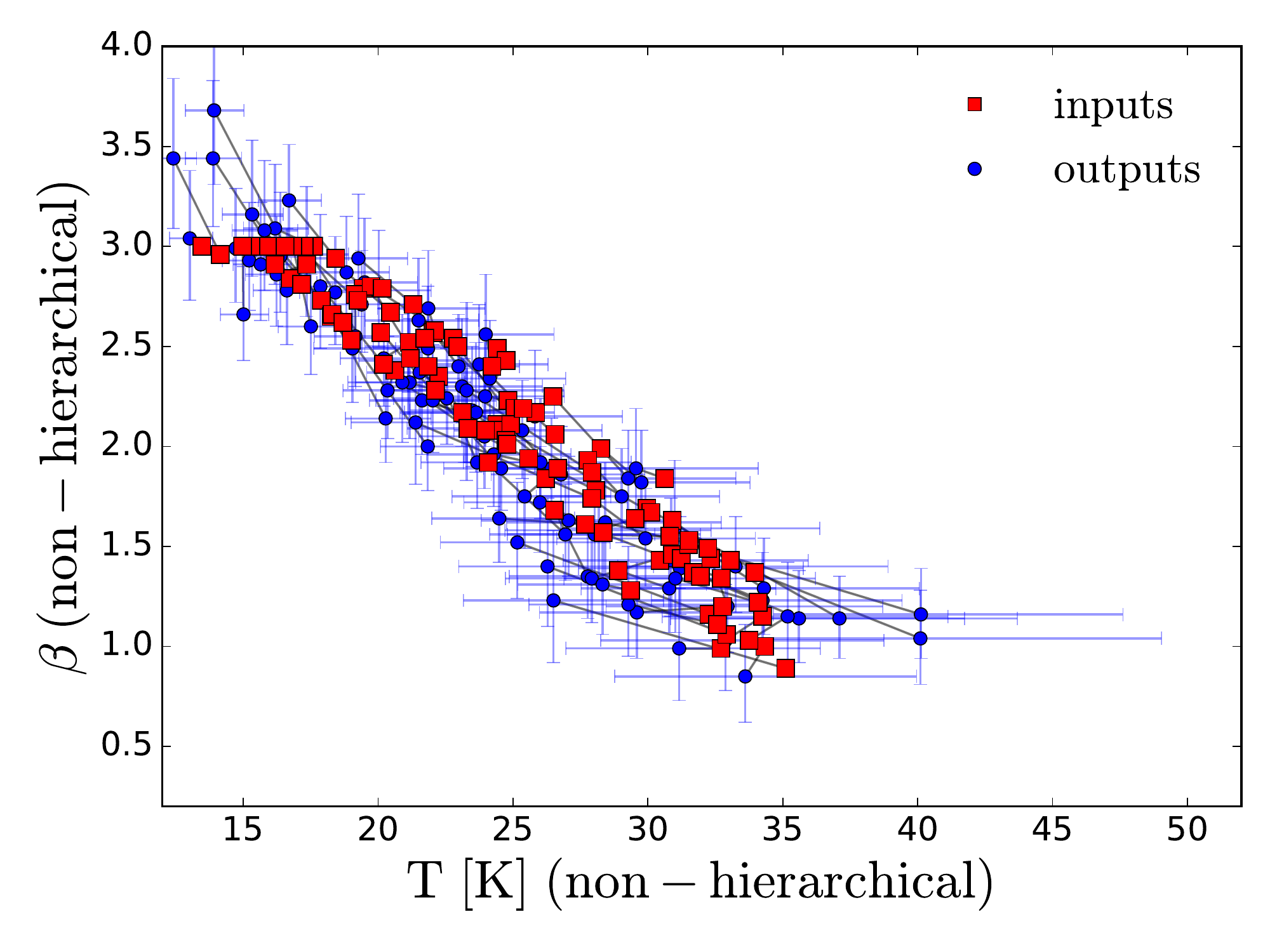}}
\subfigure{\includegraphics[width=0.48\textwidth]
{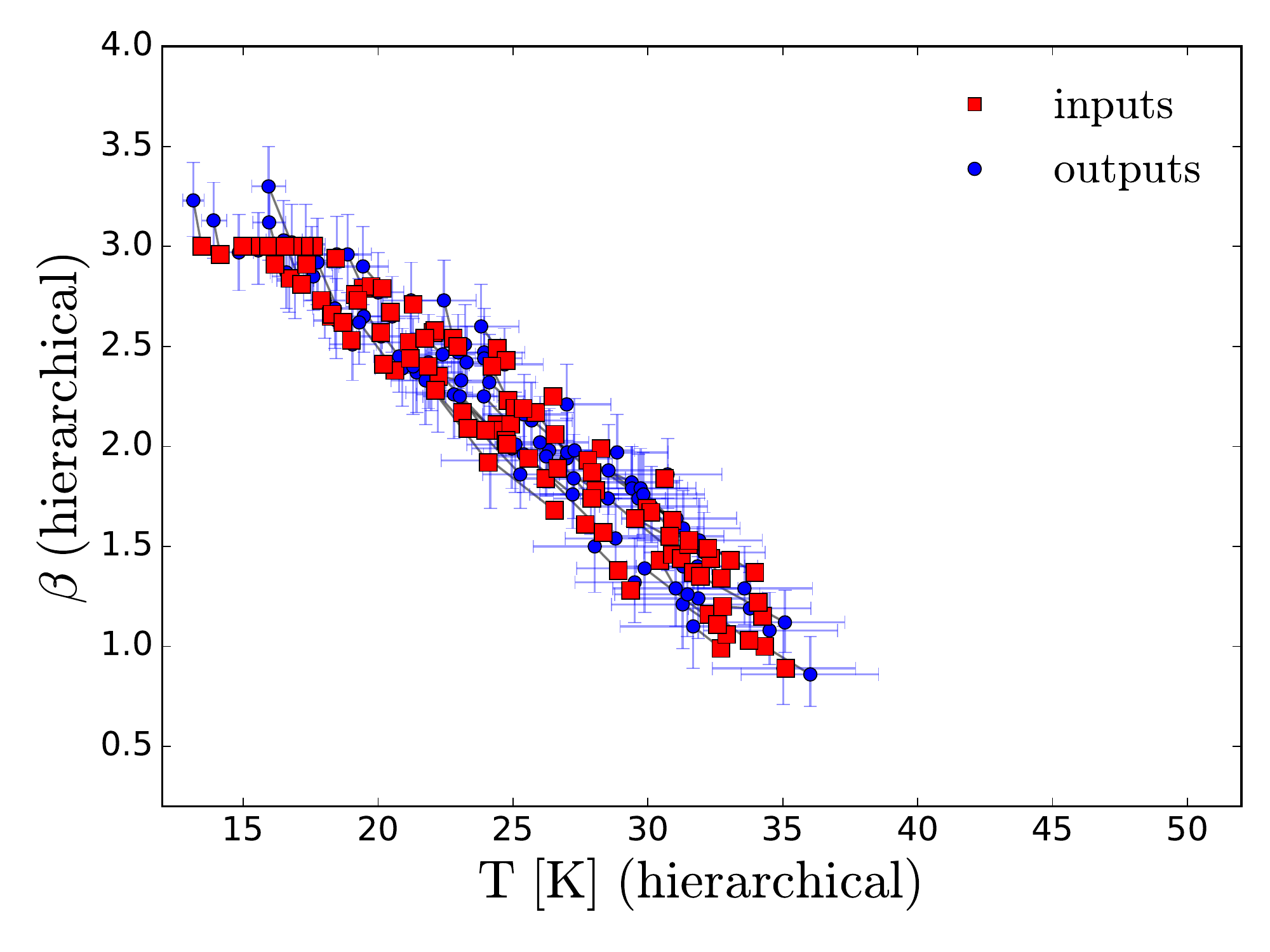}}
\caption{Results of temperature and $\beta$ derived from the fit of 100 simulated SMBB SEDs with T-$\beta$ anti-correlated and the same dust mass ($\log M_{dust}=8$ \Msun .).  The temperature are linearly distributed and some scatter is added around a linear $T-\beta$ relation. We added to every band  Gaussian noise with an amplitude proportional to the level of noise present in our data in that band. The output values are derived with the non-hierarchical (left panel) and  hierarchical (right panel) fitting approach. In red are shown the input values and in blue are the measured values (outputs), the grey lines connect the corresponding inputs and outputs.  
}
\label{fig:T-beta_anticorr}
\end{figure*}

 Fig. \ref{fig:T-beta_anticorr} shows the results from the hierarchical and non-hierarchical method. The non-hierarchical method  points move in the $T-\beta$ anti-correlation direction. Thus even if the $T-\beta$ anti-correlation is maintained, the differences between input values and measured values can be up to 8.6K in temperature and 0.7 in $\beta$.
Also the hierarchical code is able to recover the $T-\beta$ anti-correlation. The difference between input and measured 
are a bit smaller than in the non-hierarchical case (< 5.8K in temperature and 0.5 in $\beta$)

Comparing directly the input and output parameters, we see that the largest discrepancies  between input and output temperatures happen for high temperature values. This is due to the fact that  the FIR SED moves to lower wavelengths with increasing temperature. Thus for the high temperature models (T > 30 K), the peak of the SED is at wavelengths < 100\micron, which are not sampled by our data points/bands.
 This problem affects also the measurements of $\beta$: if the temperature is not well constrained, also $\beta$ will not be determined with high precision, due to the degeneracy between the two parameters. 
Additionally, due to the assumed $T$-$\beta$ anti-correlation,  high $T$ values correspond to low $\beta$ values, i.e. shallower slopes of the SED. This will also contribute to the difficulties of accurately measure $T$ and $\beta$. 

\begin{figure*}
\centering
\subfigure{\includegraphics[width=0.48\textwidth]
{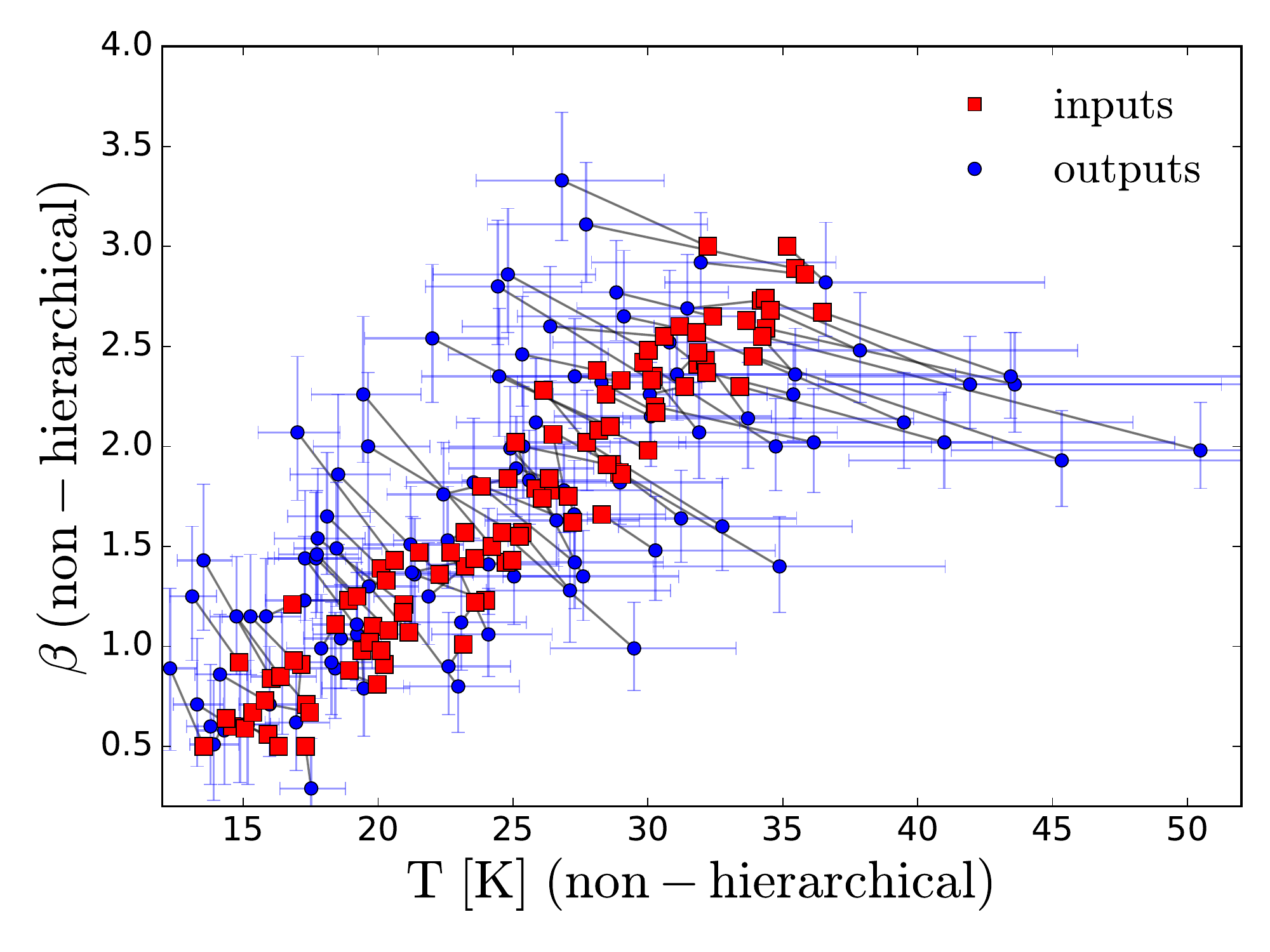}}
\subfigure{\includegraphics[width=0.48\textwidth]
{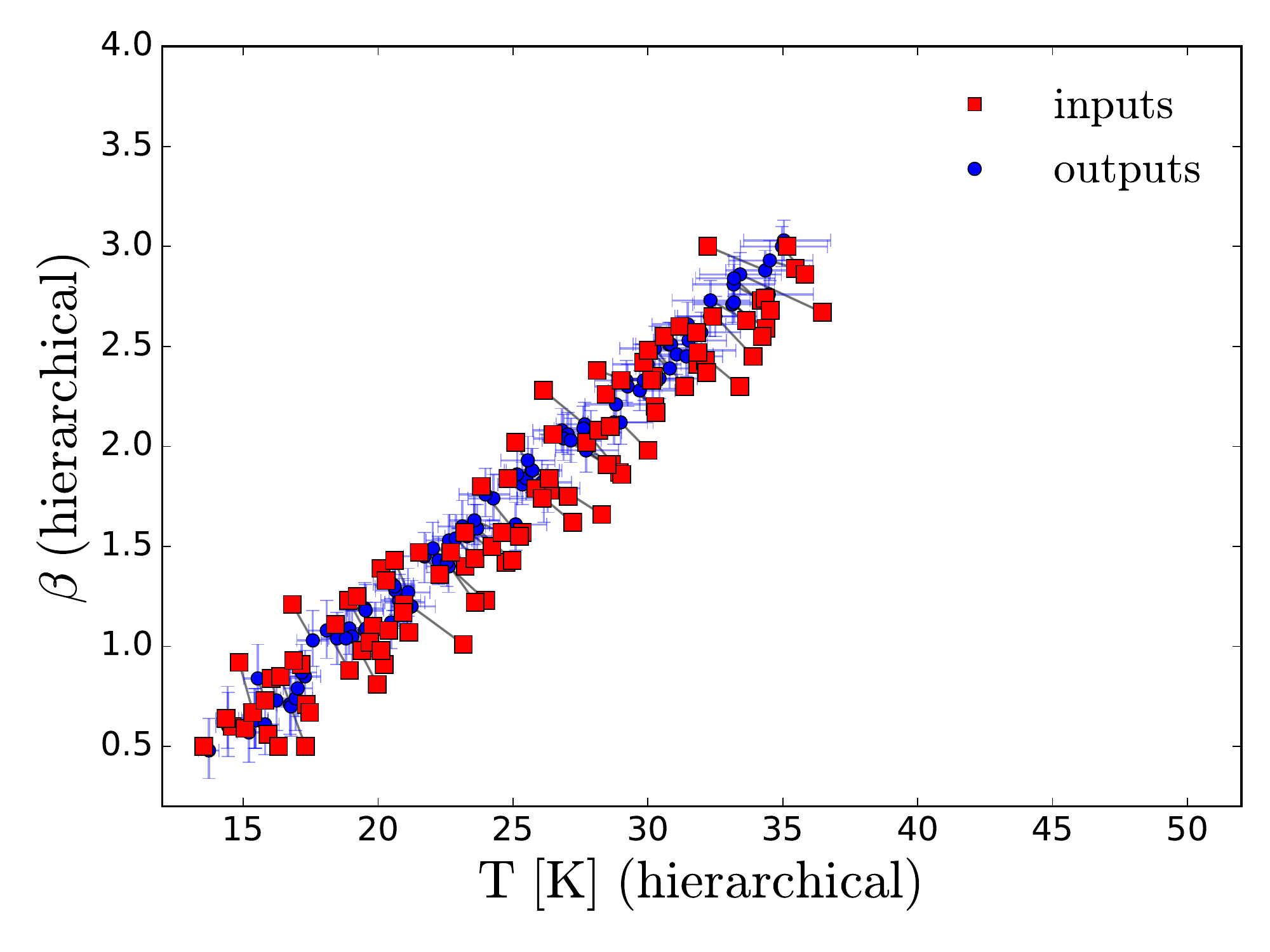}}
\caption{Results of temperature and $\beta$ derived from the fit of 100 simulated SMBB SEDs with T-$\beta$ correlated and the same dust mass ($\log M_{dust}=8$ \Msun .).  The temperature are linearly distributed and some scatter is added around a linear $T-\beta$ relation. We added to every band  Gaussian noise with an amplitude proportional to the level of noise present in our data in that band. The output values are derived with the non-hierarchical (left panel) and  hierarchical (right panel) fitting approach. In red are shown the input values and in blue are the measured values (outputs), the grey lines connect the corresponding inputs and outputs.  
}
\label{fig:T-beta_corr}
\end{figure*}

\textbf{T-$\beta$ correlated:}
We did the same test for positive correlation between $T$ and $\beta$, parametrized by the relation:
\begin{equation}
\beta = 0.121 \cdot T - 1.325.
\end{equation} 
As we can see from the left panel of Fig. \ref{fig:T-beta_corr}, the non-hierarchical method is not able to recover the positive correlation. The results of the fitting move away from the input values along diagonal lines in the $T-\beta$ plane, following the anti-correlation line. 
The right panel of Fig. \ref{fig:T-beta_corr} shows the results from the hierarchical SED fitting. The code can recover the input values and the trend quite well. 
We note that the difference between input and output is often larger than the error bars.
The points tend to move along diagonal lines in the $T-\beta$ plane, following the anti-correlation line. Therefore some points move outside the input correlation. However, the difference between input and output are small enough, that  the $T-\beta$ positive correlation is visible also in the outputs value.

From these tests we can conclude that the hierarchical approach performs better than the non-hierarchical approach in all three cases of single input, $T-\beta$ correlation, and anti-correlation.
 In the case of a positive correlation, we note that even in the hierarchical approach the difference between input and output values can sometimes be larger than our errorbars. The differences in temperature are <3~K , and the difference in $\beta$ are < 0.3. For comparison, in the non-hierarchical case, the differences in temperature are < 16 K, and the difference in $\beta$ are < 0.8.

\section{Plots of the fitted SED}

\begin{figure*}
\centering
\subfigure{\includegraphics[width=0.44\textwidth]
{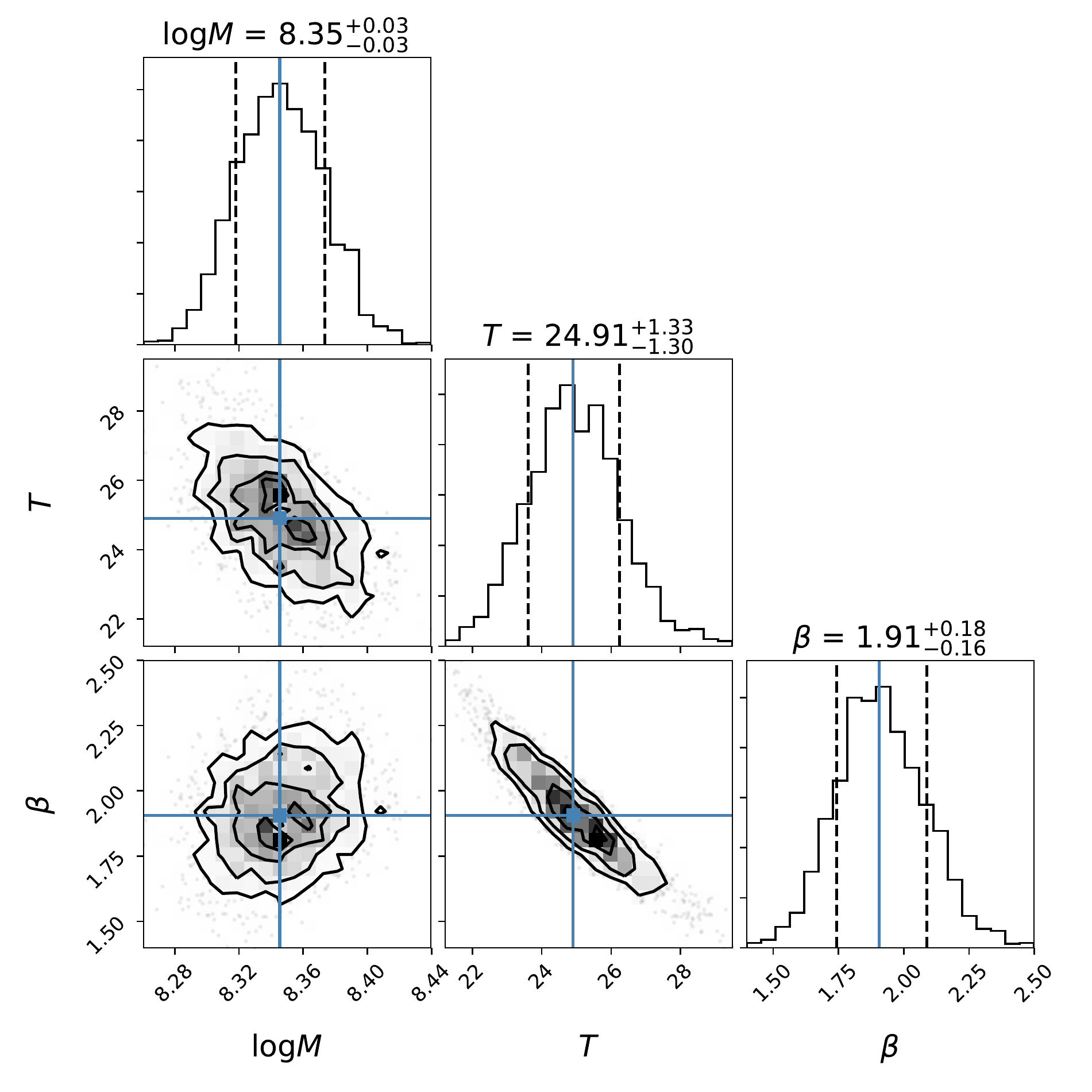}}
\subfigure{\includegraphics[width=0.44\textwidth]
{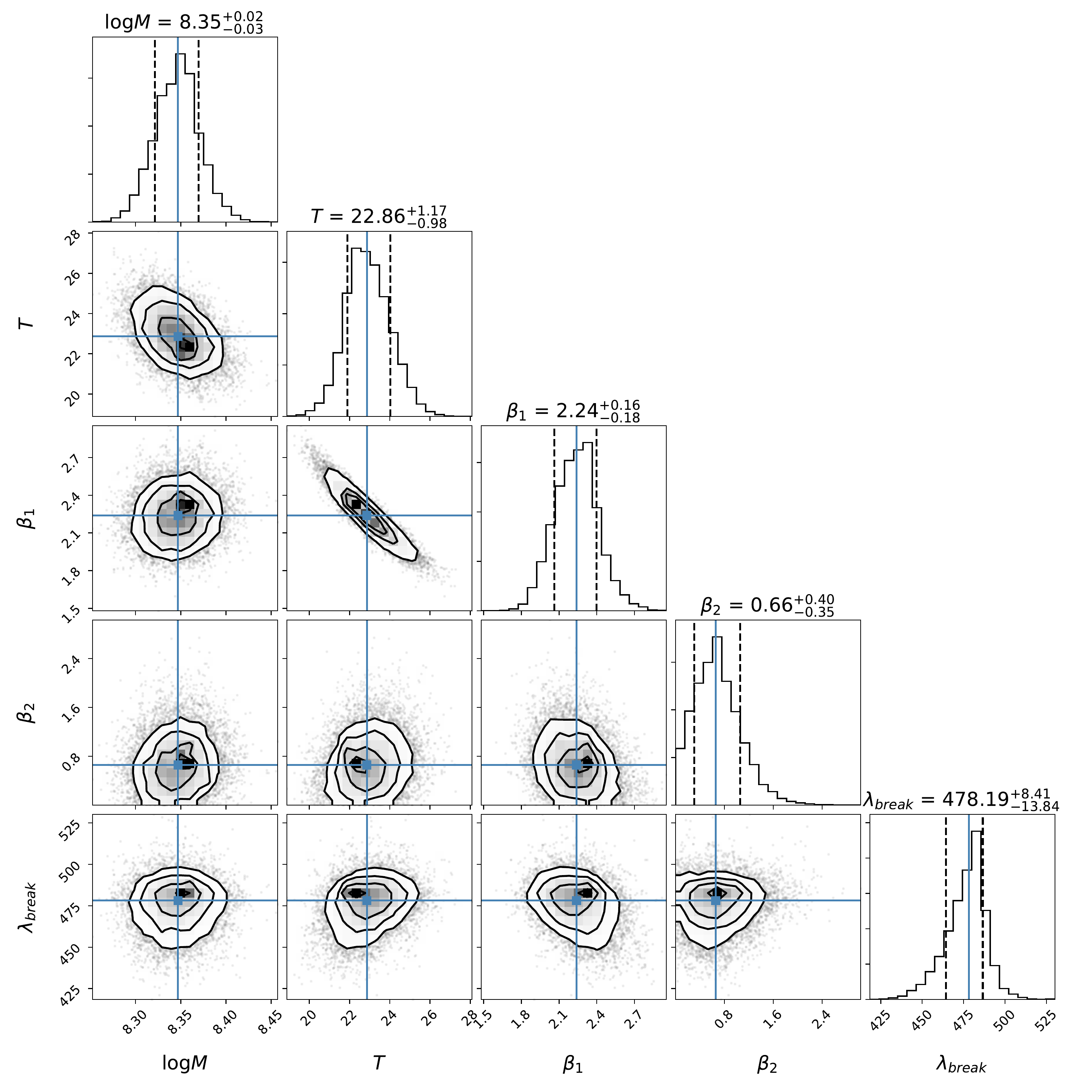}}
\subfigure{\includegraphics[width=0.44\textwidth]
{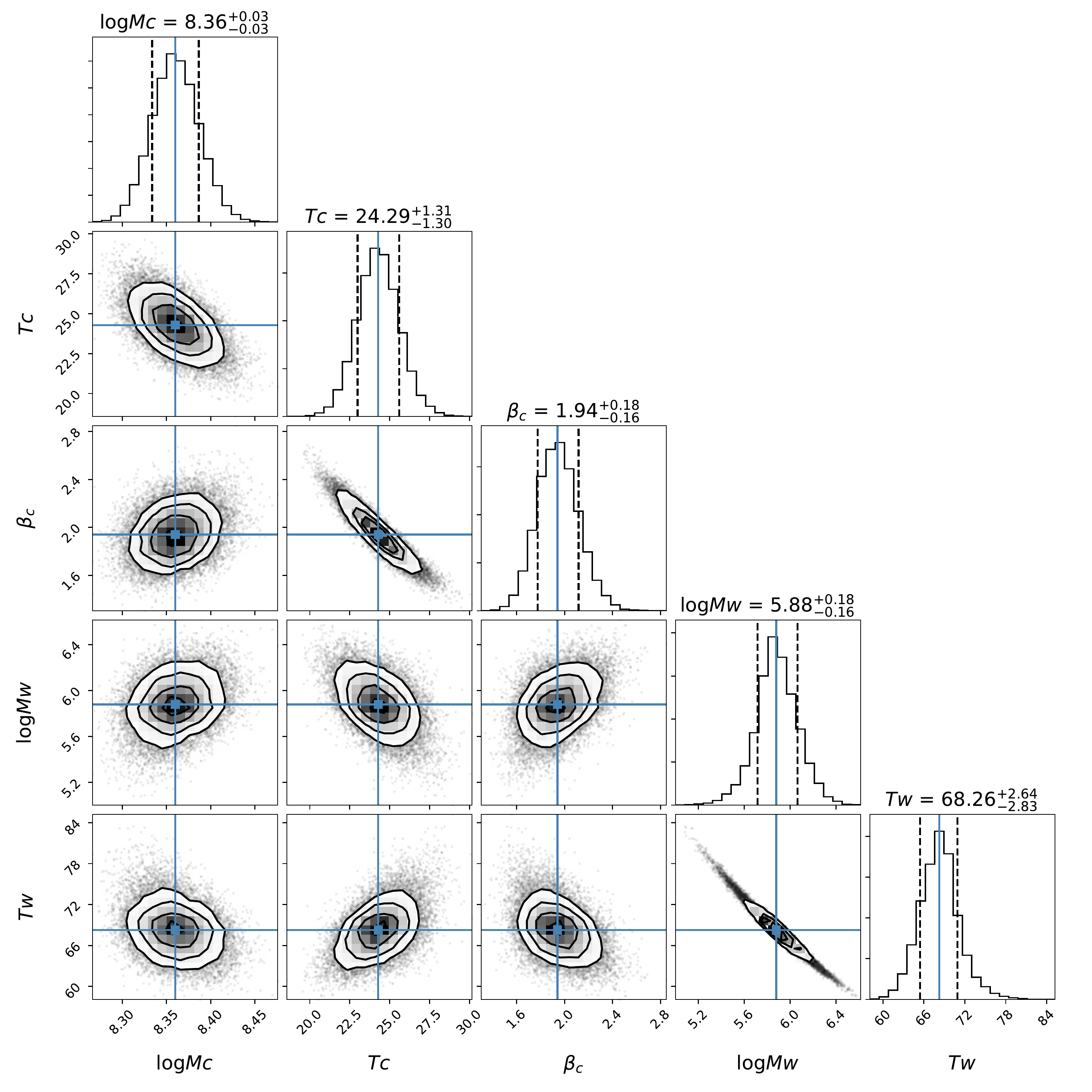}}
\caption{Example of the posterior probability density functions (PDFs) of the model parameters obtained using the hierarchical method for the fit of one galaxy (JINGLE 147). The three panels show the results of fit using the SMBB (upper left), BMBB (upper right) and TMBB (bottom) models. The blue line indicates the median values, the dotted lines show the 16th and 84th percentiles, that indicate the one-sigma uncertainties on the parameters.}
\label{fig:post_triangle_plot}
\end{figure*}

\begin{figure*}
\centering
\subfigure{\includegraphics[width=0.3\textwidth]
{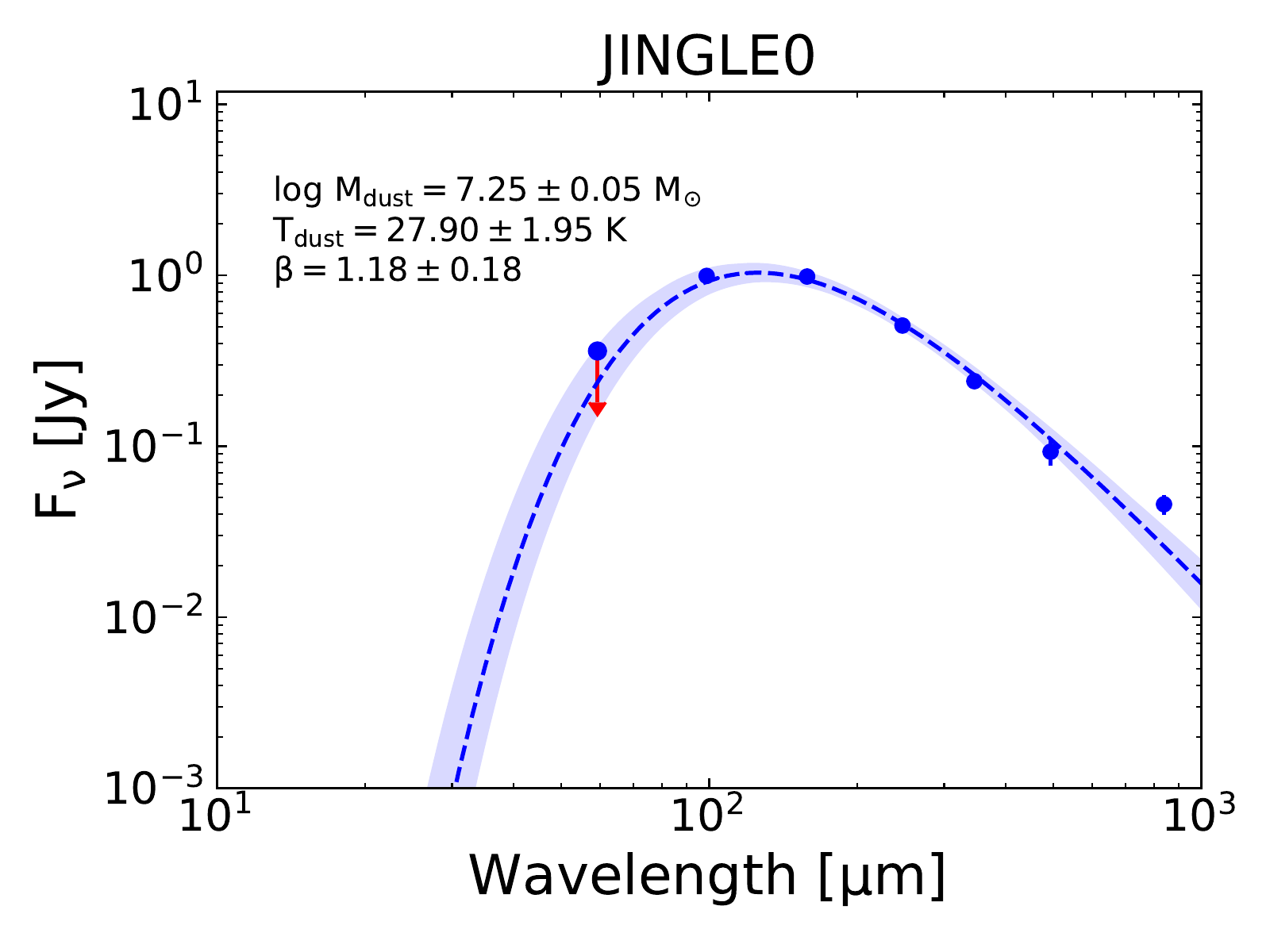}}
\subfigure{\includegraphics[width=0.3\textwidth]
{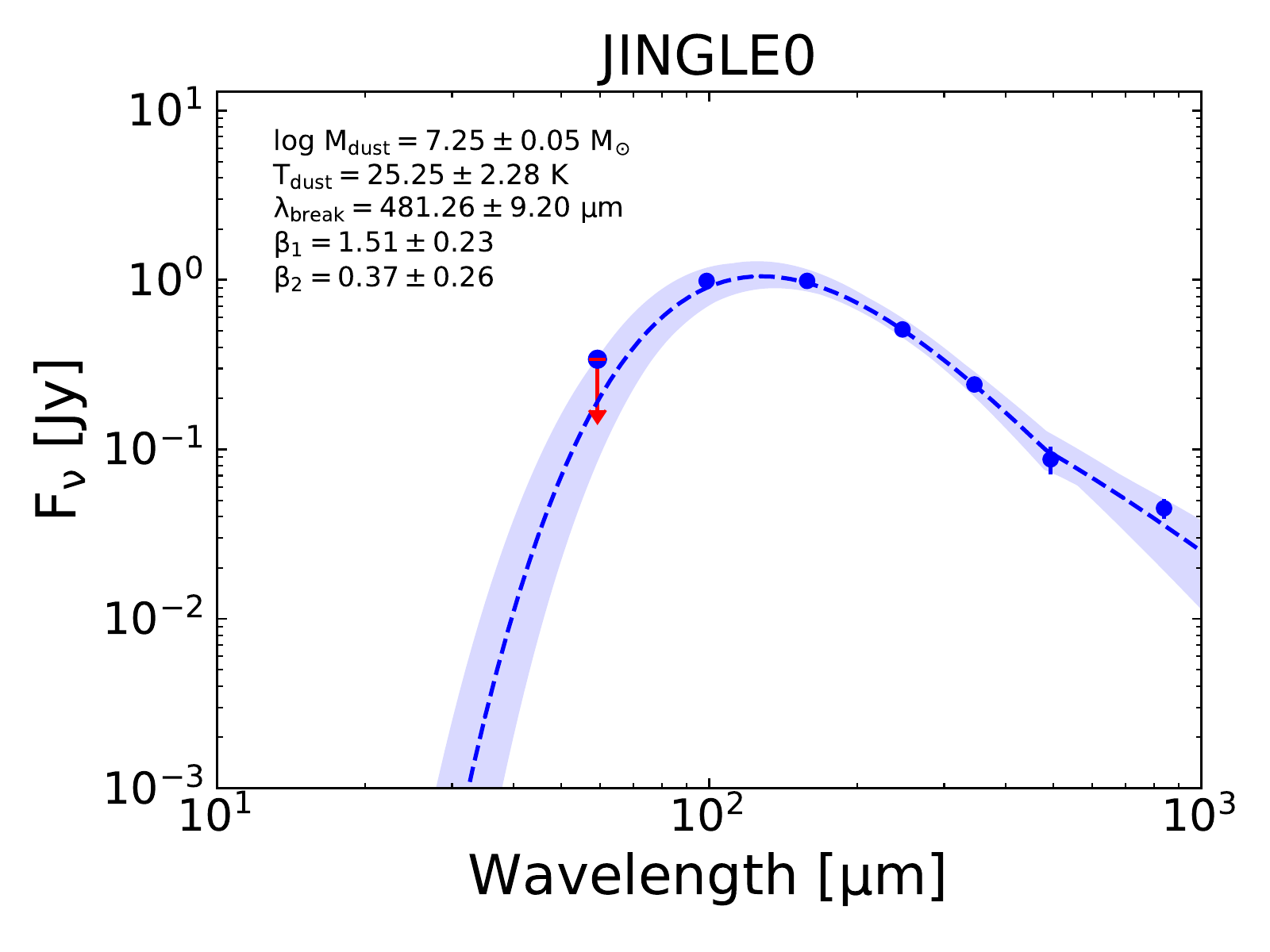}}
\subfigure{\includegraphics[width=0.3\textwidth]
{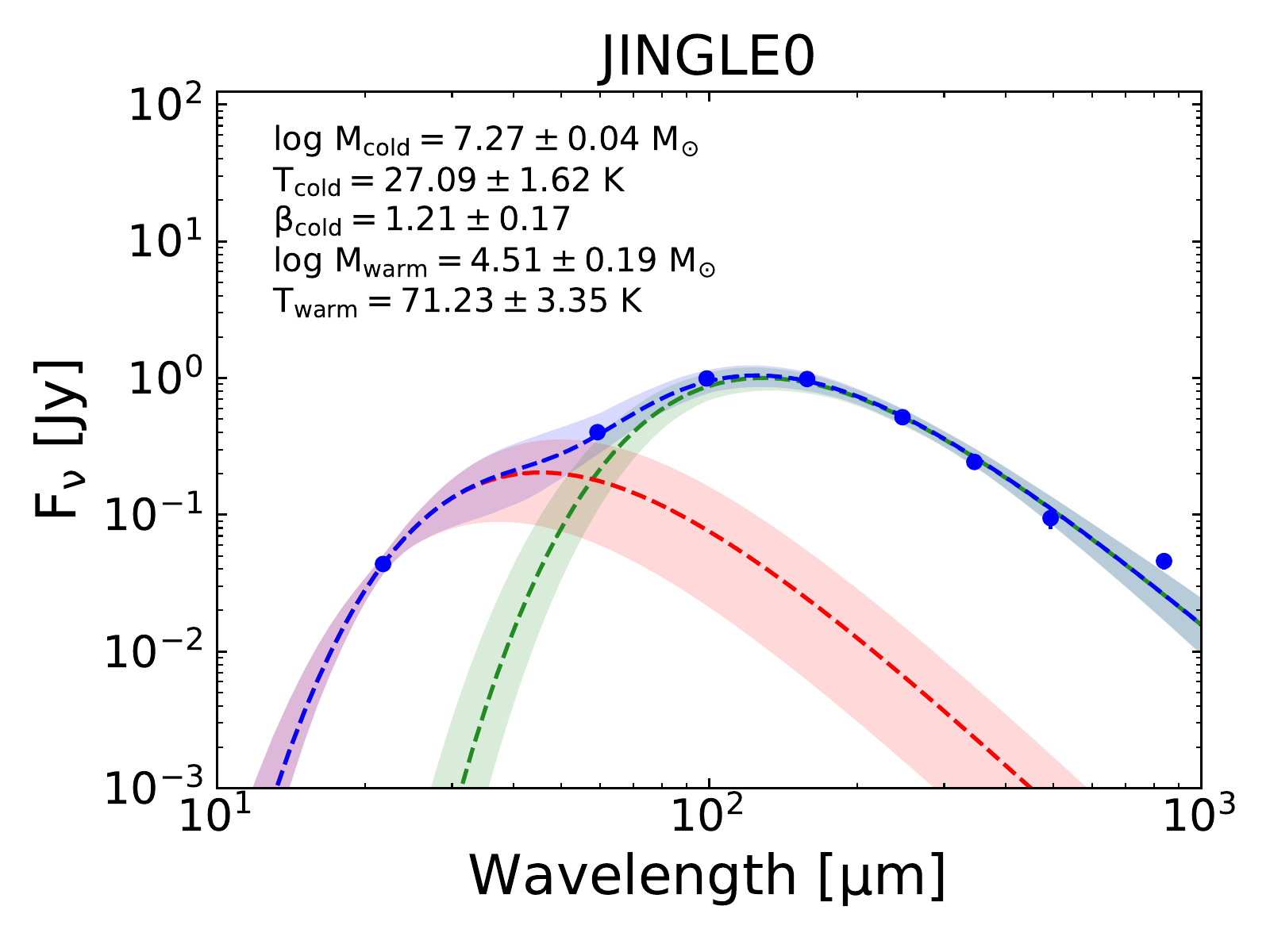}}

\subfigure{\includegraphics[width=0.3\textwidth]
{Figures/SED_models/JINGLE/SMBB/JINGLE1_SMBB_hier_SED_fit}}
\subfigure{\includegraphics[width=0.3\textwidth]
{Figures/SED_models/JINGLE/BMBB/JINGLE1_BMBB_hier_SED_fit}}
\subfigure{\includegraphics[width=0.3\textwidth]
{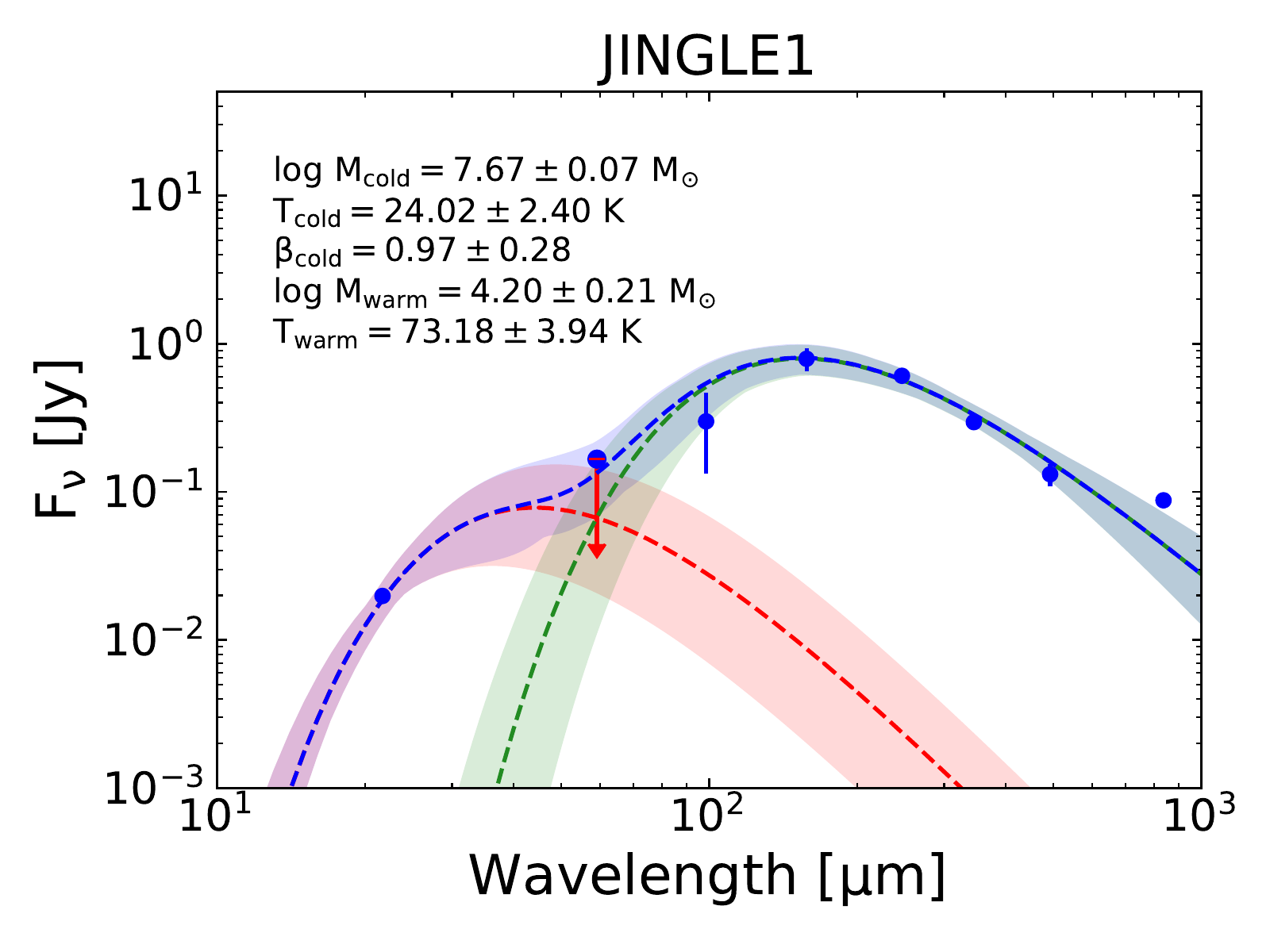}}

\subfigure{\includegraphics[width=0.3\textwidth]
{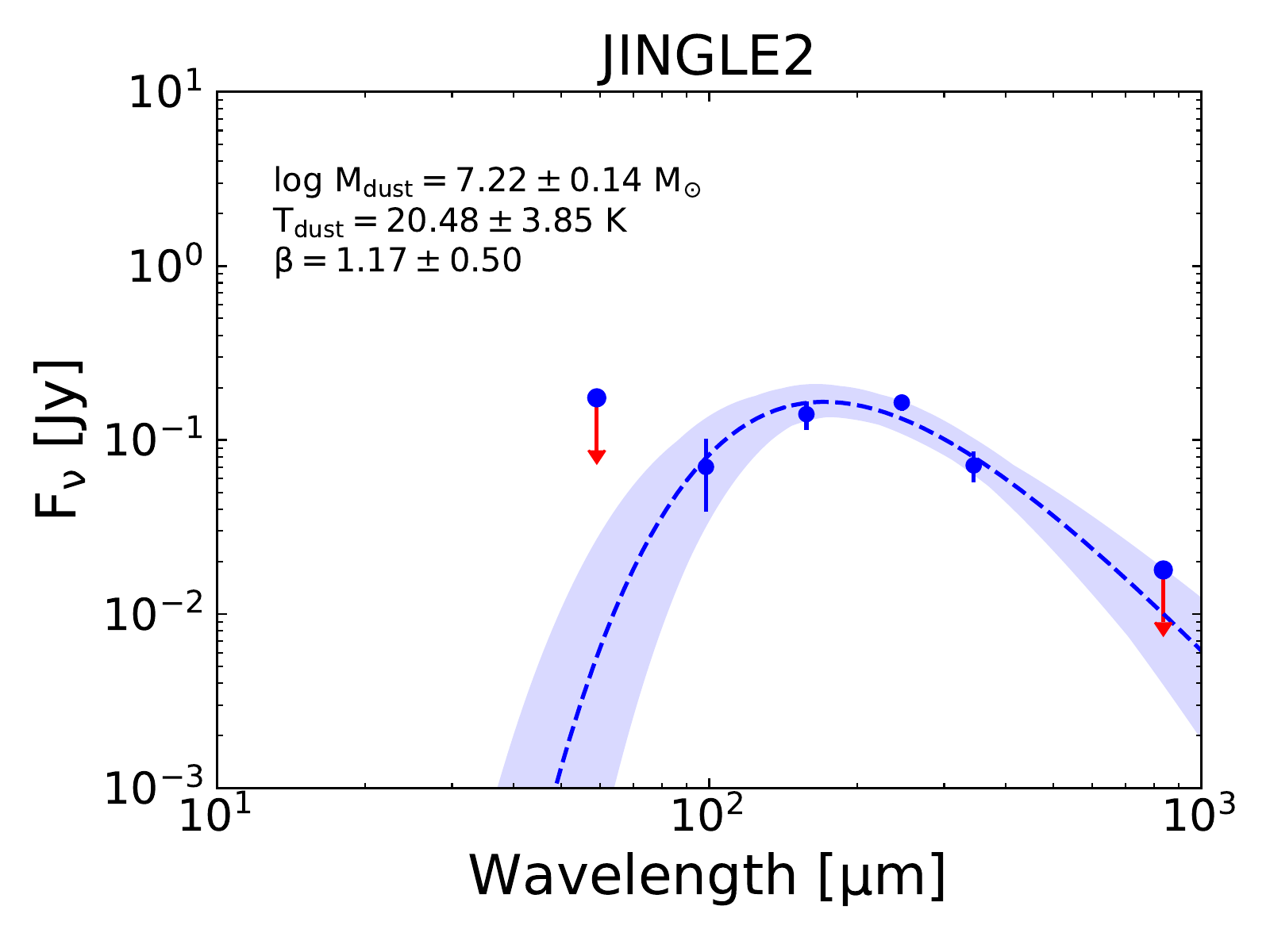}}
\subfigure{\includegraphics[width=0.3\textwidth]
{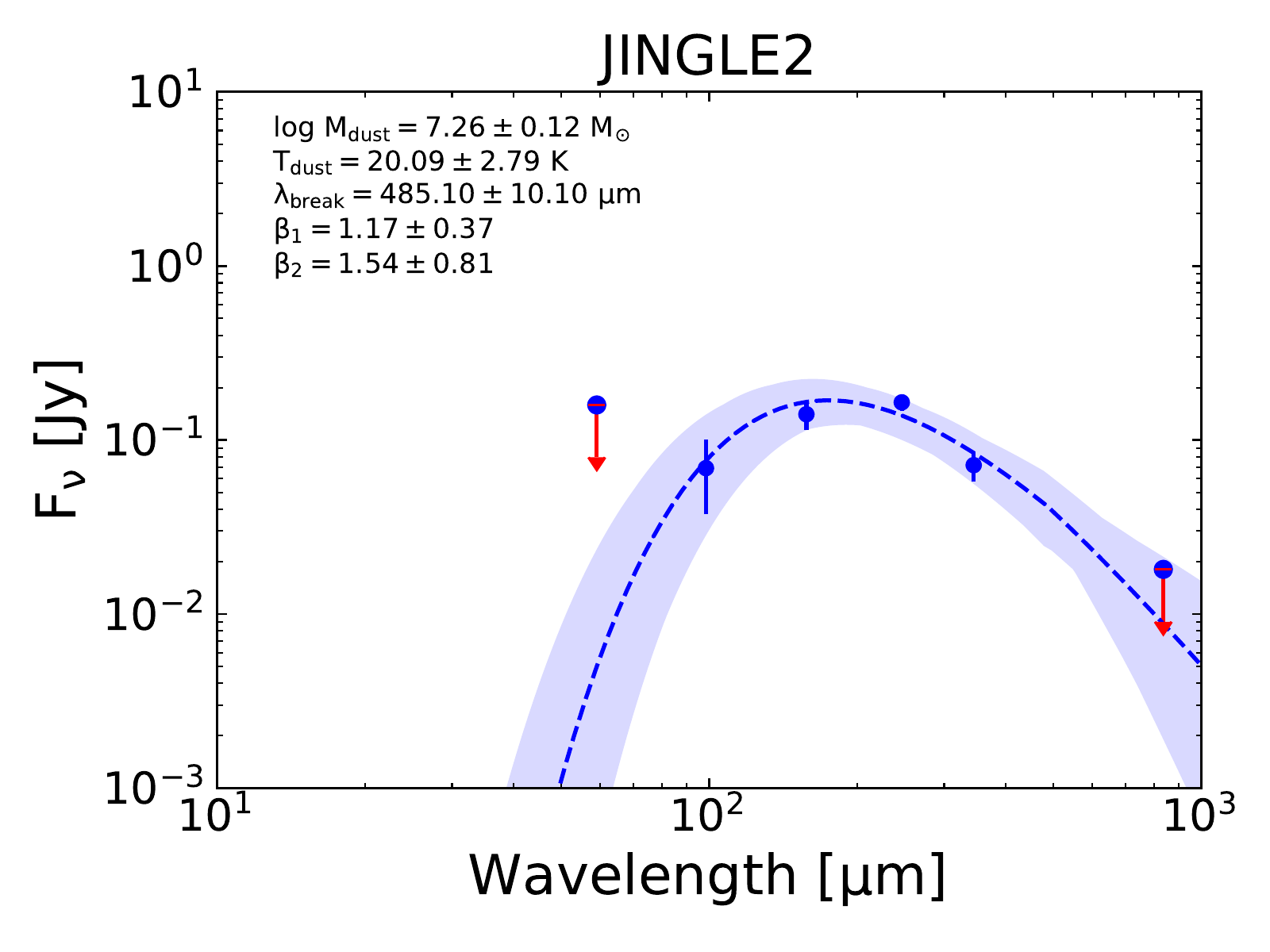}}
\subfigure{\includegraphics[width=0.3\textwidth]
{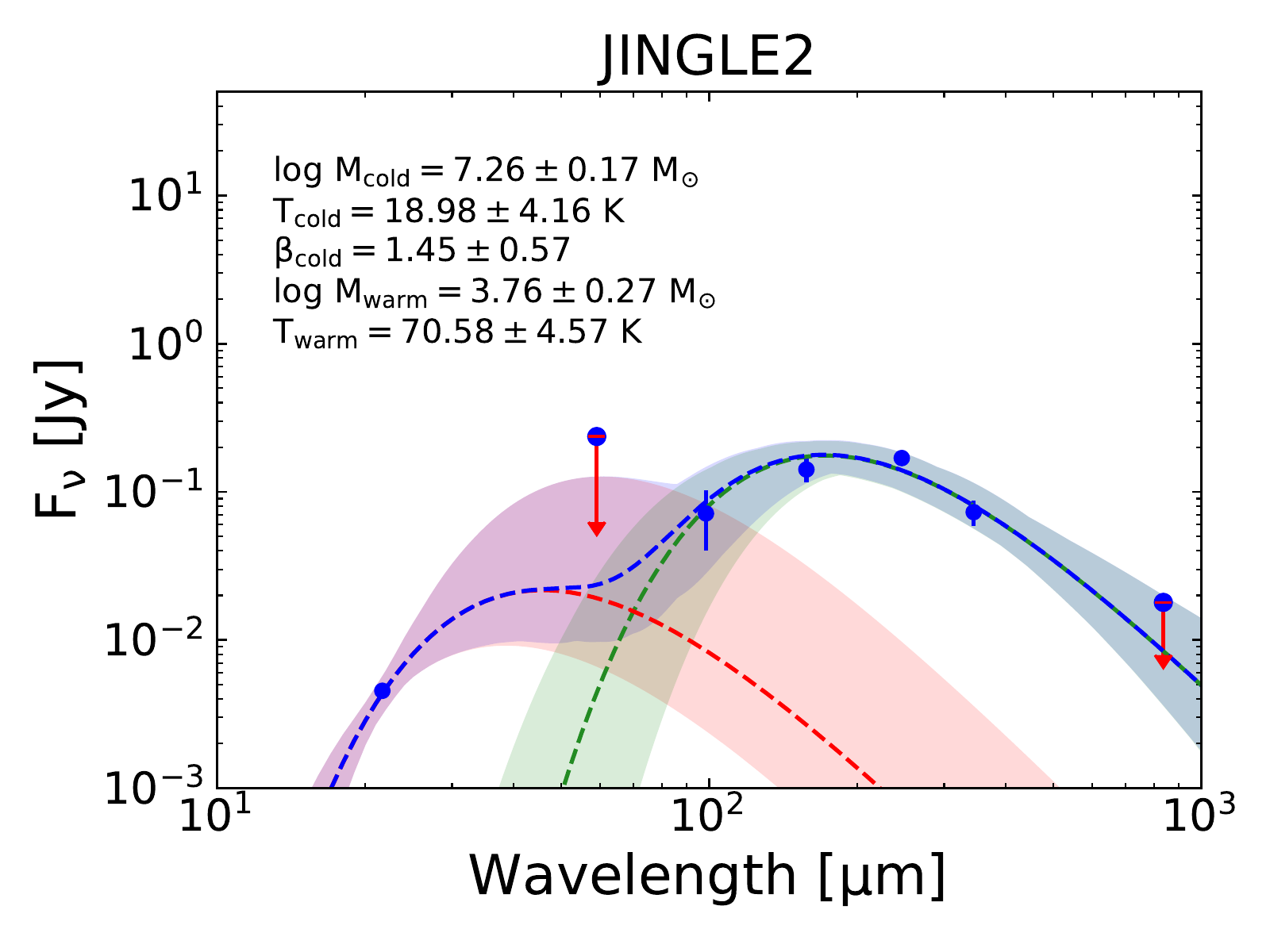}}

\subfigure{\includegraphics[width=0.3\textwidth]
{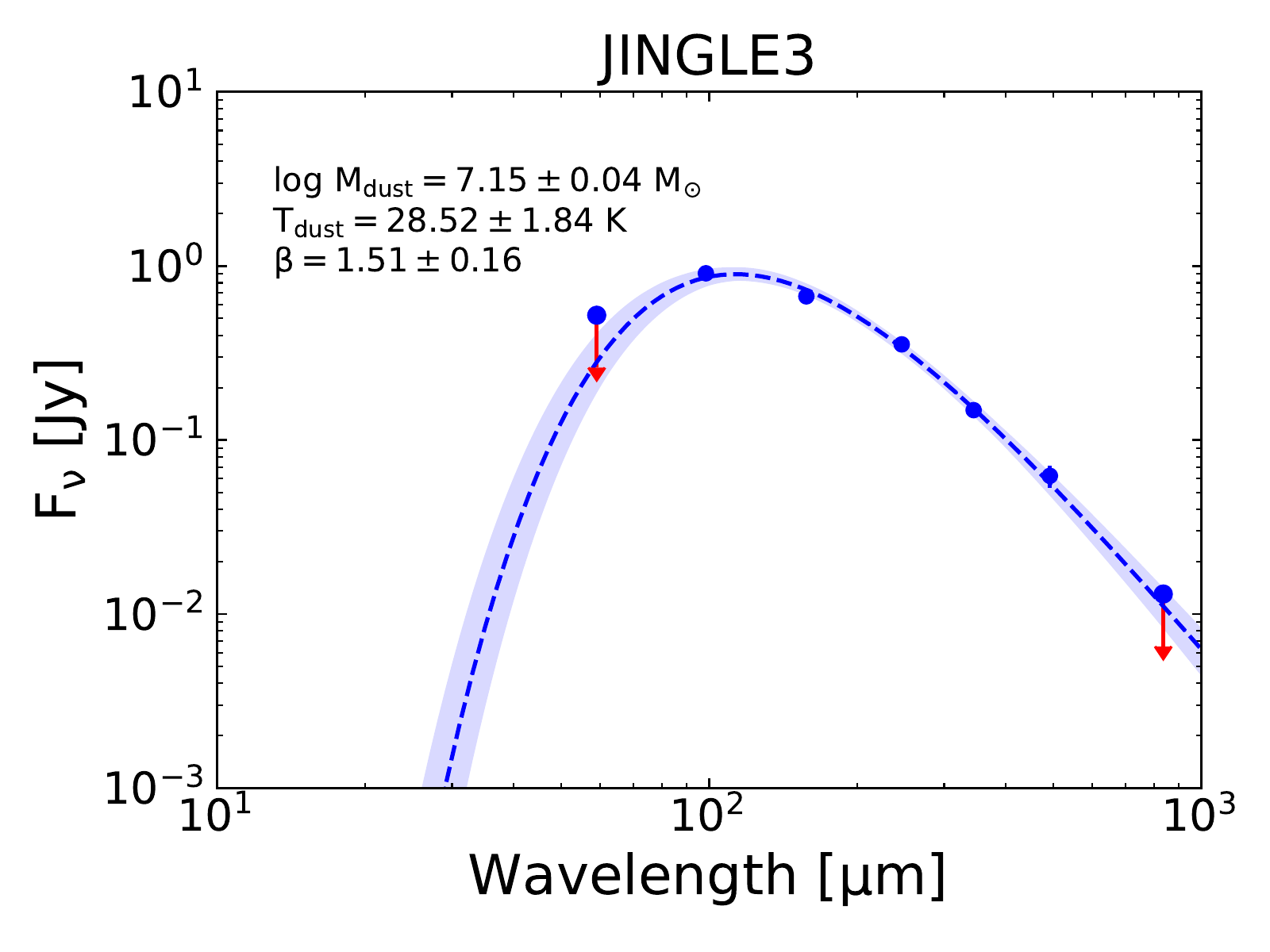}}
\subfigure{\includegraphics[width=0.3\textwidth]
{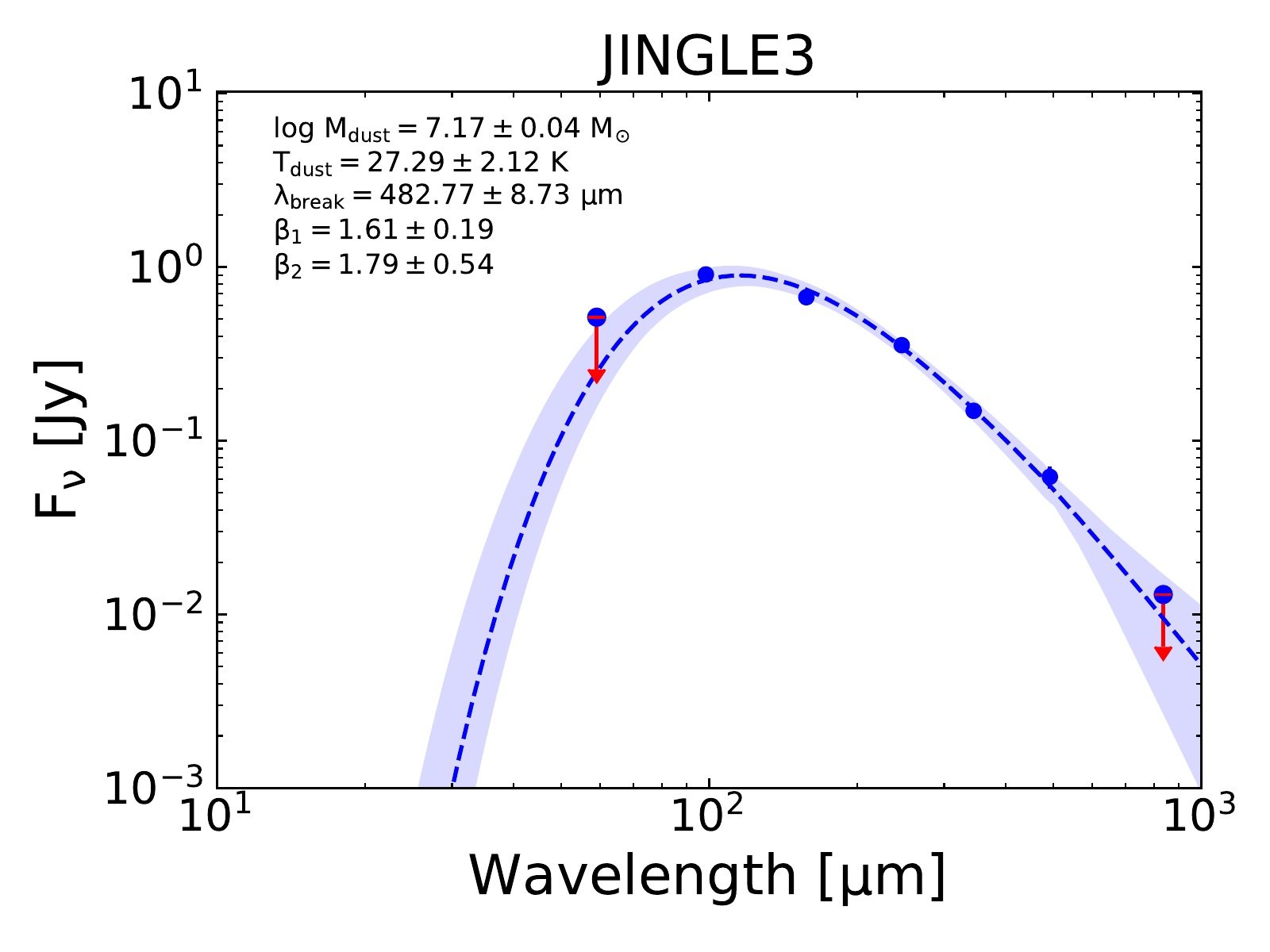}}
\subfigure{\includegraphics[width=0.3\textwidth]
{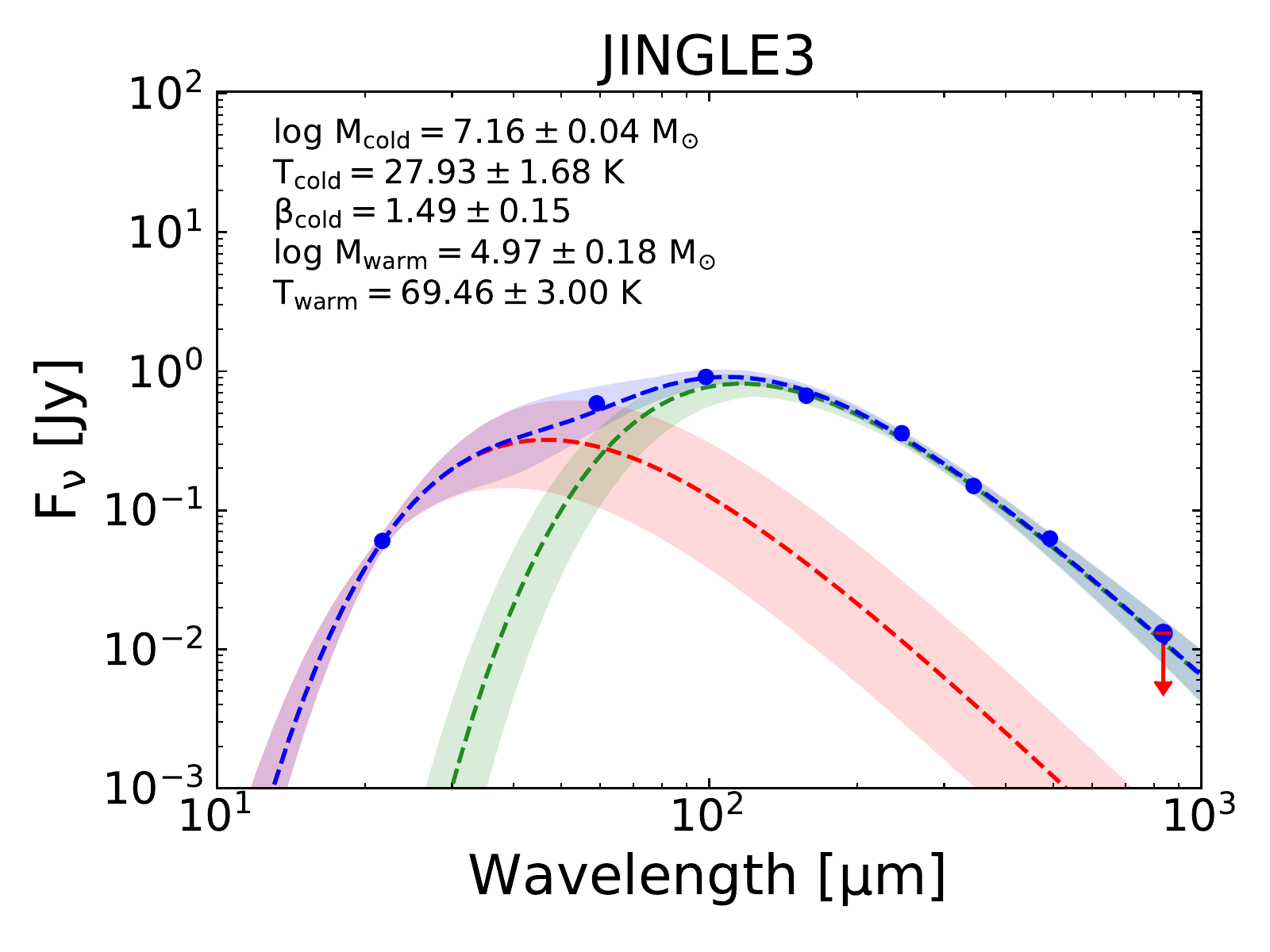}}

\subfigure{\includegraphics[width=0.3\textwidth]
{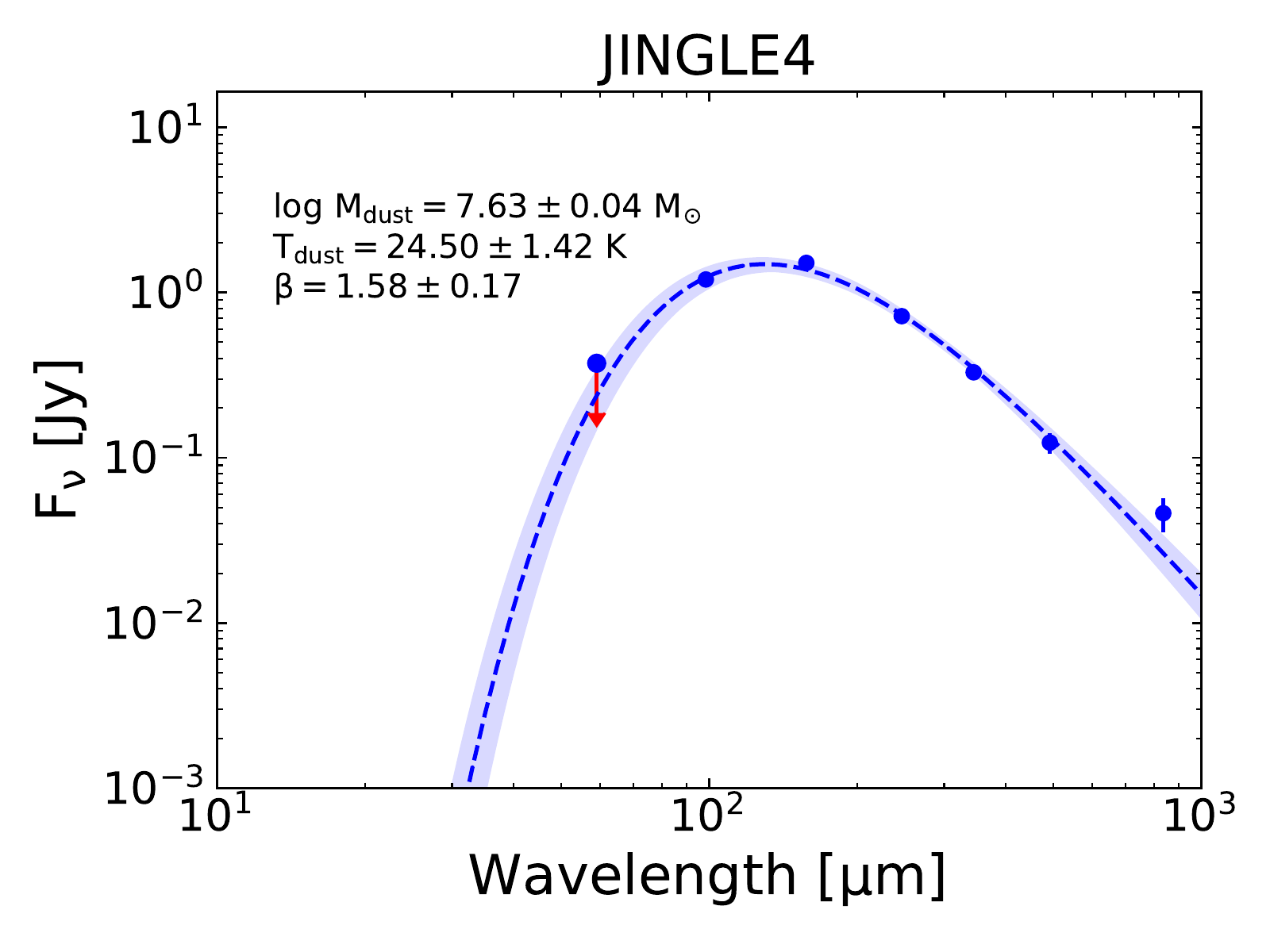}}
\subfigure{\includegraphics[width=0.3\textwidth]
{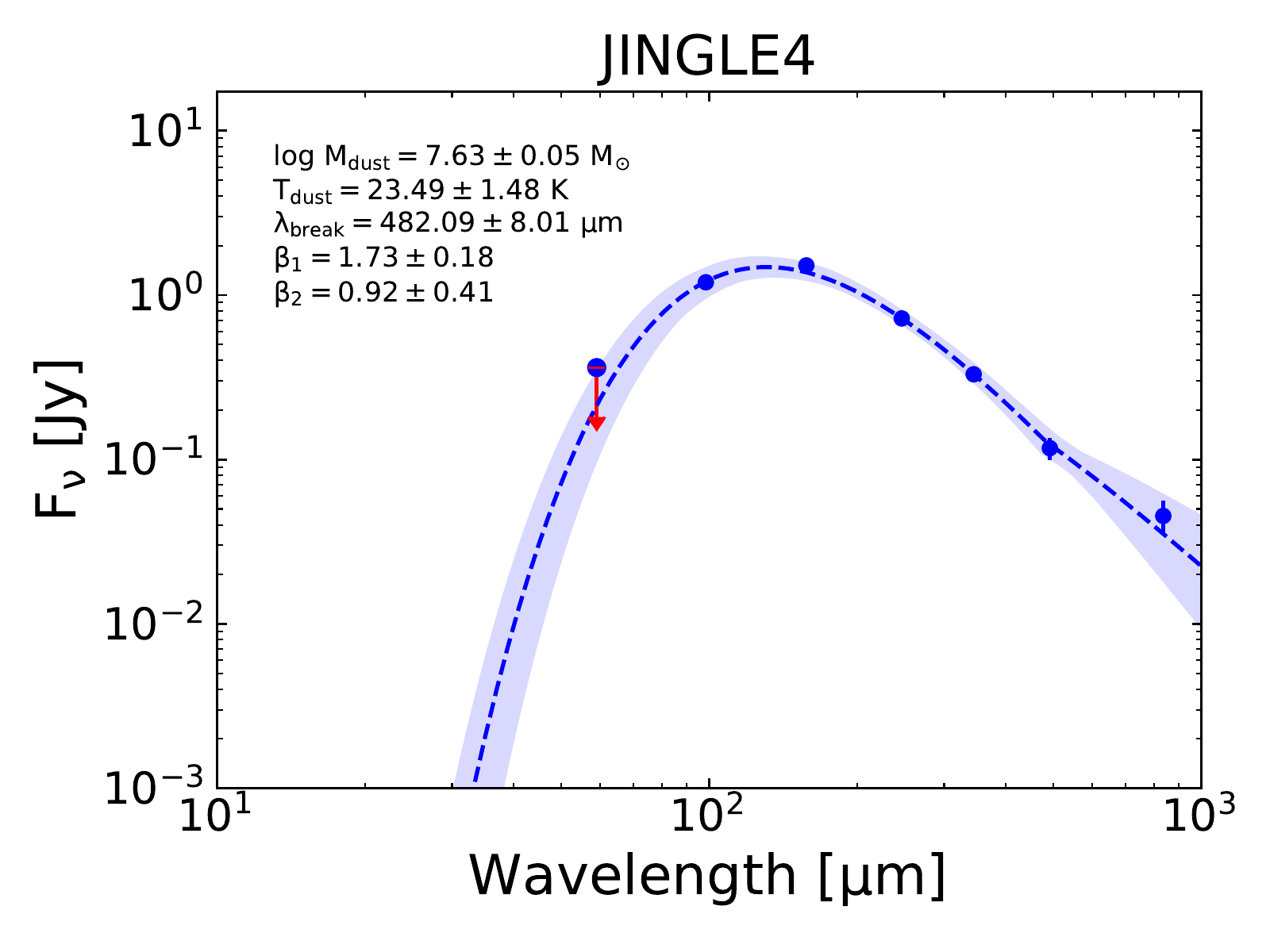}}
\subfigure{\includegraphics[width=0.3\textwidth]
{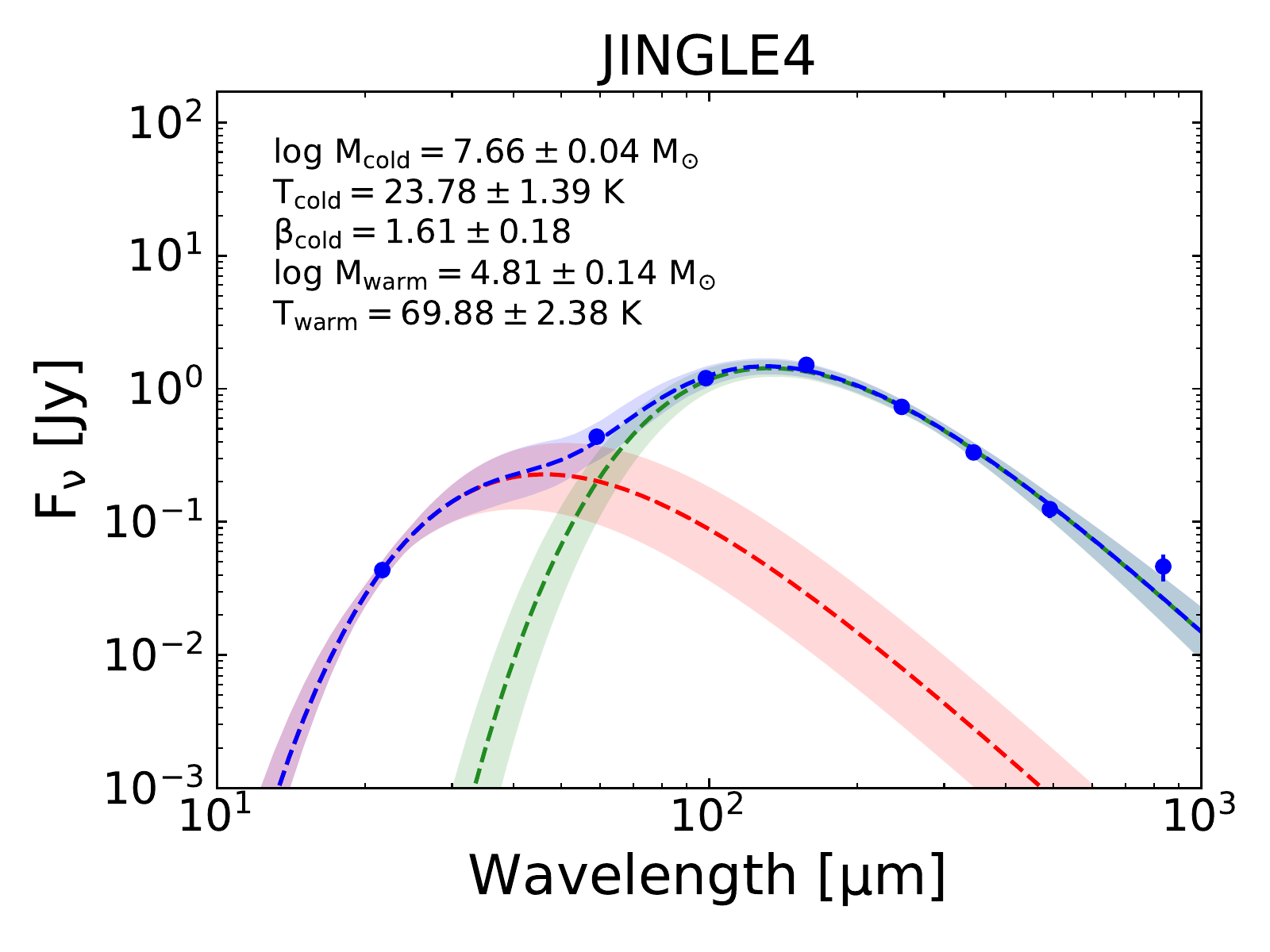}}

\caption{FIR SED of the galaxies of the JINGLE sample, fitted with the hierarchical approach using the three models: SMBB (left panel), BMBB (middle panel) and TMBB(right panel). The shaded regions show the lower and upper 1-sigma uncertainties on the SED models , defined by taking the maximum and minimum flux values of the models with likelihood values in the highest 68th percentile. Additional figures showing the entire sample of 192 JINGLE galaxies are available online.}
\label{fig:three_SED_models} 
\end{figure*}

\section{Tables}

\begin{table*}
\centering
\caption{Result parameters from the hierarchical SED fitting using the single modified black-body (SMBB) model. The parameters of the model are the dust mass ($\log M_{c}$), temperature ($T_{c}$), and emissivity index ($\beta_c$). The last column is the natural logarithm of the likelihood, i.e. the probability of the observed fluxes given the model parameters ($p \left(\vec{F}|\vec{\theta}\right)$). This table is available in its entirety in a machine-readable form in the online journal. A portion is shown here for guidance regarding its form and content.}
\label{tab:results_SMBB}
\begin{tabular}{cccccc} 
\hline
JINGLE ID & SDSS name & $\log M_{c}$  & $T_{c}$ & $\beta_{c}$ &  $\ln L$ \\
& & [\Msun]&  [K] & &
 \\ 
\hline
\hline
0 & J131616.82+252418.7 & 7.25$\pm$0.05 & 27.92$\pm$1.98 & 1.17$\pm$0.18 &11.82 \\
1 & J131453.43+270029.2 & 7.66$\pm$0.07 & 24.33$\pm$2.48 & 0.91$\pm$0.26 &7.28 \\
2 & J131526.03+330926.0 & 7.22$\pm$0.14 & 20.69$\pm$3.78 & 1.15$\pm$0.51 &11.54 \\
3 & J125606.09+274041.1 & 7.15$\pm$0.04 & 28.57$\pm$1.84 & 1.51$\pm$0.16 &14.33 \\
4 & J132134.91+261816.8 & 7.63$\pm$0.04 & 24.50$\pm$1.56 & 1.57$\pm$0.18 &12.13 \\
5 & J091728.99-003714.1 & 7.91$\pm$0.04 & 24.40$\pm$1.47 & 1.52$\pm$0.16 &10.40 \\
6 & J132320.14+320349.0 & 7.40$\pm$0.11 & 21.51$\pm$2.77 & 1.29$\pm$0.29 &16.99 \\
7 & J132051.75+312159.8 & 7.50$\pm$0.05 & 23.84$\pm$1.57 & 1.37$\pm$0.18 &16.89 \\
8 & J091642.17+001220.0 & 7.29$\pm$0.06 & 25.76$\pm$2.15 & 1.24$\pm$0.23 &12.67 \\
9 & J131547.11+315047.1 & 7.60$\pm$0.05 & 23.76$\pm$1.70 & 1.34$\pm$0.19 &15.55 \\
\hline
\end{tabular}
\end{table*}

\begin{table*}
\centering
\caption{Result parameters from the hierarchical SED fitting using the broken emissivity law modified black-body (BMBB) model. The parameters are the dust mass ($\log M_{c}$), temperature ($T_{c}$), emissivity index before the break  ($\beta_{1}$) and after the break ($\beta_{2}$), and the wavelength of the break ($\lambda_{break}$). The last column is the natural logarithm of the likelihood, i.e. the probability of the observed fluxes given the model parameters ($p \left(\vec{F}|\vec{\theta}\right)$). This table is available in its entirety in a machine-readable form in the online journal. A portion is shown here for guidance regarding its form and content.}
\label{tab:results_BMBB}
\begin{tabular}{cccccccc} 
\hline
JINGLE ID & SDSS name & $\log M_{c}$  & $T_{c}$ & $\beta_{1}$ & $\beta_{2}$ & $\lambda_{break}$ & $\ln L$\\
& & [\Msun]&  [K] & & & [\micron]& 
 \\ 
\hline
\hline
0 & J131616.82+252418.7 & 7.25$\pm$0.05 & 25.25$\pm$2.28 & 1.51$\pm$0.23 &0.37$\pm$0.26 & 481.26$\pm$9.20 & 14.59 \\
1 & J131453.43+270029.2 & 7.78$\pm$0.09 & 18.99$\pm$2.18 & 1.67$\pm$0.29 &0.26$\pm$0.23 & 481.72$\pm$13.12 & 10.86 \\
2 & J131526.03+330926.0 & 7.26$\pm$0.12 & 20.09$\pm$2.79 & 1.17$\pm$0.37 &1.54$\pm$0.81 & 485.10$\pm$10.10 & 11.55 \\
3 & J125606.09+274041.1 & 7.17$\pm$0.04 & 27.29$\pm$2.12 & 1.61$\pm$0.19 &1.79$\pm$0.54 & 482.77$\pm$8.73 & 14.52 \\
4 & J132134.91+261816.8 & 7.63$\pm$0.05 & 23.49$\pm$1.48 & 1.73$\pm$0.18 &0.92$\pm$0.41 & 482.09$\pm$8.01 & 13.42 \\
5 & J091728.99-003714.1 & 7.93$\pm$0.04 & 23.01$\pm$1.14 & 1.71$\pm$0.15 &1.37$\pm$0.36 & 483.26$\pm$6.95 & 10.59 \\
6 & J132320.14+320349.0 & 7.44$\pm$0.08 & 22.05$\pm$2.16 & 1.08$\pm$0.25 &2.04$\pm$0.60 & 485.77$\pm$8.11 & 17.54 \\
7 & J132051.75+312159.8 & 7.51$\pm$0.04 & 23.08$\pm$1.25 & 1.46$\pm$0.17 &1.47$\pm$0.47 & 483.62$\pm$7.52 & 16.85 \\
8 & J091642.17+001220.0 & 7.28$\pm$0.10 & 25.23$\pm$2.14 & 1.34$\pm$0.19 &1.19$\pm$0.77 & 483.69$\pm$7.83 & 12.65 \\
9 & J131547.11+315047.1 & 7.62$\pm$0.05 & 23.09$\pm$1.33 & 1.44$\pm$0.22 &1.29$\pm$0.36 & 483.35$\pm$7.59 & 15.65 \\

\hline
\end{tabular}
\end{table*}

\begin{table*}
\centering
\caption{Result parameters from the hierarchical SED fitting using the two modified black-bodies (TMBB) model. The parameters are the dust mass ($\log M_{c}$), temperature ($T_{c}$), and emissivity index ($\beta_{c}$) of the cold dust component, and the dust mass ($\log M_{w}$) and temperature ($T_{w}$) of the warm dust component. The emissivity index of the warm component has been fixed to $\beta_w=1.5$. The last column is the natural logarithm of the likelihood, i.e. the probability of the observed fluxes given the model parameters ($p \left(\vec{F}|\vec{\theta}\right)$). This table is available in its entirety in a machine-readable form in the online journal. A portion is shown here for guidance regarding its form and content.}
\label{tab:results_TMBB}
\begin{tabular}{cccccccc} 
\hline
JINGLE ID & SDSS name & $\log M_{c}$  & $T_{c}$ & $\beta_{c}$ &  $\log M_{w}$  & $T_{w}$ & $\ln L$ \\
& & [\Msun]&  [K] & &[\Msun]&  [K]& 
 \\ 
\hline
\hline
0 & J131616.82+252418.7 & 7.27$\pm$0.04 & 27.09$\pm$1.62 & 1.21$\pm$0.17 &4.51$\pm$0.19 & 71.23$\pm$3.35 & 18.37 \\
1 & J131453.43+270029.2 & 7.67$\pm$0.07 & 24.02$\pm$2.40 & 0.97$\pm$0.28 &4.20$\pm$0.21 & 73.18$\pm$3.94 & 11.78 \\
2 & J131526.03+330926.0 & 7.26$\pm$0.17 & 18.98$\pm$4.16 & 1.45$\pm$0.57 &3.76$\pm$0.27 & 70.58$\pm$4.57 & 18.19 \\
3 & J125606.09+274041.1 & 7.16$\pm$0.04 & 27.93$\pm$1.68 & 1.49$\pm$0.15 &4.97$\pm$0.18 & 69.46$\pm$3.00 & 20.09 \\
4 & J132134.91+261816.8 & 7.66$\pm$0.04 & 23.78$\pm$1.39 & 1.61$\pm$0.18 &4.81$\pm$0.14 & 69.88$\pm$2.38 & 18.57 \\
5 & J091728.99-003714.1 & 7.93$\pm$0.04 & 23.71$\pm$1.30 & 1.55$\pm$0.15 &4.97$\pm$0.14 & 70.26$\pm$2.39 & 15.79 \\
6 & J132320.14+320349.0 & 7.44$\pm$0.12 & 20.53$\pm$2.85 & 1.40$\pm$0.33 &4.00$\pm$0.23 & 70.68$\pm$3.66 & 22.82 \\
7 & J132051.75+312159.8 & 7.54$\pm$0.05 & 22.68$\pm$1.40 & 1.45$\pm$0.18 &4.32$\pm$0.16 & 70.96$\pm$2.82 & 22.50 \\
8 & J091642.17+001220.0 & 7.27$\pm$0.05 & 26.87$\pm$2.29 & 1.14$\pm$0.26 &4.51$\pm$0.19 & 72.35$\pm$3.40 & 17.26 \\
9 & J131547.11+315047.1 & 7.64$\pm$0.05 & 22.74$\pm$1.49 & 1.41$\pm$0.18 &4.40$\pm$0.16 & 71.10$\pm$2.79 & 21.20 \\
\hline
\end{tabular}
\end{table*}

\begin{table*}
\centering
\caption{Results of the analysis of the correlation between  dust emissivity index $\beta$ and combinations of other galaxy properties. The table shows the coefficients $a_j$ of the best polynomial expression $\beta_{model}(x_1, ..., x_k) = \sum_{j=1}^{k} a_j \log (x_j) +b$, to estimate $\beta$  using combinations of two or three galaxy properties.}
\label{tab:corr_beta} 
\begin{tabular}{|l|ccccccccc|}
\hline
\multicolumn{10}{c}{\bf emissivity index $\beta$}\\  
Nr. of param. & $\log M_*$  &  log SFR  & log Area & 12+log(O/H) &  $ \log M_{dust}$&  $ \log M_{HI}$ & intercept & BIC & R \\
 & [M$_\odot$] &[M$_\odot$ yr$^{-1}$]  & [kpc$^2$] &  & [M$_\odot$]  &[M$_\odot$]   &  & (1) & (2) \\
\hline \hline
 2 param. &0.22 $\pm$ 0.03 & -0.01 $\pm$ 0.03 &  &  &  &  & -0.44 $\pm$ 0.25 & 226.29 & 0.53 \\
 &0.42 $\pm$ 0.02 &  & -0.37 $\pm$ 0.03 &  &  &  & -1.97 $\pm$ 0.18 & 53.19 & 0.64 \\
 &0.08 $\pm$ 0.02 &  &  & 0.73 $\pm$ 0.10 &  &  & -5.34 $\pm$ 0.66 & 164.7 & 0.61 \\
 &0.49 $\pm$ 0.04 &  &  &  & -0.31 $\pm$ 0.04 &  & -0.79 $\pm$ 0.15 & 172.34 & 0.55 \\
 &0.39 $\pm$ 0.02 &  &  &  &  & -0.3 $\pm$ 0.02 & 0.64 $\pm$ 0.15 & 68.67 & 0.63 \\
 & & 0.32 $\pm$ 0.02 & -0.22 $\pm$ 0.03 &  &  &  & 2.02 $\pm$ 0.04 & 228.43 & 0.47 \\
 & & 0.04 $\pm$ 0.02 &  & 0.88 $\pm$ 0.08 &  &  & -5.89 $\pm$ 0.7 & 172.41 & 0.61 \\
 & & 0.15 $\pm$ 0.04 &  &  & 0.04 $\pm$ 0.03 &  & 1.43 $\pm$ 0.26 & 299.18 & 0.45 \\
 & & 0.44 $\pm$ 0.03 &  &  &  & -0.35 $\pm$ 0.03 & 5.01 $\pm$ 0.27 & 138.35 & 0.54 \\
 & &  & -0.21 $\pm$ 0.02 & 1.39 $\pm$ 0.07 &  &  & -10.03 $\pm$ 0.63 & 91.25 & 0.62 \\
 & &  & -0.43 $\pm$ 0.03 &  & 0.45 $\pm$ 0.03 &  & -1.15 $\pm$ 0.17 & 151.77 & 0.55 \\
 & &  & 0.05 $\pm$ 0.03 &  &  & -0.01 $\pm$ 0.03 & 1.65 $\pm$ 0.22 & 435.98 & 0.21 \\
 & &  &  & 1.07 $\pm$ 0.09 & -0.03 $\pm$ 0.02 &  & -7.35 $\pm$ 0.65 & 173.7 & 0.61 \\
 & &  &  & 1.19 $\pm$ 0.07 &  & -0.13 $\pm$ 0.02 & -7.42 $\pm$ 0.55 & 131.88 & 0.62 \\
 & &  &  &  & 0.48 $\pm$ 0.03 & -0.43 $\pm$ 0.03 & 2.05 $\pm$ 0.16 & 117.7 & 0.58 \\

 \hline
 3 param. &0.37 $\pm$ 0.03 & 0.08 $\pm$ 0.03 & -0.38 $\pm$ 0.03 &  &  &  & -1.46 $\pm$ 0.26 & 51.04 & 0.65 \\
 &0.09 $\pm$ 0.03 & -0.01 $\pm$ 0.03 &  & 0.73 $\pm$ 0.09 &  &  & -5.46 $\pm$ 0.74 & 169.97 & 0.61 \\
 &0.55 $\pm$ 0.04 & 0.26 $\pm$ 0.04 &  &  & -0.59 $\pm$ 0.06 &  & 0.78 $\pm$ 0.27 & 132.11 & 0.56 \\
 &0.28 $\pm$ 0.02 & 0.23 $\pm$ 0.03 &  &  &  & -0.41 $\pm$ 0.03 & 2.76 $\pm$ 0.32 & 21.40 & 0.66 \\
 &0.28 $\pm$ 0.03 &  & -0.38 $\pm$ 0.03 & 0.8 $\pm$ 0.09 &  &  & -7.48 $\pm$ 0.64 & -14.37 & 0.70 \\
 &0.41 $\pm$ 0.04 &  & -0.37 $\pm$ 0.03 &  & 0.01 $\pm$ 0.05 &  & -1.97 $\pm$ 0.19 & 58.63 & 0.64 \\
 &0.46 $\pm$ 0.02 &  & -0.25 $\pm$ 0.03 &  &  & -0.18 $\pm$ 0.03 & -0.85 $\pm$ 0.26 & 18.08 & 0.67 \\
 &0.39 $\pm$ 0.04 &  &  & 0.85 $\pm$ 0.09 & -0.37 $\pm$ 0.04 &  & -6.67 $\pm$ 0.68 & 96.94 & 0.64 \\
 &0.28 $\pm$ 0.03 &  &  & 0.54 $\pm$ 0.1 &  & -0.28 $\pm$ 0.02 & -3.14 $\pm$ 0.67 & 41.97 & 0.67 \\
 &0.32 $\pm$ 0.04 &  &  &  & 0.11 $\pm$ 0.06 & -0.34 $\pm$ 0.03 & 0.92 $\pm$ 0.22 & 70.94 & 0.63 \\
 & & 0.19 $\pm$ 0.02 & -0.32 $\pm$ 0.03 & 1.15 $\pm$ 0.08 &  &  & -7.77 $\pm$ 0.69 & 33.69 & 0.66 \\
 & & 0.04 $\pm$ 0.04 & -0.42 $\pm$ 0.03 &  & 0.41 $\pm$ 0.05 &  & -0.89 $\pm$ 0.35 & 156.46 & 0.55 \\
 & & 0.46 $\pm$ 0.02 & -0.06 $\pm$ 0.03 &  &  & -0.32 $\pm$ 0.03 & 4.77 $\pm$ 0.28 & 140.08 & 0.54 \\
 & & 0.23 $\pm$ 0.04 &  & 1.17 $\pm$ 0.09 & -0.25 $\pm$ 0.04 &  & -6.49 $\pm$ 0.65 & 142.45 & 0.62 \\
 & & 0.29 $\pm$ 0.03 &  & 0.8 $\pm$ 0.08 &  & -0.33 $\pm$ 0.03 & -2.14 $\pm$ 0.72 & 38.58 & 0.66 \\
 & & 0.23 $\pm$ 0.04 &  &  & 0.3 $\pm$ 0.04 & -0.46 $\pm$ 0.03 & 3.7 $\pm$ 0.31 & 85.63 & 0.60 \\
 & &  & -0.40 $\pm$ 0.03 & 1.00 $\pm$ 0.09 & 0.25 $\pm$ 0.03 &  & -8.25 $\pm$ 0.7 & 38.36 & 0.65 \\
 & &  & -0.21 $\pm$ 0.03 & 1.38 $\pm$ 0.08 &  & 0.00 $\pm$ 0.03 & -10.02 $\pm$ 0.67 & 96.77 & 0.62 \\
 & &  & -0.31 $\pm$ 0.03 &  & 0.63 $\pm$ 0.03 & -0.35 $\pm$ 0.03 & 0.56 $\pm$ 0.22 & 42.80 & 0.63 \\
 & &  &  & 0.63 $\pm$ 0.10 & 0.29 $\pm$ 0.04 & -0.34 $\pm$ 0.03 & -2.89 $\pm$ 0.78 & 84.11 & 0.63 \\

\hline
\hline

\end{tabular}
(1)~BIC: Bayesian Information Criterion \citep{Schwarz1978}, calculated as BIC $= -2\cdot \ln(L) + q\cdot \ln(m)$
where $L$ is the likelihood (i.e. the probability of the data given the parameter $p(\vec{F}|\vec{\theta})$),
 $q$ is the number of free parameters of the model, and $m$ is the number of data points (wavebands).
(2)~Pearson correlation coefficient.

\end{table*}

\begin{table*}
\centering
\caption{Same as table \ref{tab:corr_beta}, but for the correlation between dust temperature $T$ and galaxy properties.}
\label{tab:corr_T}
\begin{tabular}{|l|ccccccccc|}
\hline
\multicolumn{10}{c}{\bf Dust temperature $T$}\\  
Nr. of param. & $\log M_*$  &  log SFR  & log Area & 12+log(O/H) &  $ \log M_{dust}$&  $ \log M_{HI}$ & intercept & BIC & R \\
 & [M$_\odot$] &[M$_\odot$ yr$^{-1}$]  & [kpc$^2$] &  & [M$_\odot$]  &[M$_\odot$]   &  & (1) & (2) \\
\hline \hline

2 param. &-2.14 $\pm$ 0.20 & 2.50 $\pm$ 0.22 &  &  &  &  & 44.24 $\pm$ 1.93 & 914.52 & 0.54 \\
 &0.10 $\pm$ 0.18 &  & -0.59 $\pm$ 0.24 &  &  &  & 22.62 $\pm$ 1.53 & 1046.44 & 0.24 \\
 &0.24 $\pm$ 0.17 &  &  & -2.86 $\pm$ 0.77 &  &  & 45.26 $\pm$ 5.44 & 1038.07 & 0.18 \\
 &-1.07 $\pm$ 0.34 &  &  &  & 0.89 $\pm$ 0.35 &  & 26.50 $\pm$ 1.20 & 1045.98 & 0.14 \\
 &-0.59 $\pm$ 0.17 &  &  &  &  & 0.53 $\pm$ 0.19 & 23.62 $\pm$ 1.27 & 1045.69 & 0.17 \\
 & & 1.93 $\pm$ 0.18 & -2.17 $\pm$ 0.22 &  &  &  & 26.21 $\pm$ 0.33 & 931.29 & 0.53 \\
 & & 1.50 $\pm$ 0.15 &  & -5.74 $\pm$ 0.61 &  &  & 72.73 $\pm$ 5.45 & 948.15 & 0.41 \\
 & & 4.19 $\pm$ 0.29 &  &  & -3.73 $\pm$ 0.30 &  & 51.88 $\pm$ 2.20 & 849.60 & 0.68 \\
 & & 1.78 $\pm$ 0.22 &  &  &  & -1.52 $\pm$ 0.23 & 37.27 $\pm$ 2.12 & 985.74 & 0.36 \\
 & &  & -0.21 $\pm$ 0.18 & -1.69 $\pm$ 0.59 &  &  & 37.74 $\pm$ 5.05 & 1038.7 & 0.22 \\
 & &  & -1.19 $\pm$ 0.30 &  & 0.65 $\pm$ 0.22 &  & 19.53 $\pm$ 1.40 & 1038.33 & 0.25 \\
 & &  & -1.38 $\pm$ 0.24 &  &  & 1.00 $\pm$ 0.22 & 15.42 $\pm$ 1.78 & 1026.52 & 0.29 \\
 & &  &  & -3.47 $\pm$ 0.72 & 0.45 $\pm$ 0.18 &  & 49.50 $\pm$ 5.74 & 1032.98 & 0.17 \\
 & &  &  & -2.65 $\pm$ 0.60 &  & 0.33 $\pm$ 0.15 & 42.65 $\pm$ 4.65 & 1035.02 & 0.18 \\
 & &  &  &  & -0.54 $\pm$ 0.22 & 0.54 $\pm$ 0.25 & 21.84 $\pm$ 1.24 & 1051.14 & 0.14 \\

\hline

3 param. &-1.56 $\pm$ 0.23 & 2.83 $\pm$ 0.22 & -1.40 $\pm$ 0.25 &  &  &  & 40.58 $\pm$ 2.08 & 887.83 & 0.61 \\
 &-1.65 $\pm$ 0.23 & 2.52 $\pm$ 0.21 &  & -2.89 $\pm$ 0.74 &  &  & 64.51 $\pm$ 5.56 & 905.66 & 0.55 \\
 &0.19 $\pm$ 0.37 & 4.24 $\pm$ 0.29 &  &  & -3.96 $\pm$ 0.48 &  & 51.72 $\pm$ 2.11 & 854.86 & 0.68 \\
 &-1.92 $\pm$ 0.21 & 3.04 $\pm$ 0.25 &  &  &  & -0.97 $\pm$ 0.24 & 51.26 $\pm$ 2.56 & 903.19 & 0.58 \\
 &0.64 $\pm$ 0.22 &  & -0.65 $\pm$ 0.24 & -2.98 $\pm$ 0.72 &  &  & 43.22 $\pm$ 5.51 & 1036.39 & 0.25 \\
 &-1.24 $\pm$ 0.34 &  & -1.3 $\pm$ 0.29 &  & 1.9 $\pm$ 0.41 &  & 22.31 $\pm$ 1.50 & 1030.77 & 0.26 \\
 &-0.20 $\pm$ 0.19 &  & -1.25 $\pm$ 0.27 &  &  & 1.07 $\pm$ 0.24 & 16.49 $\pm$ 1.95 & 1031.0 & 0.29 \\
 &-0.64 $\pm$ 0.36 &  &  & -3.11 $\pm$ 0.79 & 1.02 $\pm$ 0.33 &  & 48.37 $\pm$ 5.39 & 1035.13 & 0.18 \\
 &-0.04 $\pm$ 0.24 &  &  & -2.57 $\pm$ 0.75 &  & 0.35 $\pm$ 0.21 & 42.16 $\pm$ 5.75 & 1040.49 & 0.19 \\
 &-0.94 $\pm$ 0.36 &  &  &  & 0.53 $\pm$ 0.46 & 0.32 $\pm$ 0.28 & 25.04 $\pm$ 1.74 & 1050.03 & 0.16 \\
 & & 2.43 $\pm$ 0.19 & -1.85 $\pm$ 0.22 & -4.66 $\pm$ 0.63 &  &  & 66.19 $\pm$ 5.50 & 885.12 & 0.61 \\
 & & 4.09 $\pm$ 0.29 & -0.50 $\pm$ 0.29 &  & -3.32 $\pm$ 0.38 &  & 49.48 $\pm$ 2.51 & 851.99 & 0.69 \\
 & & 2.15 $\pm$ 0.22 & -1.94 $\pm$ 0.26 &  &  & -0.47 $\pm$ 0.27 & 30.33 $\pm$ 2.38 & 933.86 & 0.54 \\
 & & 4.06 $\pm$ 0.29 &  & -1.85 $\pm$ 0.75 & -3.31 $\pm$ 0.31 &  & 64.70 $\pm$ 5.44 & 848.76 & 0.68 \\
 & & 2.85 $\pm$ 0.23 &  & -6.20 $\pm$ 0.62 &  & -1.71 $\pm$ 0.22 & 92.99 $\pm$ 6.09 & 897.60 & 0.57 \\
 & & 4.19 $\pm$ 0.31 &  &  & -3.70 $\pm$ 0.32 & -0.06 $\pm$ 0.27 & 52.12 $\pm$ 2.55 & 855.07 & 0.68 \\
 & &  & -1.42 $\pm$ 0.29 & -4.09 $\pm$ 0.71 & 1.49 $\pm$ 0.26 &  & 48.84 $\pm$ 4.95 & 1013.67 & 0.30 \\
 & &  & -1.10 $\pm$ 0.27 & -1.74 $\pm$ 0.60 &  & 1.01 $\pm$ 0.22 & 30.02 $\pm$ 5.51 & 1023.57 & 0.29 \\
 & &  & -1.46 $\pm$ 0.29 &  & 0.13 $\pm$ 0.26 & 0.93 $\pm$ 0.26 & 15.14 $\pm$ 1.93 & 1031.79 & 0.29 \\
 & &  &  & -3.47 $\pm$ 0.81 & 0.46 $\pm$ 0.34 & -0.01 $\pm$ 0.29 & 49.36 $\pm$ 6.84 & 1038.49 & 0.17 \\

\hline
\hline

\end{tabular}
(1)~BIC: Bayesian Information Criterion \citep{Schwarz1978}, calculated as BIC $= -2\cdot \ln(L) + q\cdot \ln(m)$
where $L$ is the likelihood (i.e. the probability of the data given the parameter $p(\vec{F}|\vec{\theta})$),
 $q$ is the number of free parameters of the model, and $m$ is the number of data points (wavebands).
(2)~Pearson correlation coefficient.
\end{table*}

\bsp	
\label{lastpage}
\end{document}